\RequirePackage{rotating}
\documentclass[twocolumn, tighten]{aastex631}
\usepackage{amsmath}
\usepackage[caption=false]{subfig}
\usepackage{xfrac}
\usepackage{CJK}

\shorttitle{DESTINYS: DO Tau}
\shortauthors{Huang et al.}

\begin{document}
\begin{CJK*}{UTF8}{gbsn} 
\title{Disk Evolution Study Through Imaging of Nearby Young Stars
(DESTINYS): A Panchromatic View of DO Tau's Complex Kilo-au Environment}

\correspondingauthor{Jane Huang}
\email{jnhuang@umich.edu}
\author[0000-0001-6947-6072]{Jane Huang}
\altaffiliation{NASA Hubble Fellowship Program Sagan Fellow}
\affiliation{Department of Astronomy, University of Michigan, 323 West Hall, 1085 S. University Avenue, Ann Arbor, MI 48109, United States of America}
\author[0000-0002-4438-1971]{Christian Ginski}
\affiliation{Anton Pannekoek Institute for Astronomy, University of Amsterdam, Science Park 904, 1098XH Amsterdam, The Netherlands}
\affiliation{Leiden Observatory, Leiden University, PO Box 9513, 2300 RA, Leiden, The Netherlands}
\author[0000-0002-7695-7605]{Myriam Benisty}
\affiliation{Unidad Mixta Internacional Franco-Chilena de Astronom\'{i}a, CNRS/INSU UMI 3386, Departamento de Astronom\'ia, Universidad de Chile, Camino El Observatorio 1515, Las Condes, Santiago, Chile}
\affiliation{Univ. Grenoble Alpes, CNRS, IPAG, 38000 Grenoble, France}
\author[0000-0003-1698-9696]{Bin Ren (任彬)}
\affiliation{Department of Astronomy, California Institute of Technology, MC 249-17, 1200 East California Boulevard, Pasadena, CA 91125, USA}
\author[0000-0003-1401-9952]{Alexander J. Bohn}
\affiliation{Leiden Observatory, Leiden University, PO Box 9513, 2300 RA, Leiden, The Netherlands}
\author[0000-0002-9173-0740]{\'Elodie Choquet}
\affiliation{Aix Marseille Univ, CNRS, CNES, LAM, Marseille, France}
\author[0000-0001-8798-1347]{Karin I. \"Oberg} \affiliation{Center for Astrophysics \textbar\ Harvard \& Smithsonian, 60 Garden St., Cambridge, MA 02138, United States of America}
\author[0000-0003-3133-3580]{\'Alvaro Ribas}
\affiliation{European Southern Observatory, Alonso de C\'ordova 3017, Vitacura, Santiago, Chile}
\author[0000-0001-7258-770X]{Jaehan Bae}
\affiliation{Department of Astronomy, University of Florida, Gainesville, FL 32611, United States of America}
\author[0000-0003-4179-6394]{Edwin A. Bergin}
\affiliation{Department of Astronomy, University of Michigan, 323 West Hall, 1085 S. University Avenue, Ann Arbor, MI 48109, United States of America}
\author[0000-0002-1899-8783
]{Til Birnstiel}
\affiliation{University Observatory, Faculty of Physics, Ludwig-Maximilians-Universit\"at M\"unchen, Scheinerstr. 1, 81679 Munich, Germany}
\affiliation{Exzellenzcluster ORIGINS, Boltzmannstr. 2, D-85748 Garching, Germany}
\author[0000-0002-8692-8744]{Yann Boehler}
\affiliation{Univ. Grenoble Alpes, CNRS, IPAG, 38000 Grenoble, France}
\author[0000-0003-4689-2684]{Stefano Facchini}
\affiliation{Universit\`a degli Studi di Milano, via Giovanni Celoria 16, 20133 Milano, Italy}
\affiliation{European Southern Observatory, Karl-Schwarzschild-Strasse 2, 85748 Garching bei M\"unchen, Germany}
\author[0000-0001-6307-4195]{Daniel Harsono}
\affiliation{Institute of Astronomy, Department of Physics, National Tsing Hua
University, Hsinchu, Taiwan}
\author[0000-0001-5217-537X]{Michiel Hogerheijde}
\affiliation{Leiden Observatory, Leiden University, PO Box 9513, 2300 RA, Leiden, The Netherlands}
\affiliation{Anton Pannekoek Institute for Astronomy, University of Amsterdam, Science Park 904, 1098XH Amsterdam, The Netherlands}
\author[0000-0002-7607-719X]{Feng Long}
\affiliation{Center for Astrophysics \textbar\ Harvard \& Smithsonian, 60 Garden St., Cambridge, MA 02138, United States of America}
\author[0000-0003-3562-262X]{Carlo F. Manara}
\affiliation{European Southern Observatory, Karl-Schwarzschild-Strasse 2, 85748 Garching bei M\"unchen, Germany}
\author[0000-0002-1637-7393]{Fran\c{c}ois M\'enard}
\affiliation{Univ. Grenoble Alpes, CNRS, IPAG, 38000 Grenoble, France}
\author[0000-0001-8764-1780]{Paola Pinilla}
\affiliation{Max-Planck-Institut f\"ur Astronomie, K\"onigstuhl 17, 69117, Heidelberg, Germany}
\affiliation{Mullard Space Science Laboratory, University College London, Holmbury St Mary, Dorking, Surrey RH5 6NT, UK}
\author[0000-0001-5907-5179]{Christophe Pinte}
\affiliation{School of Physics and Astronomy, Monash University, Clayton, Vic 3800, Australia}
\affiliation{Univ. Grenoble Alpes, CNRS, IPAG, 38000 Grenoble, France}
\author[0000-0003-1817-6576]{Christian Rab}
\affiliation{University Observatory, Faculty of Physics, Ludwig-Maximilians-Universit\"at M\"unchen, Scheinerstr. 1, 81679 Munich, Germany}
\affiliation{Max-Planck-Institut f\"ur extraterrestrische Physik, Giessenbachstrasse 1, 85748 Garching, Germany}
\author[0000-0001-5058-695X]{Jonathan P. Williams}
\affiliation{Institute for Astronomy, University of Hawaii, Honolulu, HI 96822, USA}
\author[0000-0002-5903-8316]{Alice Zurlo}
\affiliation{N\'ucleo de Astronom\'ia, Facultad de Ingenier\'ia y Ciencias, Universidad Diego Portales, Av. Ejercito 441, Santiago, Chile}
\affiliation{Escuela de Ingenier\'ia Industrial, Facultad de Ingenier\'ia y Ciencias, Universidad Diego Portales, Av. Ejercito 441, Santiago, Chile}
\affiliation{Aix Marseille Univ, CNRS, CNES, LAM, Marseille, France}

\begin{abstract}
 While protoplanetary disks are often treated as isolated systems in planet formation models, observations increasingly suggest that vigorous interactions between Class II disks and their environments are not rare. DO Tau is a T Tauri star that has previously been hypothesized to have undergone a close encounter with the HV Tau system. As part of the DESTINYS ESO Large Programme, we present new VLT/SPHERE polarimetric observations of DO Tau and combine them with archival HST scattered light images and ALMA observations of CO isotopologues and CS to map a network of complex structures. The SPHERE and ALMA observations show that the circumstellar disk is connected to arms extending out to several hundred au. HST and ALMA also reveal stream-like structures northeast of DO Tau, some of which are at least several thousand au long. These streams appear not to be gravitationally bound to DO Tau, and comparisons with previous Herschel far-IR observations suggest that the streams are part of a bridge-like structure connecting DO Tau and HV Tau. We also detect a fainter redshifted counterpart to a previously known blueshifted CO outflow. While some of DO Tau's complex structures could be attributed to a recent disk-disk encounter, they might be explained alternatively by interactions with remnant material from the star formation process. These panchromatic observations of DO Tau highlight the need to contextualize the evolution of Class II disks by examining processes occurring over a wide range of size scales. \end{abstract}


\section{Introduction} \label{sec:intro}

Planet formation has often been modelled as taking place in isolated protoplanetary disks \citep[e.g.,][]{2012AA...544A..32L, 2012AA...547A.111M, 2018ApJ...869L..47Z}. However, observations are increasingly challenging the general applicability of such an assumption. Gaps and rings have been detected in the millimeter continuum emission of Class I disks, potentially indicative of planet-disk interactions in these young embedded systems \citep[e.g.,][]{2018ApJ...857...18S, 2020ApJ...904L...6A, 2020Natur.586..228S}. Meanwhile, molecular and scattered light observations have revealed large-scale streams, spirals, and tails associated with the ostensibly more evolved Class II protoplanetary disks, suggestive of ongoing infall or close encounters with other stars \citep[e.g.,][]{1995AJ....109.1181N, 2006AA...452..897C, 2012AA...547A..84T, 2019AJ....157..165A, 2021ApJS..257...19H}. Together, these observations indicate that many disks are perturbed by their surroundings while planet formation is ongoing. Key theorized consequences of these external perturbations include modification of the total mass budget available for planet formation, disk truncation, inducement of misalignments or instabilities, formation of disk substructures, and stellar outbursts \citep[e.g.,][]{1993MNRAS.261..190C,2015ApJ...805...15B, 2011MNRAS.413..423H, 2018AA...618L...3M, 2019AA...628A..20D, 2019MNRAS.483.4114C, 2020AA...633A...3K,  2021AA...656A.161K, 2022ApJ...928...92K}. Thus, mapping disk environments is crucial for understanding what processes control protoplanets' formation location and timescale, mass, orbital behavior, and survival. 

High-resolution panchromatic observations are essential for characterizing interactions between disks and their environments. Millimeter continuum emission primarily traces the distribution of millimeter-sized dust grains and is therefore a good probe of the bulk of the disk mass reservoir. However, because grain growth is most effective in dense regions and larger dust grains are more susceptible to inward radial drift and vertical settling, millimeter continuum does not trace extended, more tenuous structures well. Optical and infrared scattered light trace the distribution of sub-micron-sized dust grains. In the context of disk observations, they have primarily been used to characterize the surface layers of disks \citep[e.g.,][]{2000ApJ...538..793K, 2004ApJ...605L..53F, 2018ApJ...863...44A}. However, scattered light imaging has also proven effective for identifying large-scale structures that do not have detected millimeter continuum counterparts \citep[e.g.,][]{1999ApJ...523L.151G,2020AA...633A..82G}. Furthermore, because the strength of the scattered light signal and the degree of linear polarization depend in part on the scattering angle, they can be used to infer how structures are oriented relative to the illuminating star \citep[e.g.,][]{2016AA...596A..70S,2021ApJ...908L..25G}. Although small dust grains and gas are thought to be well-coupled, molecular line observations can still reveal material not visible in scattered light because the detectability of structures with scattered light is affected by disk self-shadowing and the rapid decrease in stellar illumination with distance \citep[e.g.,][]{2020ApJ...898..140H, 2022AA...658A.137G}. Molecular line observations are also critical for probing kinematics and gas mass. 

The Disk Evolution Study Through Imaging of Nearby Young Stars (DESTINYS) Large Programme has been conducting a near-infrared polarimetric survey of young stars with the SPHERE instrument on the Very Large Telescope \citep{2020AA...642A.119G, 2021ApJ...908L..25G}. Whereas early scattered light studies disproportionately targeted Herbig Ae stars \citep[see, e.g.,][and references therein]{2018AA...620A..94G}, DESTINYS aims to provide a comprehensive understanding of the disk structures of lower-mass pre-main sequence stars. While the original primary aims of DESTINYS were to characterize dust evolution in disks, identify small-scale substructures that may be signs of planet-disk interactions, and search for stellar and planetary-mass companions, DESTINYS has serendipitously provided insights into interactions between disks and their environments, as demonstrated by the detection of SU Aur's spectacular infalling tail structures and misaligned disk in \citet{2021ApJ...908L..25G}. 

Another DESTINYS target exhibiting signs of interactions with its surroundings is DO Tau, an M0.3 T Tauri star located $138.4\pm0.7$ pc away in the Taurus star-forming region \citep{2014ApJ...786...97H, 2021AA...649A...1G, 2021AJ....161..147B}. It hosts a Class II protoplanetary disk associated with a jet and molecular outflow \citep[e.g.,][]{1994ApJ...427L..99H, 1995ApJS..101..117K, 2020AJ....159..171F}. Coronagraphic Hubble Space Telescope optical and Subaru infrared images revealed an arc-like structure extending up to a few hundred au north of the star \citep{2004ASPC..321..244G, 2008PASJ...60..223I}. In projection, DO Tau is situated $\sim91''$ (12,600 au) northwest of the T Tauri triple star system HV Tau. HV Tau A and B have a spectral type of M1 and have a projected separation from one another of 10 au \citep{1988cels.book.....H, 1996ApJ...469..890S}. HV Tau C is an M0.5 star that hosts an edge-on protoplanetary disk and has a projected separation of $\sim550$ au from HV Tau AB \citep[e.g.,][]{1998AA...338..122W, 2010ApJ...712..112D}. DO Tau and HV Tau are surrounded by nebulosities visible at both optical and far-infrared wavelengths \citep[e.g.,][]{1962PASP...74..474S, 2004AA...420..975M, 2013ApJ...776...21H}. Based on the morphology of these nebulosities, \citet{2018MNRAS.479.5522W} hypothesized that DO Tau and HV Tau originally constituted a quadruple system and underwent a close encounter that stripped material from their disks and ejected DO Tau. 

In this work, we combine new SPHERE observations of DO Tau with ALMA molecular line observations and archival optical and infrared images in order to investigate the spatial and kinematic relationship between DO Tau's disk and its complex surroundings. The observations and data reduction are summarized in Section \ref{sec:observations}. The scattered light structures are described in Section \ref{sec:scatteredlight}, and the molecular emission structures are described and compared to the scattered light structures in Section \ref{sec:molecularlines}. The properties and potential origins of the structures are discussed in Section \ref{sec:discussion}, and the results are summarized in Section \ref{sec:summary}.  

\section{Observations and Data Reduction}\label{sec:observations}

In this section, we describe the various new and archival observations analyzed in this work. The observations are also listed in Table \ref{tab:allobservations}. 

\begin{deluxetable*}{cccc}
\tablecaption{DO Tau observations analyzed in this work \label{tab:allobservations}}
\tablehead{
\colhead{Facility}&\colhead{Program}&\colhead{Observation type\tablenotemark{a}}&\colhead{Reference}}
\startdata
 \multicolumn{4}{c}{Main text} \\
 \hline
VLT/SPHERE & 0104.C-0850(A) & $H-$band polarimetric and total intensity imaging & 1 \\
VLT/SPHERE & 1104.C-0415(E) (DESTINYS)& $H-$band polarimetric and total intensity imaging & 1 \\
ALMA & 2016.1.00627.S & $^{12}$CO, $^{13}$CO, C$^{18}$O, and CS line imaging& 1, 2, 3, 4 \\
HST/STIS & HST-GO-9136 & Broadband optical imaging & 5, 6\\
Apache Point Observatory & Sloan Digital Sky Survey & $g$, $r$, $i$-band imaging & 7, 8\\
Herschel PACS & KPOT\_bdent\_1 (GASPS) & 160 $\mu$m imaging & 9, 10 \\
\hline
 \multicolumn{4}{c}{Appendix only} \\
 \hline
 HST/NICMOS & HST-GO-7418 & F110W and F160W imaging & 11, 12\\
 Subaru/CIAO & o05146 & $H$-band imaging & 13
\enddata
\tablenotetext{a}{This column describes only the observations that appear in this work, not necessarily all observations obtained by the program.}
\tablerefs{(1) This work (2) \citealt{2019ApJ...876...25B} (3) \citealt{2019ApJ...876...72L} (4) \citealt{2020ApJ...890..142P}, (5) \citealt{2004ASPC..321..244G}} (6) \citealt{2017SPIE10400E..21R}, (7) \citealt{2000AJ....120.1579Y}, (8) \citealt{2004AJ....128.2577F}, (9) \citealt{2013PASP..125..477D}, (10) \citealt{2013ApJ...776...21H}, (11) \citealt{2014SPIE.9143E..57C}, (12) \citealt{2018AJ....155..179H}, (13) \citealt{2008PASJ...60..223I}
\end{deluxetable*}

\subsection{SPHERE/IRDIS Observations}

DO Tau was observed with the Very Large Telescope's SPHERE/IRDIS instrument \citep{2019AA...631A.155B} using the broadband $H$ filter (central wavelength: 1.6255 $\mu$m) on 2019 November 27 (program I.D. 0104.C-0850(A), P.I.: Y. Boehler) and 2019 December 20 (corresponding to the DESTINYS Large Programme with I.D. 1104.C-0415(E), P.I.: C. Ginski). Both programs used dual-beam polarimetric imaging (DPI) with pupil tracking \citep{2020AA...633A..63D, 2020AA...633A..64V}. In each epoch, the main observing sequence consisted of 56 frames with the DO Tau star behind an apodized Lyot coronagraph, which has an inner working angle of 92.5 mas \citep{2011ExA....30...39C, 2011ExA....30...59G}. The integration time of each frame was 64 seconds, for a total time of 59.7 minutes per epoch. Each coronagraphic sequence was preceded and followed by 5 flux calibration frames, in which DO Tau was observed offset from the coronagraph and with the ND1.0 neutral density filter to prevent saturation. Each flux calibration frame lasted for 0.84 seconds, for a total integration time of 8.4 seconds per epoch. Weather conditions were excellent on both nights. On 2019 November 27, the typical seeing was $0.6''$ and the typical coherence time was 8 ms. On 2019 December 20, the typical seeing was $0.5''$ and the typical coherence time was 10 ms. 

The data from each night were individually reduced with the publicly available IRDIS Data reduction for Accurate Polarimetry (IRDAP) pipeline \citep{2017SPIE10400E..15V, 2020AA...633A..64V, 2020ascl.soft04015V}, yielding polarized intensity images through polarimetric differential imaging (PDI) and total intensity images through classical angular differential imaging (cADI). After checking that results from the two nights were consistent, the reduced images were then averaged together to increase the signal-to-noise ratio of the faint extended features. 

Since ADI results in self-subtraction artifacts, we also produced total intensity images of DO Tau using reference differential imaging (RDI). DESTINYS Large Programme observations of IP Tau (taken on 2019 December 15) were used to create a PSF reference library. The observational setup and weather conditions for IP Tau were similar to those for DO Tau. As an M0.6 T Tauri star located in the Taurus star-forming region \citep{2014ApJ...786...97H}, IP Tau provides a close color match to DO Tau. However, since IP Tau also has a disk, using it as a PSF reference is only useful for recovering structures beyond an arcsecond or so from the star. PSF subtraction was performed with the Karhunen-Lo\`eve Image Projection (KLIP) algorithm described in \citet{2012ApJ...755L..28S}. Since the standard KLIP procedure results in overestimation of the stellar signal when extended structures are present, we applied the iterative disk feedback reference differential imaging (IDF-RDI) approach presented in Vaendel et al. (in prep) and \citet{2021ApJ...908L..25G}. In brief, the PSF-subtracted image produced with KLIP is treated as an initial estimate of the non-stellar signal. This image is subtracted from the original stack of science images of DO Tau (i.e., the non-PSF-subtracted images) in order to remove as much non-stellar signal as possible. KLIP is then used on the modified science image stack to obtain an improved estimate of the stellar signal, and PSF subtraction is performed on this modified stack. If the resulting PSF-subtracted image still shows significant non-stellar structures, additional iterations are performed until the stellar signal estimate no longer improves. For DO Tau, 10 iterations of IDF-RDI were performed for each epoch, and the results for each epoch were averaged together. Appendix \ref{sec:RDIcomparison} shows a comparison of the total intensity images of DO Tau produced with standard RDI with KLIP (equivalent to the zeroth iteration of IDF-RDI) and after 10 iterations of IDF-RDI.

\begin{deluxetable*}{ccccc}
\tablecaption{ALMA Imaging Summary\label{tab:imageproperties}}
\tablehead{
\colhead{Transition}&\colhead{Rest frequency}&\colhead{Synthesized beam}&\colhead{Peak $I_\nu$}&\colhead{RMS noise\tablenotemark{a}}\\
&(GHz)&(arcsec $\times$ arcsec ($^\circ$))&(mJy beam$^{-1}$)&(mJy beam$^{-1}$)}
\startdata
$^{12}$CO $J=2-1$ &230.5380000&0.84 $\times$ 0.53 (33.7$^\circ$)&576.9&4.7 \\
$^{13}$CO $J=2-1$ &220.3986842&0.87 $\times$ 0.56 (32.9$^\circ$)&201.6&4.7 \\
C$^{18}$O $J=2-1$ &219.5603541& 0.88 $\times$ 0.56 (33.3$^\circ$) &61.9&3.9\\
CS $J=5-4$ & 244.9355565 & 0.69 $\times$ 0.55 (21.3$^\circ$)&56.2&3.3\\
\enddata
\tablenotetext{a}{With channel widths of 0.25 km s$^{-1}$.}
\end{deluxetable*}

\subsection{ALMA Observations}
DO Tau was observed with ALMA in two Band 6 spectral settings (henceforth referred to as the 1.3 mm and 1.1 mm settings) as part of program 2016.1.00627.S (PI: K. \"Oberg). Portions of DO Tau observations from this program have been published previously: C$^{18}$O $J=2-1$ and upper limits for H$^{13}$CN $J=3-2$ and C$_2$H $N=3-2$ in \citet{2019ApJ...876...25B}, H$_2$CO $J_{K_aK_c} = 3_{03}-2_{02}$ in \citet{2020ApJ...890..142P}, CS $J=5-4$ in \citet{2019ApJ...876...72L}, and upper limits on DCN $J=3-2$ and HC$^{15}$N $J=3-2$ in \citet{2020ApJ...898...97B}. In this article, we present the $^{12}$CO and $^{13}$CO $J=2-1$ observations from this program for the first time. We also reprocess the C$^{18}$O and CS observations for consistency; whereas the aforementioned works only analyzed the Keplerian disk, the present article examines the extended structures in further detail. (The extended structures are not detected in the other previously published line observations.)

Details of the observations, including the dates, array and correlator setups, and calibrators, are provided in Appendix \ref{sec:ALMAdetails}. Because of the narrow spectral windows used in both spectral settings, bandwidth switching was employed, which required manual calibration by ALMA/NAASC staff. The associated calibration scripts are available on the ALMA archive. Following data delivery from ALMA, we performed additional processing and imaging steps in \texttt{CASA 5.6.1} \citep{2007ASPC..376..127M}. For each individual execution block, a set of pseudo-continuum visibilities was produced by flagging channels with strong line emission and then averaging together the remaining channels. A preliminary continuum image was produced for each execution block using the Clark CLEAN algorithm as implemented in the  \texttt{tclean} task. After measuring the location of the continuum peak by fitting a two-dimensional Gaussian to the disk emission, each set of visibilities was shifted to align the continuum peak with the phase center. We then compared the fluxes of different execution blocks taken with the same spectral setting. For the 1.1 mm setting, the continuum flux of the third execution block was $\sim10\%$ lower than the earlier two execution blocks. For the 1.3 mm setting, the continuum flux of the second execution block was $\sim15\%$ lower than the first execution block. These discrepancies are comparable to ALMA's nominal absolute flux calibration accuracy in Band 6 \citep[e.g.,][]{ALMACycle4Guide}. For the 1.3 mm setting, we rescaled the visibilities of the first execution block such that the continuum flux matched that of the second execution block, and for the 1.1 mm setting, we rescaled the visibilities of the third execution block such that the continuum flux matched that of the first two. The choice of which execution block served as the reference for rescaling is arbitrary, but the interpretation of the data in this work is not affected by this choice. 

For each spectral setting, the pseudo-continuum visibilities of execution blocks were imaged together with Clark CLEAN to form an initial model for self-calibration. One round of phase self-calibration and then one round of amplitude self-calibration was applied to the pseudo-continuum visibilities. The self-calibration tables derived from continuum imaging were applied to the full-resolution spectral windows containing line data. Continuum subtraction was then performed in the $uv$ plane using the \texttt{uvcontsub} task. Image cubes were produced for the $^{12}$CO $J=2-1$, $^{13}$CO $J=2-1$, C$^{18}$O $J=2-1$, and CS $J=5-4$ transitions using multi-scale \texttt{CLEAN} (scales of [0, $0.8''$, $1.6''$, $3.2''$]) and a Briggs robust parameter of 1.0. Because the line emission is extended and irregular, we used CASA's \texttt{auto-multithresh} automasking algorithm \citep{2020PASP..132b4505K} to draw CLEAN masks. After some initial experimentation, we set the automasking parameter values to \texttt{sidelobethreshold=2.0}, \texttt{noisethreshold=3.5}, \texttt{lownoisethreshold=1.5}, \texttt{minbeamfrac=0.4}, and \texttt{negativethreshold=15.0}. The channel spacing was set to 0.25 km s$^{-1}$, to balance between achieving adequate sensitivity in individual channels and resolving kinematic detail. A primary beam correction was applied to all CLEANed image cubes using the  \texttt{impbcor} task.  The rms of each image cube was measured within line-free channels of the non-primary beam corrected images. The synthesized beam size, rms, and peak intensity of each image cube are given in Table \ref{tab:imageproperties}.

\subsection{Archival Optical/Infrared Observations}
\subsubsection{HST STIS}
DO Tau was imaged with the Hubble Space Telescope STIS instrument as part of program HST-GO-9136 (P.I.: C. Grady) on 2001 December 08 and 2003 February 12, with a total exposure time of 2268 s per epoch. The STIS CCD has a broad bandpass that extends from 2000 to 10,300 \AA, with a central wavelength of 5740 \AA. For each set of observations, DO Tau was placed under the coronagraphic mask at the WEDGEA1.0 location, which has a width of $1''$.  These STIS data were first presented in \citet{2004ASPC..321..244G}.

The calibrated, flat-fielded science files for DO Tau were retrieved from the Mikulski Archive for Space Telescopes. Following the procedure described in \citet{2017SPIE10400E..21R}, a median filter was applied to correct bad pixels. The two rolls were rotated such that north was up and east was to the left, and then aligned using the \texttt{centerRadon} package \citep{2019ascl.soft06021R}. The two rolls were then combined, with pixels averaged in regions where the two rolls overlapped. However, if a pixel from one roll fell within a diffraction spike, the combined image only used the pixel value from the other roll. The portion of the combined image lying within $8''$ of DO Tau's stellar position was then replaced with the KLIP PSF-subtracted image presented in \citet{2017SPIE10400E..21R}. PSF subtraction over the entire field of view of the DO Tau observations was not feasible due to limitations in the field of view of the PSF reference library. The PSF wings are negligible outside a radius of $8''$, but the intensities are discontinuous at the boundary between the PSF-subtracted portion of the combined DO Tau image and the non-PSF subtracted portion. Thus, caution should be applied in interpreting features at this boundary. 

\subsubsection{SDSS}
The field around DO Tau was imaged by the Sloan Digital Sky Survey (SDSS) on 2002 December 31 (MJD 52639). The original SDSS run-camcol-field identifier for the data is 3559-5-55. Details of the SDSS observations and data reduction are provided in \citet{2000AJ....120.1579Y, 2004AJ....128.2577F, 2012ApJS..203...21A}. We downloaded pipeline-processed SDSS $g$, $r$, and $i$ band cutouts from NASA Skyview \citep{1998IAUS..179..465M} and produced a color composite using the \texttt{Astropy} \citep{2013AA...558A..33A} implementation of the \citet{2004PASP..116..133L} algorithm. 

\subsubsection{Herschel PACS}
DO Tau was imaged at 160 $\mu$m with the Herschel PACS instrument \citep{2010AA...518L...2P} on 2011 March 30 as part of the GASPS Open Key Time Project \citep{2013PASP..125..477D, 2013ApJ...776...21H}. We retrieved the pipeline-processed Level 2.5 Unimap map (which combined data from observation IDs 1342217478, 1342217479, 1342217480, and 1342217481) from the Herschel Science Archive.   

\section{Morphology in Scattered Light \label{sec:scatteredlight}}

 \begin{figure*}
\begin{center}
\includegraphics{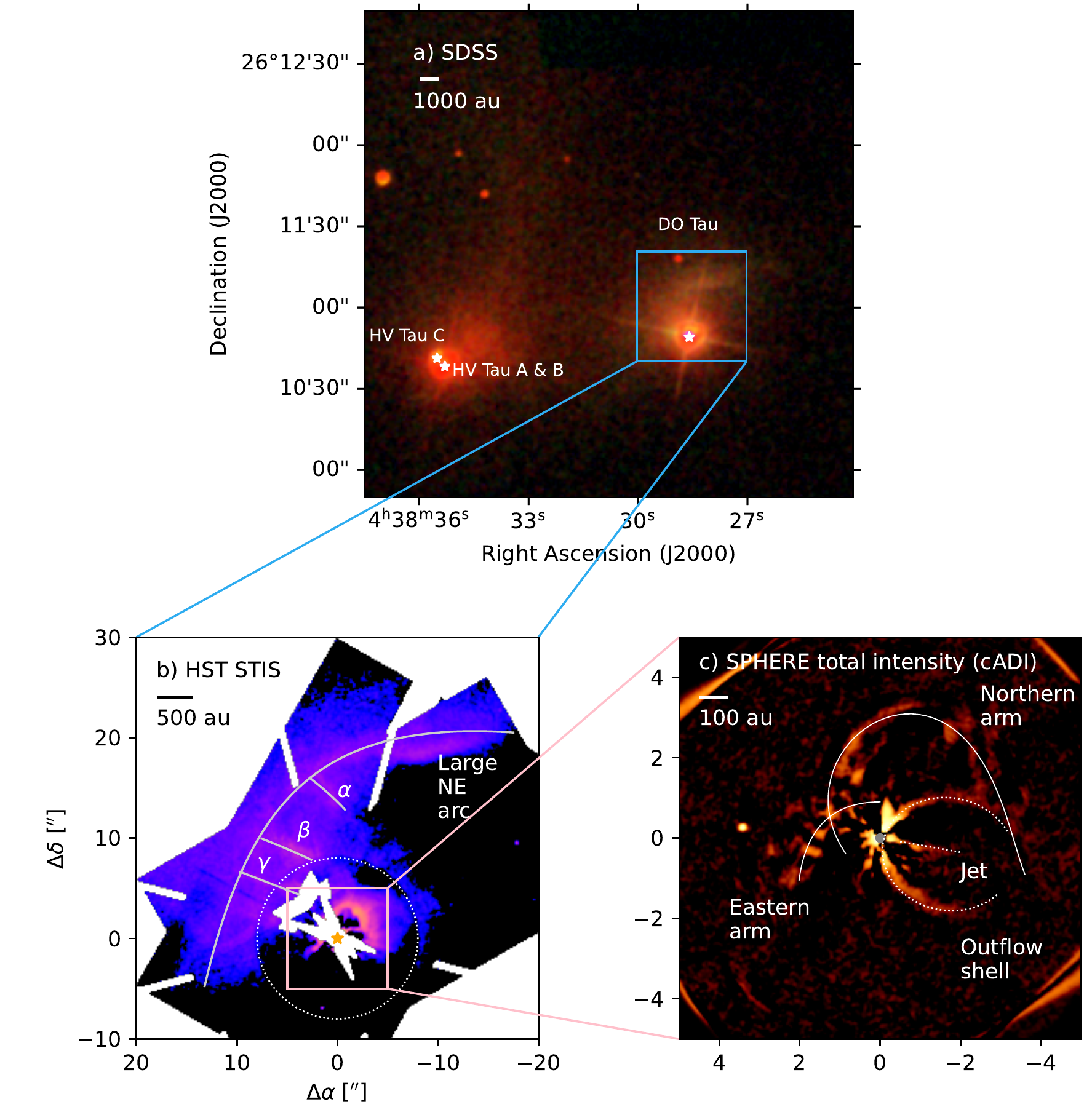}
\end{center}
\caption{a) SDSS color composite image of DO Tau and HV Tau made from $g$, $r$, and $i$-band images. The blue box shows the relative size of the STIS image shown in part b). The white stars mark the locations of HV Tau A and B, HV Tau C, and DO Tau (only a single star is used to mark HV Tau A and B because of their close proximity). b) HST STIS image of DO Tau, smoothed by convolving with a 2-D Gaussian with a standard deviation of $0.1''$ (equal to two pixels). The gray curves mark the locations of the large northeastern arc and the $\alpha$, $\beta$, and $\gamma$ streams. The yellow star shows the location of DO Tau. Pixels falling within the coronagraphic wedges or stellar diffraction spikes are masked. The dotted white circle denotes the region where PSF subtraction was performed with KLIP. The pink box shows the relative size of the SPHERE image in part c). The axes are marked with the angular offsets from DO Tau. c) SPHERE $H-$band total intensity image of DO Tau, produced with cADI and smoothed by convolving with a 2-D Gaussian with a standard deviation of $0.0368''$ (equal to three pixels). The solid curves mark the northern and eastern arms, the dotted white curve marks the outflow shell, and the dotted line marks the jet. The gray circle denotes the extent of the coronagraph. Versions of the STIS and SPHERE images with colorbars and no annotations are presented in Appendix \ref{sec:noannotations}.} \label{fig:SDSS}
\end{figure*}

DO Tau is surrounded by a series of complex structures spanning several orders of magnitude in size scales, as shown in the SDSS, HST STIS, and SPHERE images presented in Figure \ref{fig:SDSS}. The SDSS optical color image shows the broad nebulosity surrounding DO Tau and HV Tau, which are separated by $\sim12,600$ au. This nebula is relatively well-known, having previously been listed in catalogs such as those by \citet{1962PASP...74..474S} and \citet{2003AA...399..141M}. Within this broad nebulosity, an arc-like structure (which we call the ''large northeastern arc'') is visible to the northeast of DO Tau in the SDSS image, but can be seen in finer detail in the HST STIS optical image. The southeastern end of the arc is located at a position angle (P.A.) east of north of $\sim110^\circ$ and a projected separation of $\sim14''$ ($\sim1900$ au) from DO Tau, while the northwestern end is located at a P.A. of $\sim310^\circ$ and a separation of $\sim27''$ ($\sim3700$ au) from the star. There are also at least three stream-like structures that are oriented nearly perpendicularly to the large northeastern arc. To our knowledge, these stream-like structures have not been explicitly identified in the literature; previous publications of this HST data \citep{2004ASPC..321..244G, 2008PASJ...60..223I, 2017SPIE10400E..21R} only presented cropped versions of the image. We label the stream-like structures as $\alpha$, $\beta$, and $\gamma$ (in order from north to south). The southwestern ends of these structures appear to point toward a smaller arc north of DO Tau. This smaller arc was first detected by \citet{2004ASPC..321..244G} in the same set of STIS observations. 

Although the STIS wedges obscure much of the material within several hundred au of DO Tau, SPHERE provides a clearer view of these structures. We identify a counterpart to the HST image's smaller arc, which we call the ``northern arm.'' We also more clearly identify an ``eastern arm'' that is only partially visible in the HST STIS image. Although the eastern and northern arms appear as though they could be a single structure in the total intensity images, we identify them as separate structures because they are kinematically distinct and because they appear to connect separately to the disk in polarized scattered light, as discussed later in this section and in Section \ref{sec:molecularlines}. In addition, a narrow structure that is west of DO Tau and partially obscured by the wedges in the STIS image turns out to be part of a larger arc-like structure southwest of DO Tau. Based on visual inspection of the SPHERE cADI image in Figure \ref{fig:SDSS}, it can be approximated as an elliptical arc centered $\sim1.7''$ ($\sim240$ au)  west and $\sim0.4''$ ($\sim55$ au) south of the star, with a P.A. of $80^\circ$, an angular extent of $\sim300^\circ$, a semi-major axis of $\sim1.65''$ ($\sim230$ au), and a semi-minor axis of $\sim1.4''$ ($\sim190$ au). This geometry matches one of the blueshifted CO outflow shells identified in \citet{2020AJ....159..171F}. In the cADI image, the outflow shell is bisected by a linear feature that has the same orientation as the blueshifted side of the [Fe II] jet imaged by \citet{2021AA...650A..46E}. \citet{2021AA...650A..46E} detected emission from the 1.53 and 1.64 $\mu$m [Fe II] lines, which fall within the wavelength range (but are not spectrally resolved by) of SPHERE's broadband $H$ filter. Outside of the arm structures, a background point source is visible $\sim3.5''$ east of DO Tau  and is discussed further in Appendix \ref{sec:pointsource}. 

\begin{figure*}
\begin{center}
\includegraphics{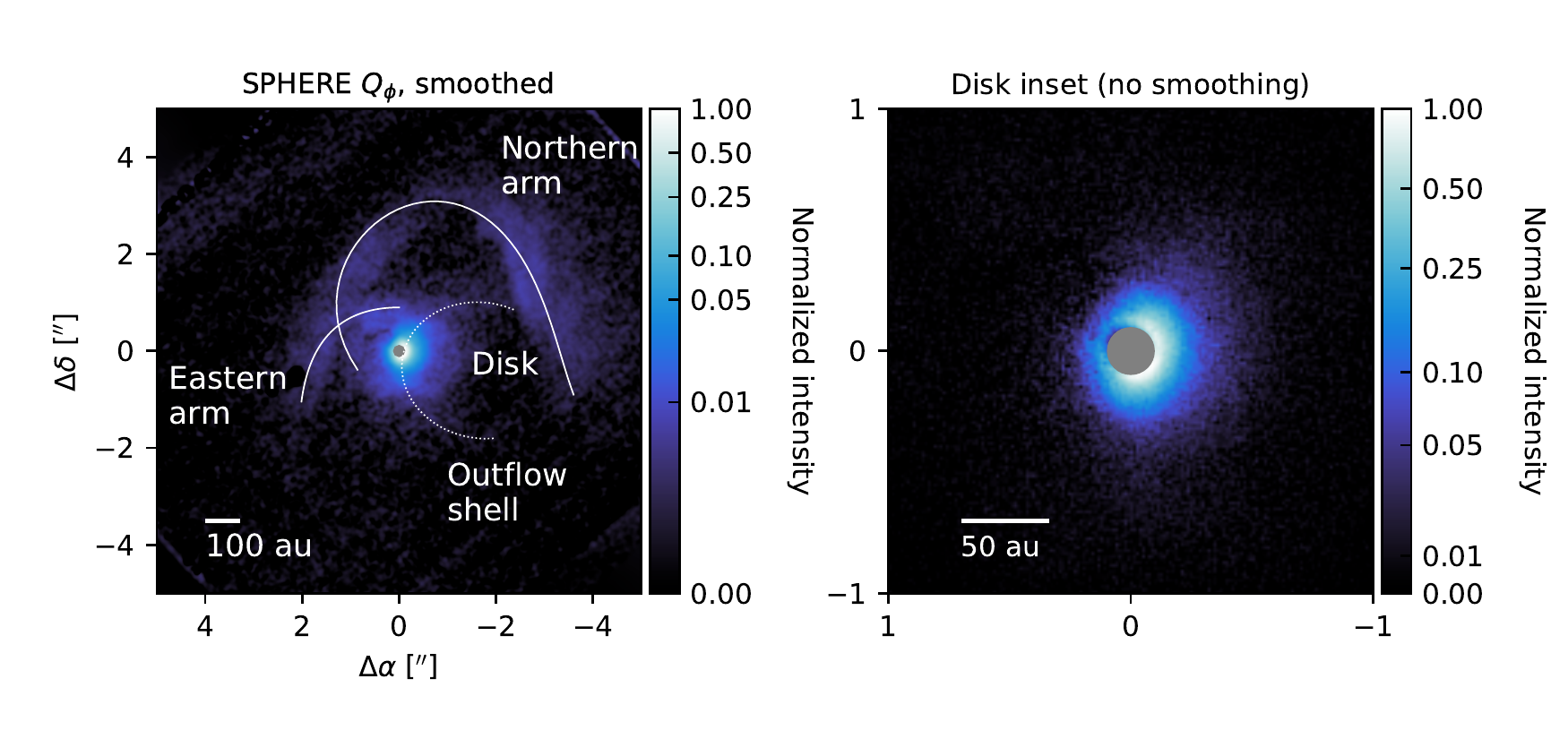}
\end{center}
\caption{Left: Smoothed SPHERE $Q_\phi$ $H-$band image of DO Tau. The gray circle marks the extent of the coronagraph. The solid white curves mark the eastern and northern arms. The dotted white curve marks part of an outflow shell. Offsets from the star are labeled on the axes. North is up and east is to the left. A version of this image without annotations is presented in Figure \ref{fig:noannotations}. Right: An inset of the SPHERE $Q_\phi$ $H-$band image of DO Tau. Unlike the lefthand panel, no smoothing is applied to the righthand panel in order to show more detail at smaller separations from the star. Note that the color stretch is also different from that of the lefthand panel in order to show brightness variations across the disk region more clearly.   \label{fig:diskinset}}
\end{figure*}

While the SPHERE cADI image is dominated by self-subtraction artifacts within $\sim1''$ of the star, the $Q_\phi$ image clarifies the relationship between some of the arm structures and the circumstellar disk (Figure \ref{fig:diskinset}). The quantity $Q_\phi$ is defined as 
\begin{equation}
Q_\phi = -Q_\text{IPS}\cos(2\phi)-U_\text{IPS}\sin(2\phi),
\end{equation}
where $\phi$ is the azimuthal angle in the observer frame and $Q_\text{IPS}$ and $U_\text{IPS}$ are the linear polarization components of the Stokes vector following correction for instrumental polarization \citep[e.g.,][]{2020AA...633A..63D}. $Q_\phi$ is related to the linearly polarized intensity $PI$ by $PI=\sqrt{Q_\phi^2+U_\phi^2}$, where $U_\phi$ is defined as
\begin{equation}
U_\phi = Q_\text{IPS}\sin(2\phi)-U_\text{IPS}\cos(2\phi).
\end{equation}
Positive signal in the $Q_\phi$ image traces azimuthal polarization, which is expected to be dominant for relatively low-inclination disks such as DO Tau (27.6$^\circ$ as measured from millimeter continuum emission in \citet{2019ApJ...882...49L}). 

In projection, the northern arm appears to emerge from the disk at a P.A. of $\sim114^\circ$ east of north. The arm curves up to a separation of $\sim3.3''$ ($\sim460$ au) northwest of DO Tau and then turns over toward the southwest, terminating at a P.A. of $\sim256^\circ$ and a separation of $\sim3.7''$ ($\sim510$ au) from the star. (These quantities are estimated via visual inspection since there is no sharp delineation between the disk and arm structures).  The eastern arm extends from the north side of the disk and terminates at a P.A. of $\sim117^\circ$ and a separation of $\sim2.3''$ ($\sim320$ au) from the star.

The circumstellar disk itself appears relatively compact in scattered light, with a radius of $\sim80$ au as determined by visual inspection. However, the size of the disk is not straightforward to quantify due to the aforementioned extended structures. Furthermore, scattered light only provides a lower bound on the disk size, since sensitivity decreases with the square of the distance from the star, and many disks are also self-shadowed \citep[e.g.,][]{2022AA...658A.137G}. DO Tau's disk appears about twice as large in scattered light compared to millimeter continuum. The former predominantly traces submicron-sized dust grains in the disk atmosphere, while the latter primarily traces millimeter-sized dust grains in the disk midplane. \citet{2019ApJ...882...49L} find that 95\% of the millimeter continuum flux is enclosed within a radius of only 36.4 au. The millimeter continuum image from \citet{2019ApJ...882...49L} exhibits no substructures at an angular resolution of $\sim0.1''$ (14 au). Likewise, no gaps or rings are visible in the scattered light image. The disk signal has a broader extent on the west side compared to the east side, which we interpret as a projection effect of viewing a flared, inclined disk surface. Some contamination from the outflow shell may also enhance the difference between the west and east sides. Since the east side appears to be foreshortened, we infer that it is inclined toward the observer. Based on obscuration of DO Tau's jet by the eastern side of the disk, \citet{2021AA...650A..46E} likewise concluded that the east side is tilted toward the observer. 

\begin{figure*}
\begin{center}
\includegraphics{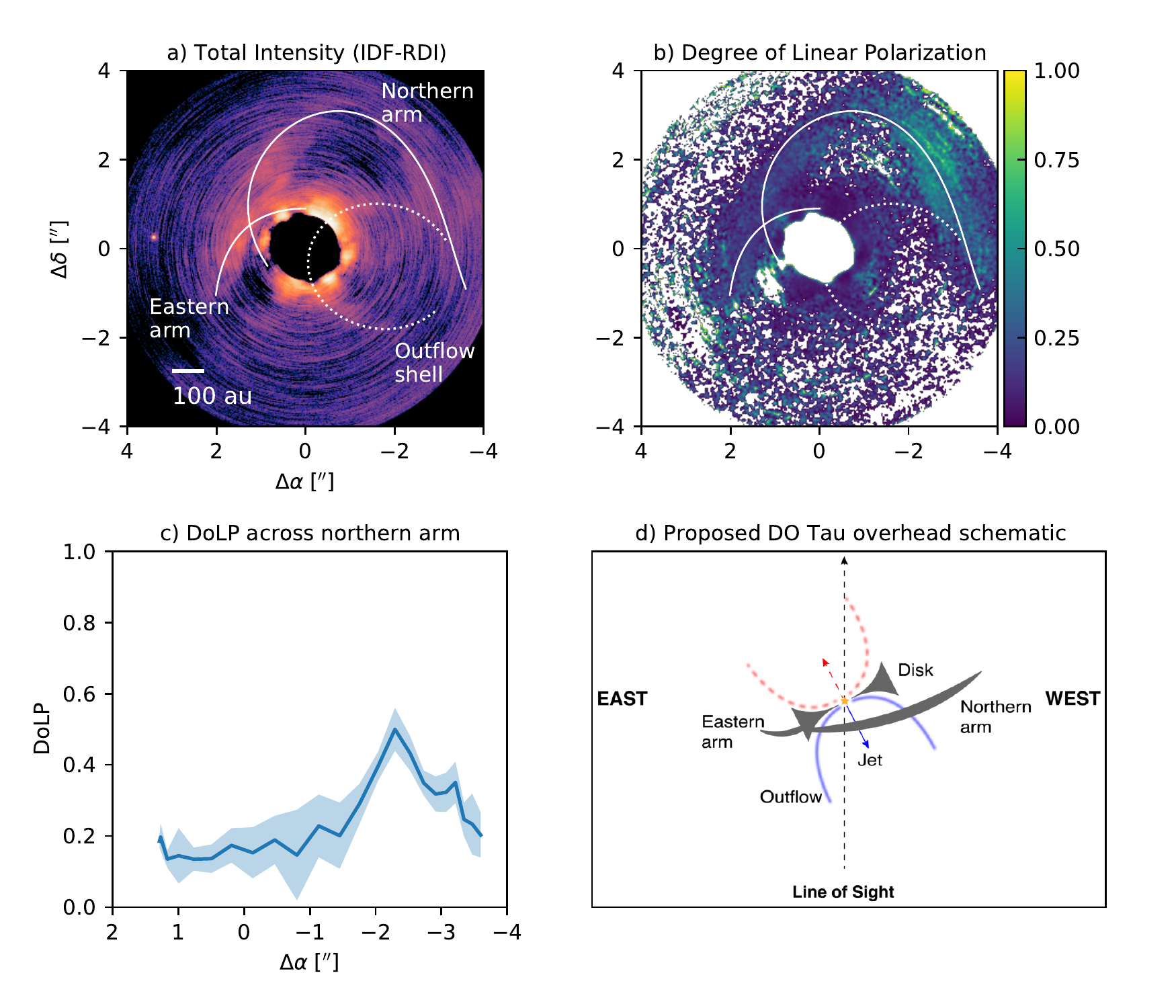}
\end{center}
\caption{a) SPHERE total intensity image of DO Tau produced with IDF-RDI. The image is binned by a factor of 2 relative to the native IRDIS pixel resolution in order to increase the SNR of the extended structures, which are marked by white curves. The central region is masked because the presence of a disk around the PSF reference star prevents accurate estimates of total intensity at small separations from DO Tau. b) Degree of linear polarization map for DO Tau. Values less than 0 or greater than 1 are masked because they are unphysical. c) DoLP values measured across the northern arm. The blue curve denotes the mean values measured within 6-pixel $\times$ 6-pixel boxes (roughly $0.15''\times0.15''$) centered on points along the white curve corresponding to the northern arm in part b). The shaded region shows the standard deviation measured within each of these boxes. d) A schematic of an overhead view of the different structures traced by SPHERE around DO Tau, showing how various structures are inclined relative to the line of sight. North points out of the page and south points into the page. The structures are not drawn to scale. The redshifted jet and outflow are drawn with dashed curves because they are not detected with SPHERE, but are known to be present based on other observations (see \citealt{2021AA...650A..46E} and Section \ref{sec:molecularlines}.) \label{fig:DoLP}}
\end{figure*}

The degree of linear polarization (DoLP), defined as the ratio of the linearly polarized intensity to total intensity, provides constraints on the three-dimensional orientation of the extended structures around DO Tau. Since the extended structures around DO Tau are not detected in the U$_\phi$ images, we assume that $PI\approx Q_\phi$, so that we can estimate the DoLP without the additional noise contribution from the U$_\phi$ component. To calculate the DoLP, we use the total intensity image produced with IDF-RDI rather than cADI because the latter procedure generally results in significant self-subtraction of the extended structures. However, since IDF-RDI does not yield a perfect subtraction of the stellar PSF, the resulting DoLP values must still be treated with caution. Therefore, our analysis focuses on relative rather than absolute DoLP values. To increase the SNR of the DoLP map, we smoothed the $Q_\phi$ and total intensity images using a Gaussian kernel with a standard deviation of one pixel (0.013$''$) prior to division and then binned the resulting DoLP map by a factor of two. 

Parts a) and b) in Figure \ref{fig:DoLP} show the IDF-RDI total intensity image and DoLP map of DO Tau. Within the DoLP map, the northern arm stands out most clearly. Part c) of Figure \ref{fig:DoLP} shows how the DoLP of the northern arm varies from east to west. As the arm emerges from the eastern side of the disk, the DoLP values are low ($<0.2$). The DoLP values generally increase from east to west across the arm, peaking at a value of $\sim0.5$, but then decrease again near the western terminus of the arm. If we assume that the DoLP has a roughly bell-shaped dependence upon the scattering angle, with a peak at $\sim90^\circ$ \citep[e.g.,][]{2005AA...432..909M}, then the DoLP pattern of the northern arm could be explained by one of several scenarios: 1) The scattering angles increase from east to west across the arm, starting at a value $<90^\circ$ and ending at a value $>90^\circ$. 2)  The scattering angles decrease from east to west across the arm, starting at a value $>90^\circ$ and ending at a value $<90^\circ$. 3) Starting from a value $<90^\circ$, the scattering angles increase from east to west up to $90^\circ$, then decrease again at the western end of the arm. 4) Starting from a value $>90^\circ$, the scattering angles decrease from east to west down to $90^\circ$, then increase again at the western end of the arm. Given that the northern arm appears to emerge from the eastern side of the disk, which is tilted toward the observer, we expect the eastern side of the northern arm to have small scattering angles. Furthermore, the eastern side of the arm is brighter in total intensity compared to the western side, suggesting stronger forward scattering on the eastern side. This brightness difference was previously noted by \citet{2008PASJ...60..223I} in Subaru/CIAO $H$-band observations of DO Tau and also ascribed to forward scattering. Scenarios 2) and 4) therefore appear unlikely. We favor Scenario 1) over Scenario 3) because we do not observe the total intensity of the arm increasing at the western terminus, as one might expect if scattering angles were decreasing from $90^\circ$ and forward scattering became stronger. However, Scenario 3 is not definitively ruled out because distance from the star also affects total intensity. The eastern arm appears to have total intensity and DoLP values comparable to the eastern side of the northern arm, suggesting that they have similar (small) scattering angles. Meanwhile, the DoLP in the vicinity of the outflow shell appears to be close to zero. Since the disk has a relatively low inclination and CO kinematics show that the outflow shell is in front of the disk \citep{2020AJ....159..171F}, the shell is expected to have low scattering angles and therefore a low DoLP. A schematic of the proposed geometry is presented in part d) of Figure \ref{fig:DoLP}.

\section{Molecular Line Observations\label{sec:molecularlines}}
\subsection{Overview}

\begin{figure*}
\begin{center}
\includegraphics{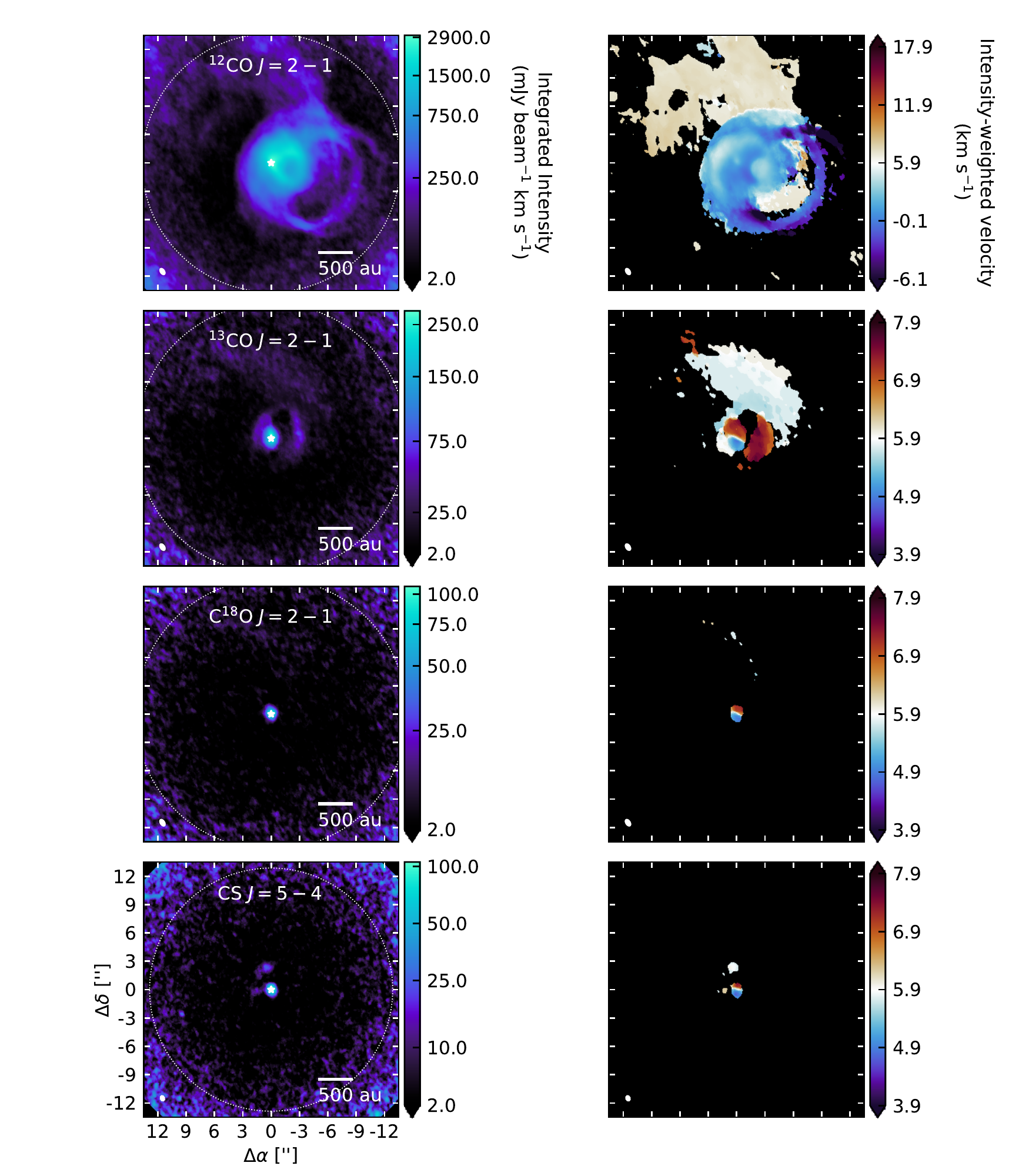}
\end{center}
\caption{Left column: Integrated intensity maps of $^{12}$CO $J=2-1$, $^{13}$CO $J=2-1$, C$^{18}$O $J=2-1$, and CS $J=5-4$ toward DO Tau (in order from top to bottom). The synthesized beam is shown as a solid white ellipse in the lower left corner of each panel. The white star marks the position of the millimeter continuum peak, which is located at the phase center. Offsets from the phase center (in arcseconds) are marked in the lower left panel. North is up and east is to the left. The dotted white ellipse marks the FWHM of the ALMA primary beam. An arcinsh stretch is applied to the color scale to highlight faint emission features. Right column: Corresponding intensity-weighted velocity maps. Note that the color scale for $^{12}$CO is different from the other molecules to accommodate the wider velocity range over which emission is detected. \label{fig:mom0maps}}
\end{figure*}

Integrated intensity maps and intensity-weighted velocity maps of $^{12}$CO $J=2-1$, $^{13}$CO $J=2-1$, C$^{18}$O $J=2-1$, and CS $J=5-4$ are shown in Figure \ref{fig:mom0maps}. The velocity integration ranges are selected based on the extent over which emission above the $3\sigma$ level is detected: $-13.75$ to 20.5 km s$^{-1}$ for $^{12}$CO, $-0.75$ to 9.25 km s$^{-1}$ for $^{13}$CO, 4.0 to 8.75 km s$^{-1}$ for C$^{18}$O, and 4.0 to 8.25 km s$^{-1}$ for CS. Because the spatial distribution of the emission changes substantially from channel to channel, most of the channels at any given spatial location are emission-free even though the overall velocity range of the emission is large. To reduce the contribution of signal-free regions to the integrated intensity maps, we only include pixels in the image cube with intensities exceeding the $2\sigma$ level. For the intensity-weighted velocity maps, a $5\sigma$ pixel clip is used instead since higher moments are more sensitive to outliers. The full set of channel maps is presented in Appendix \ref{sec:chanmaps}.

The characteristic rotation velocity pattern for a Keplerian disk is visible in the velocity-weighted intensity maps of $^{13}$CO, C$^{18}$O, and CS, but $^{12}$CO is dominated by non-disk emission. The extended emission structures traced by all four of these molecules are described in more detail below. 

\subsection{Outflow shells}

\begin{figure*}
\begin{center}
\includegraphics{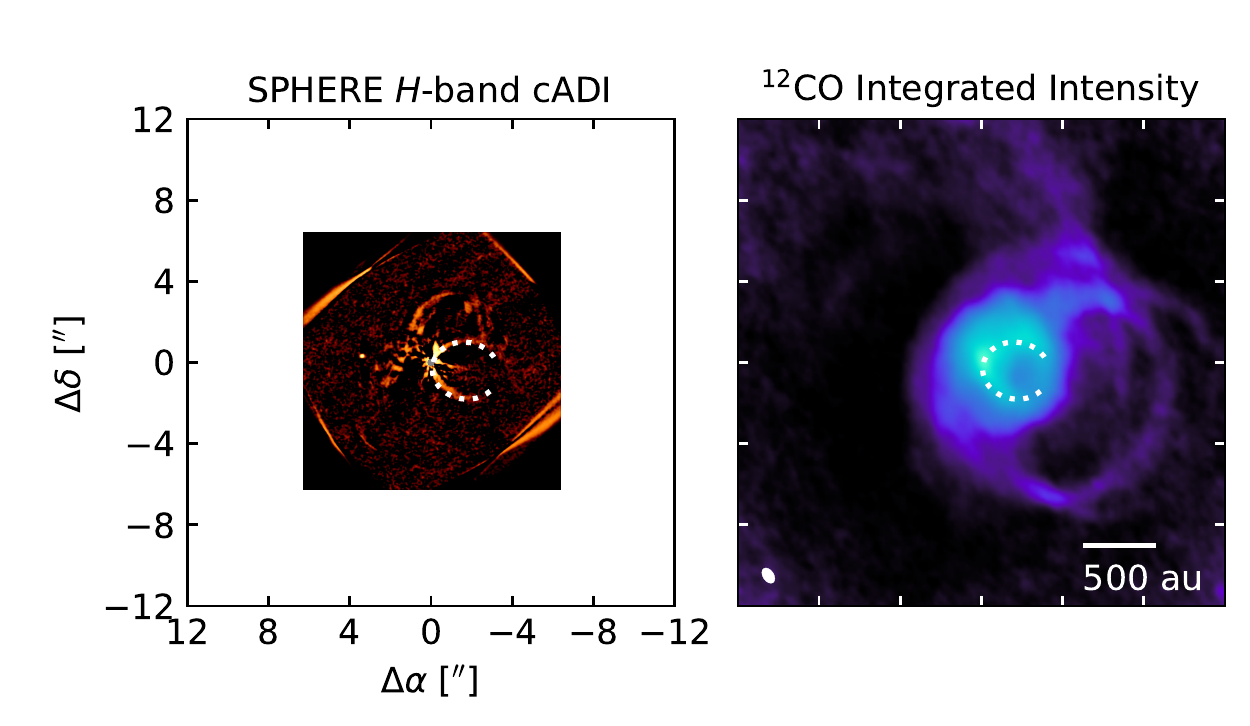}
\end{center}
\caption{A comparison between the SPHERE $H$-band cADI image of DO Tau (left) and the $^{12}$CO integrated intensity map (right) on the same relative scale. The dotted white curve marks the position of a scattered light arc that coincides with the innermost outflow shell as traced by $^{12}$CO. \label{fig:outflowcomparison}}
\end{figure*}

The $^{12}$CO emission is dominated by a series of blueshifted, overlapping ring-like structures southwest of DO Tau. These structures were previously detected in an independent set of ALMA $^{12}$CO $J=2-1$ observations by \citet{2020AJ....159..171F}, who attributed them to a series of outflow shells. We do not detect these shells in the other three molecular tracers. A comparison of the SPHERE $H$-band cADI image to the $^{12}$CO integrated intensity map shows the coincidence between the southwestern arc detected in scattered light and the innermost CO outflow shell (Figure \ref{fig:outflowcomparison}).

\begin{figure}
\begin{center}
\includegraphics{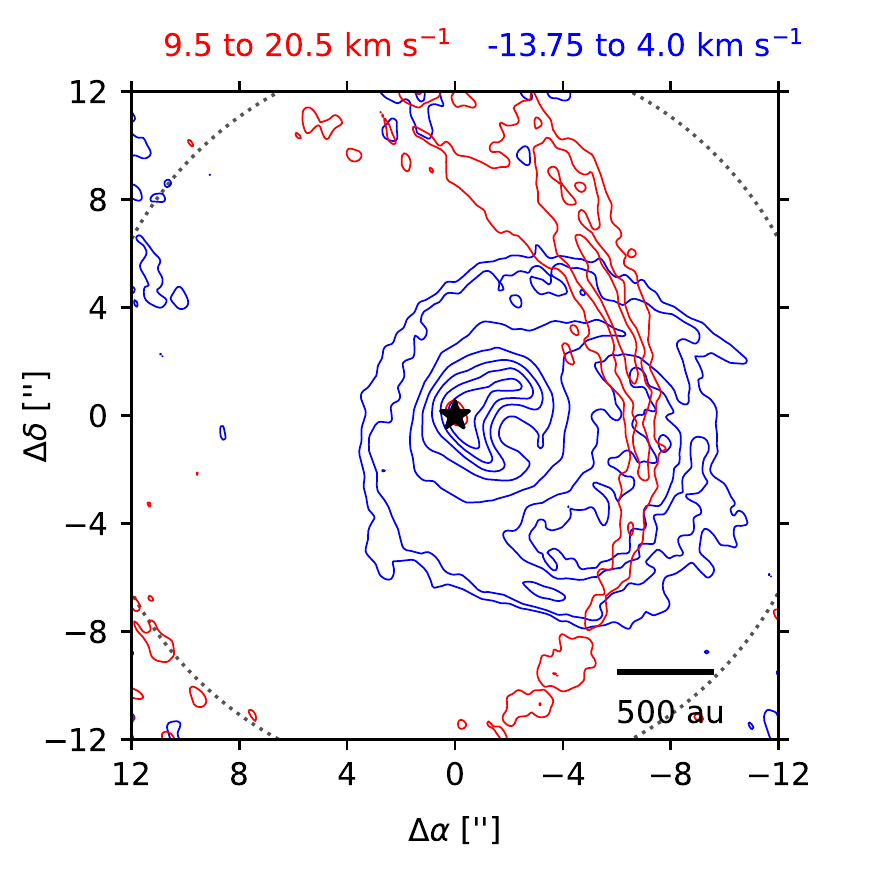}
\end{center}
\caption{Overlaid integrated intensity maps of the high-velocity $^{12}$CO emission toward DO Tau. No clipping is applied to make these integrated intensity maps because the redshifted outflow emission levels are low in individual channels. The velocity integration ranges are given at the top of the figure. The blue contours, corresponding to the blueshifted emission, are drawn at 75, 150, 400, 700, 1000, 1300, and 1600 mJy beam$^{-1}$ km s$^{-1}$. The red contours, corresponding to the redshifted emission, are drawn at 50, 100, and 150 mJy beam$^{-1}$ km s$^{-1}$. The black star marks the location of DO Tau. The dotted black circle marks the FWHM of the primary beam. Offsets from the phase center (in arcseconds) are marked in the lower left panel. North is up and east is to the left.  \label{fig:outflow}}
\end{figure}

Thanks to the longer integration time of our observations compared to those presented in \citet{2020AJ....159..171F} ($\sim4\times$ longer), we are able to detect a fainter, redshifted outflow. Figure \ref{fig:outflow} compares integrated intensity maps of the redshifted and blueshifted outflows. The blueshifted outflow is detected at  Local Standard of Rest (Kinematic), or LSRK, velocities ranging from roughly $-13.75$ to 4.0 km s$^{-1}$, while the redshifted outflow is detected from approximately 9.5 to 20.5 km s$^{-1}$. (\citet{2020AJ....159..171F} identify $^{12}$CO outflow shells at LSRK velocities between 4.4 and 8.2 km s$^{-1}$, but we attribute this emission instead largely to the northern and eastern arms based on comparison with the $^{13}$CO emission in Section \ref{sec:arms}.) The peak integrated intensity of the blueshifted outflow is 2.1 Jy beam$^{-1}$ km s$^{-1}$, while that of the redshifted outflow is only 0.21 Jy beam$^{-1}$ km s$^{-1}$. Likewise, \citet{2021AA...650A..46E} found that the blueshifted side of the [Fe II] jet is brighter than the redshifted side. Whereas the blueshifted outflow manifests as multiple rings of different sizes, the redshifted outflow appears to be a single large arc west of DO Tau (although the widening of the emission at the northern end suggests that it may trace two slightly offset rings). Because of the faintness of the redshifted outflow in individual channels (see Appendix \ref{sec:chanmaps}), it is ambiguous how the morphology varies as a function of velocity. 

\begin{figure}
\begin{center}
\includegraphics{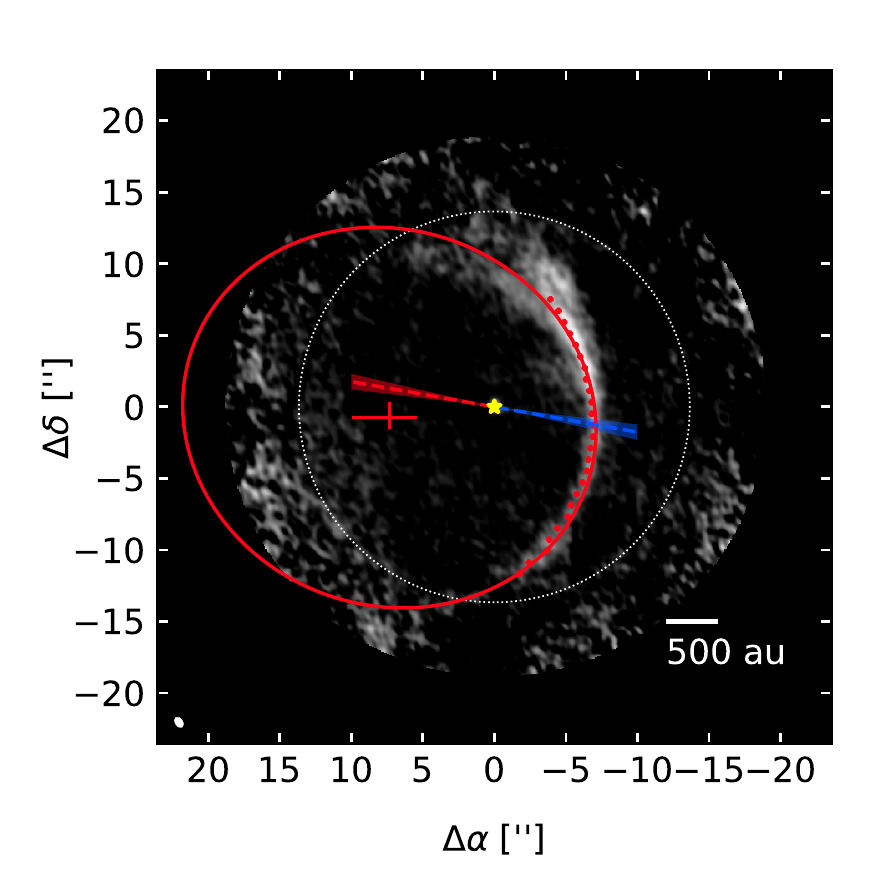}
\end{center}
\caption{Model ellipse with median parameter values (in red) overlaid on an integrated intensity map of the redshifted outflow as traced by $^{12}$CO (including velocities between 9.5 and 20.5 km s$^{-1}$). The red cross marks the center of the model ellipse and associated $1\sigma$ error bars. The dotted red points denote the positions of the local intensity maxima measured along horizontal cuts across the image. The yellow star marks the position of DO Tau. The white dotted circle denotes the FWHM of the primary beam. The dashed red and blue line segments show the position angle of DO Tau's jet as measured by \citet{2021AA...650A..46E}. The shaded regions show the $1\sigma$ uncertainties in the jet orientation. The line segments are drawn longer than the observed extent of the jet in order to show the apparent offset from the center of the redshifted outflow.  \label{fig:redoutflowmodel}}
\end{figure}

As shown in Figure \ref{fig:outflow}, the two ends of the redshifted arc extend toward the FWHM of the primary beam, and the emission blends into the noise as sensitivity decreases away from the phase center. Given the presence of ring-like blueshifted outflow emission, we consider it likely that the redshifted arc is part of a ring-like outflow shell that extends beyond ALMA's field of view. To constrain the geometry of the redshifted outflow, we first smoothed the image with a Gaussian kernel with a standard deviation of $0.3''$ to increase the signal-to-noise ratio of the faint arc and then measured the positions $x_i$, $y_i$ of local intensity maxima along horizontal (east-west) slices of the image spaced $0.8''$ apart in the vertical (north-south) direction. The northern and southern bounds of where measurements were taken were determined based on whether a distinct peak was identifiable in a given image slice. We then modelled the arc as an ellipse with the free parameters $x_0$, $y_0$, $a_0$, $f$, and $\theta$, where $x_0$ and $y_0$ are offsets (in arcseconds) from DO Tau, $a_0$ is the semimajor axis (also in arcseconds), $f$ is the ratio of the semiminor axis to semimajor axis, and $\theta$ is the angle (in radians) between the semimajor axis and the northern direction. The parameters $x_0$ and $y_0$ are defined such that the positive direction is toward the east and north, respectively. The log-likelihood function is written as

\begin{equation}
\log\mathcal{L}=-\frac{1}{2}\sum_i \left[ \frac{d_i^2}{\sigma^2}+\log(2\pi\sigma^2)\right],
\end{equation}
where $d_i$ is the smallest distance between point $i$ and the model ellipse, and $\sigma$ is the standard deviation of the major axis of the synthesized beam. We calculate $d_i$ using the analytical expressions described in \citet{zhangconic}. Flat priors were specified for all the free parameters, with bounds of $[0'',20'']$ for $x_0$, $[-20'', 20'']$ for $y_0$, $[5'', 20'']$ for $a_0$, $[0,1]$ for $f$, and $[0, \pi]$ for $\theta$. We used \texttt{emcee} to explore the posteriors, with 40 walkers evolved over 2000 steps each. After discarding the first 1000 steps as burn-in, we find the following posterior medians, with the uncertainties corresponding to the 16th and 84th percentiles: $x_0=7.3\substack{+2.6 \\ -1.9}$ arcseconds, $y_0=-0.8\substack{+1.1 \\ -0.8}$ arcseconds, $a_0=14.6\substack{+2.6\\ -1.5}$ arcseconds, $f=0.90\substack{+0.07\\ -0.08}$, and $\theta=1.2\pm0.9$ rad. The ellipse with the median parameter values is shown in Figure \ref{fig:redoutflowmodel}. The center of the ellipse is tentatively offset to the south from the axis of DO Tau's [Fe II] jet; the former has a P.A. of $96^\circ\substack{+7^\circ\\-9^\circ}$, while the latter has a P.A. of $260^\circ\pm3^\circ$ \citep{2021AA...650A..46E}. However, \citet{2021AA...650A..46E} note that the redshifted side of the [Fe II] jet may be misaligned from the blueshifted side. The redshifted outflow also appears to be slightly misaligned relative to the axis of the blueshifted molecular outflow, which has a P.A. of $253.6^\circ\pm0.1^\circ$ \citep{2020AJ....159..171F}. Deeper, mosaicked CO observations will be required to measure the full extent of the redshifted molecular outflow in order to confirm whether it is indeed misaligned. \citet{2021AA...650A..46E} inferred that DO Tau's jet was precessing, so precession may account for misalignment of the molecular outflow.  

The detected redshifted and blueshifted outflow shells have strikingly different radii. Whereas the shell radii in the blueshifted outflow range from $\sim200-1000$ au \citep{2020AJ....159..171F}, the redshifted outflow shell has a much larger radius of 2000 au. We estimate the dynamical age of the redshifted outflow shell as $t_\text{dyn} = D_\text{outflow}/v_\text{outflow}$, where $D_\text{outflow}$ is the deprojected distance from DO Tau to the edge of the outflow shell and $v_\text{outflow}$ is the deprojected velocity of the outflow relative to DO Tau ($v_\text{outflow}=(v_\text{LSRK} - v_\text{sys})/\cos i$). DO Tau's systemic velocity $v_\text{sys}$ is 5.9 km s$^{-1}$ \citep{2020ApJ...890..142P}. To compute $D_\text{outflow}$, we make the simplifying assumption that the outflow axis is perpendicular to the outer disk, for which the inclination has been constrained. Since the outflow misalignment does not seem to be extreme, this simplification should not significantly affect the dynamical timescale estimate. We take the disk inclination to be $27.6^\circ$, based on the millimeter continuum measurement in \citet{2019ApJ...882...49L}. Emission from the redshifted outflow is detected at LSRK velocities from $\sim9.5$ to 20.5 km s$^{-1}$, and does not have a well-defined peak along the spectral axis. We therefore use the upper and lower ends of the velocity range to estimate that $t_\text{dyn}$ is between $\sim900$ and 3500 years, which is comparable to or older than the range of dynamical ages that \citet{2020AJ....159..171F} estimated for the blueshifted outflow shells ($\sim460-1090$ years).  

\subsection{The northern and eastern arms\label{sec:arms}}

\begin{figure*}
\begin{center}
\includegraphics{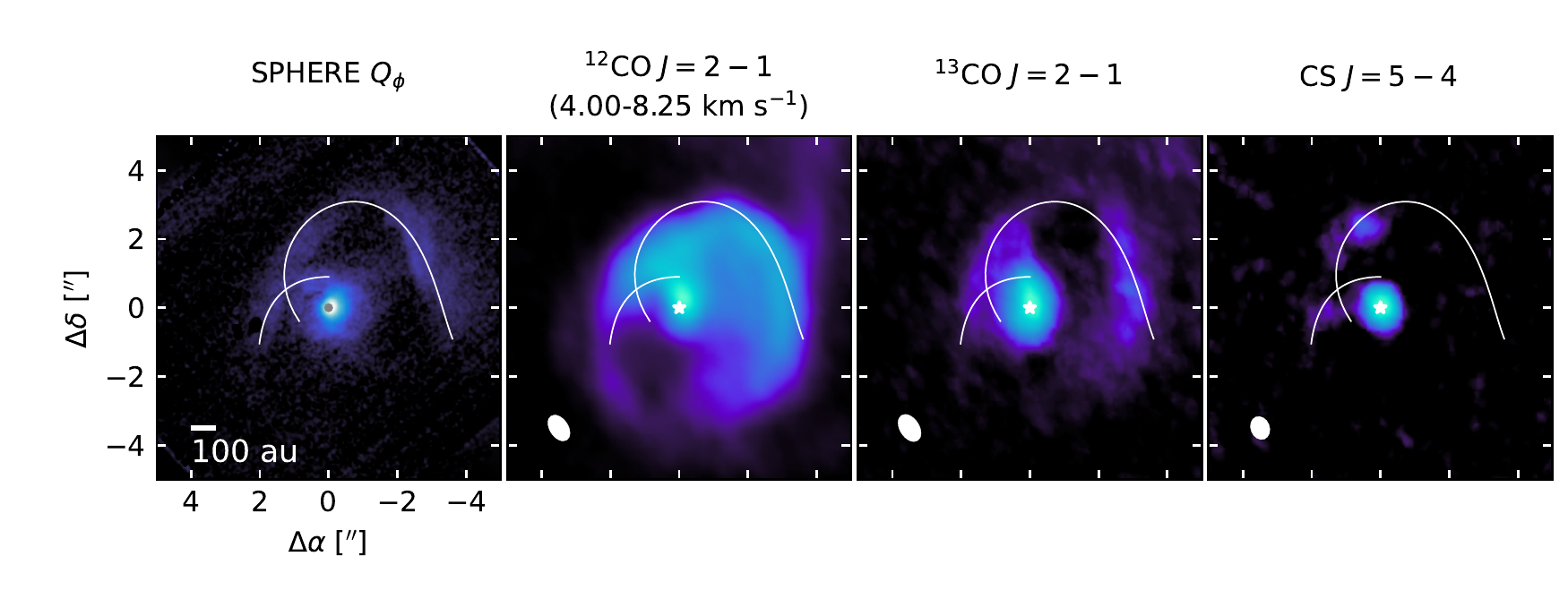}
\end{center}
\caption{A comparison of the SPHERE $Q_\phi$ image of DO Tau to insets of the $^{12}$CO, $^{13}$CO, and CS integrated intensity maps. The integration range of $^{12}$CO is truncated to 4.00-8.25 km s$^{-1}$ in order to minimize contamination from the outflows. The white curves mark the locations of the northern and eastern arms. The white star denotes the position of DO Tau. Offsets from the phase center (in arcseconds) are marked on the leftmost panel. North is up and east is to the left. \label{fig:northernarccomparison}}
\end{figure*}

\begin{figure*}
\begin{center}
\includegraphics{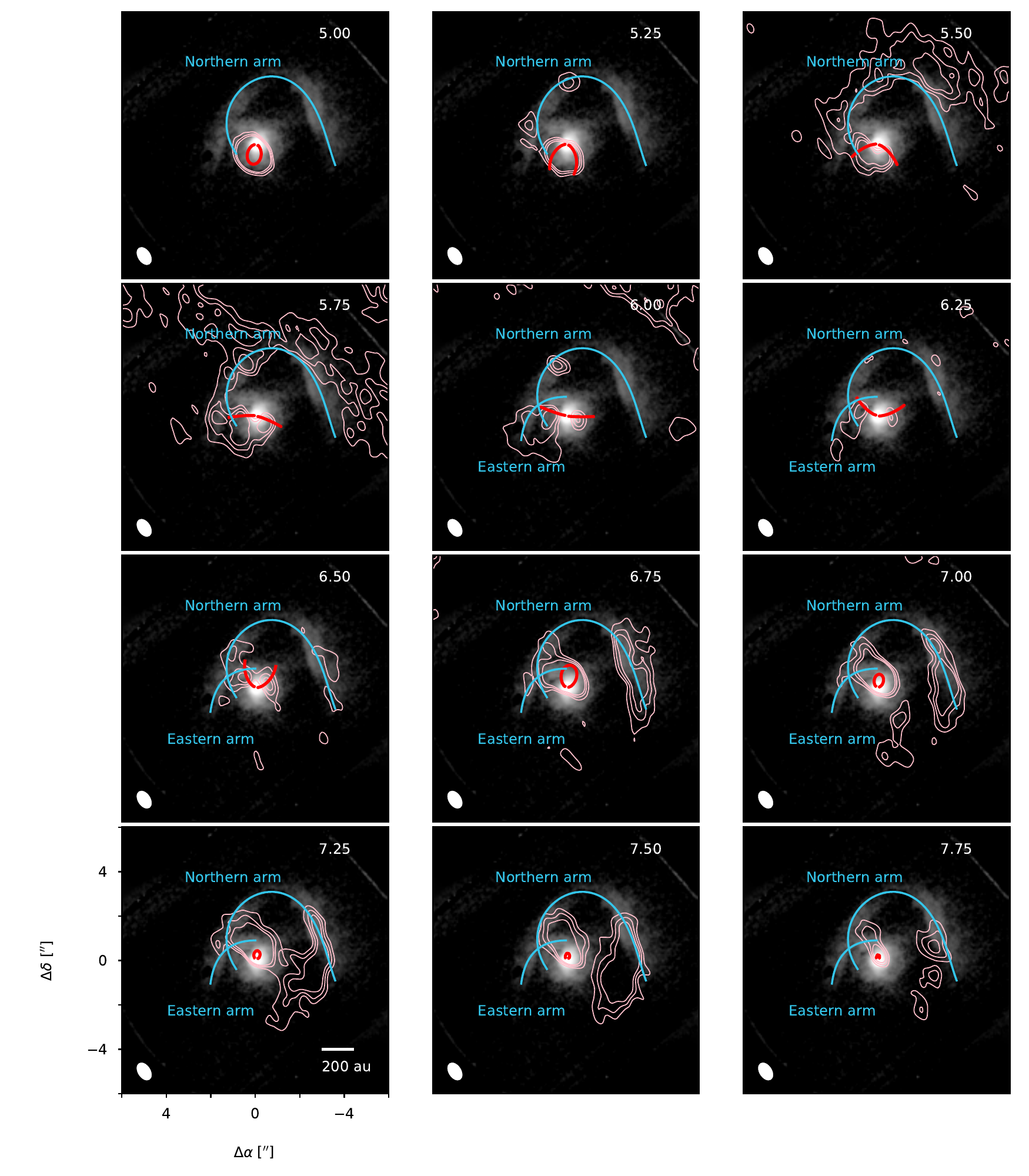}
\end{center}
\caption{A comparison of the northern and eastern arms as traced by polarized scattered light (black and white image) and $^{13}$CO emission. The $^{13}$CO contours are drawn at the 4, 6, and 8$\sigma$ levels. North is up and east is to the left. The synthesized beam is drawn as a white ellipse in the lower left corner of each panel, while the LSRK velocity (km s$^{-1}$) is in the top right corner. The blue curves mark the northern and eastern arms. The red curves mark the Keplerian disk.  \label{fig:northernarm13CO}}
\end{figure*}

The northern and eastern arms detected in scattered light also have counterparts in $^{12}$CO, $^{13}$CO and CS emission (Figure \ref{fig:northernarccomparison}). A C$^{18}$O counterpart is not visible in the moment maps, but a portion of it is tentatively visible at an LSRK velocity of 5.5 km s$^{-1}$ in the channel maps (Appendix \ref{sec:chanmaps}). In $^{12}$CO emission, the northern arm is only partially visible due to spatial filtering in channels near the systemic velocity and to contamination from other structures. The northern arm is more clearly visible in $^{13}$CO emission, and appears to extend further south compared to its scattered light counterpart. It is also possible, though, that this apparent southward extension is contamination from another structure. In CS, only the eastern side of the northern arm is detected. Peculiarly, the portion of the arm brightest in CS emission coincides with the region where the arm becomes more tenuous in $^{13}$CO emission. Observing additional transitions of CS and $^{13}$CO to better constrain the physical conditions across the arm may yield useful insight into why CS exhibits this emission pattern.  

The kinematics of the northern and eastern arms are best probed by $^{13}$CO, which is less contaminated by outflows compared to $^{12}$CO but has a higher signal-to-noise ratio compared to CS. Figure \ref{fig:northernarm13CO} compares the SPHERE $Q_\phi$ image to the $^{13}$CO channel maps. Emission from the northern arm is detected roughly between 5.00 and 7.75 km s$^{-1}$. The eastern end of this arm emerges from the side of the Keplerian disk that is blueshifted relative to the systemic velocity (5.9 km s$^{-1}$). The behavior of the western side of the northern arm is more difficult to interpret. Blueshifted emission overlaps the western side of the arm in the channels from 5.50 to 5.75 km s$^{-1}$. Little to no emission is detected on the western side of the arm in the channels from 6.00 to 6.25 km s$^{-1}$. Then, a bright redshifted component overlaps the northern arm from 6.50 to 7.75 km s$^{-1}$. The two distinct kinematic components may indicate that what appears to be a single continuous arm in the SPHERE image may actually be multiple structures overlapping in projection, or perhaps that the arm is expanding. Meanwhile, the eastern arm is detected between 6.00 and 7.75 km s$^{-1}$, emerging from the redshifted side of the Keplerian disk. 

In Section \ref{sec:scatteredlight}, we inferred from the total intensity and DoLP map that the northern arm is oriented such that the eastern side is tilted toward the observer. If this is indeed the case, then its blueshifted motion indicates that it is moving away from the disk. Meanwhile, the redshifted motion of the eastern arm indicates that it is moving opposite to the eastern side of the northern arm.

\begin{figure*}
\begin{center}
\includegraphics{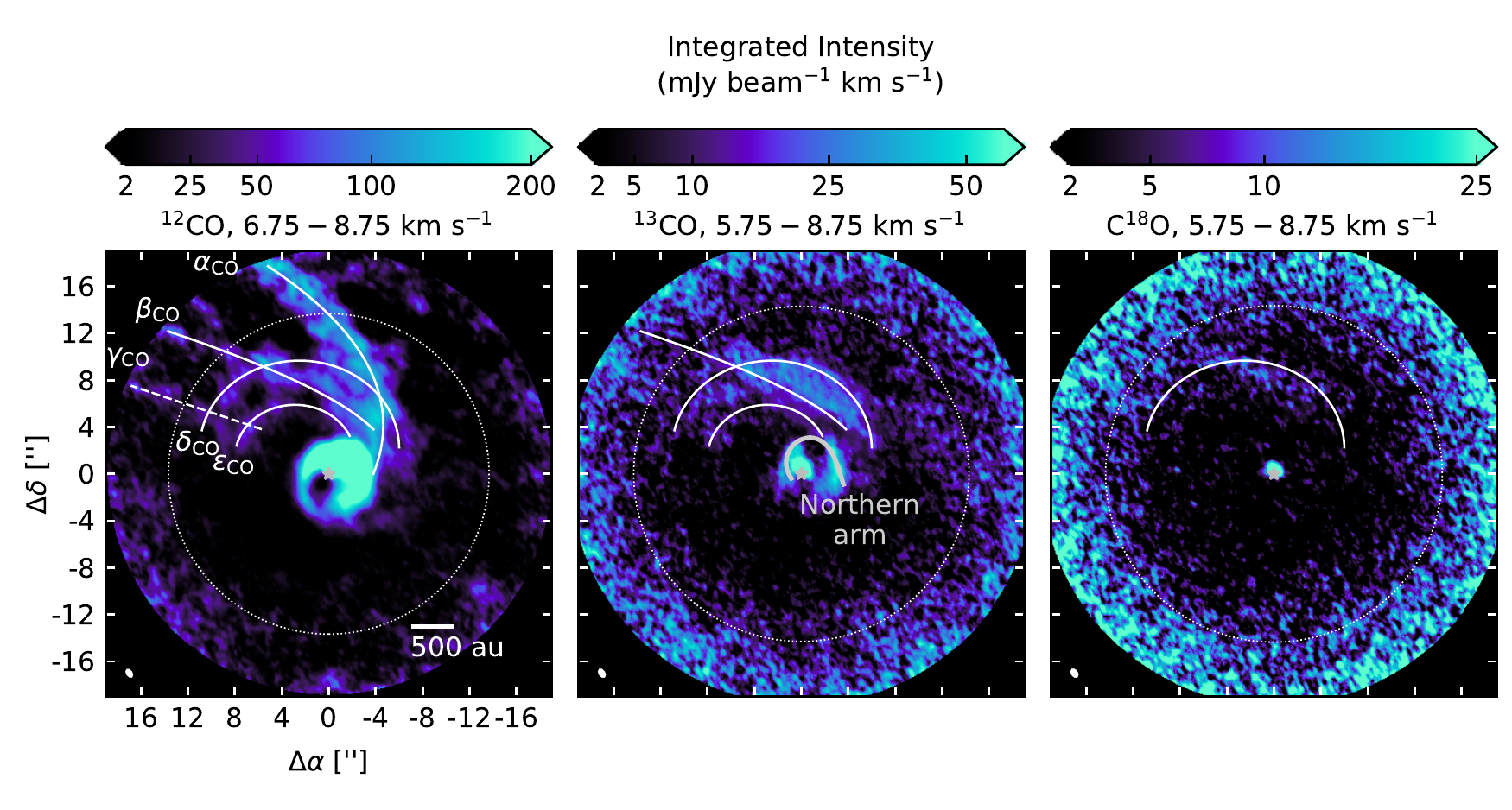}
\end{center}
\caption{A comparison of integrated intensity maps of stream-like structures detected in $^{12}$CO (left), $^{13}$CO (middle), and C$^{18}$O (right). The velocity range of the integrated intensity maps is labelled above each panel. The color scale is saturated to make faint features more visible. Due to severe spatial filtering between 5.25 and 6.5 km s$^{-1}$, $^{12}$CO is integrated over a different velocity range from the other two isotopologues. The locations of the $\alpha_\text{CO}$, $\beta_\text{CO}$, $\gamma_\text{CO}$, $\delta_\text{CO}$, and $\epsilon_\text{CO}$ streams are marked with white curves. The gray curve on the $^{13}$CO image shows the position of DO Tau's northern arm. The gray star marks the position of DO Tau in each panel. The white dotted circle marks the FWHM of the primary beam. \label{fig:isotopologuestreams}}
\end{figure*}

\subsection{Stream-like structures}
We detect several stream-like structures (labeled $\alpha_\text{CO}$, $\beta_\text{CO}$, $\delta_\text{CO}$, and $\epsilon_\text{CO}$) northeast of DO Tau in $^{12}$CO emission at velocities between 6.75 and 8.75 km s$^{-1}$ (Figure \ref{fig:isotopologuestreams}), which is redshifted relative to DO Tau's systemic velocity of 5.9 km s$^{-1}$. We also tentatively identify a $\gamma_\text{CO}$ feature. Its geometry is less certain due to low sensitivity outside the primary beam FWHM as well as confusion with the $\delta_\text{CO}$ and $\epsilon_\text{CO}$ structures. In $^{13}$CO emission, we identify counterparts to $\beta_\text{CO}$, $\delta_\text{CO}$, and $\epsilon_\text{CO}$. Much of this emission appears at a velocity range (5.75 km s$^{-1}$ to 6.5 km s$^{-1}$) where $^{12}$CO exhibits severe spatial filtering. Thus, in the intensity-weighted velocity maps (Figure \ref{fig:mom0maps}), the stream-like structures appear blueshifted in $^{13}$CO compared to $^{12}$CO. C$^{18}$O also exhibits extended emission that overlaps with the $\delta_\text{CO}$ feature. 

\begin{figure*}
\begin{center}
\includegraphics{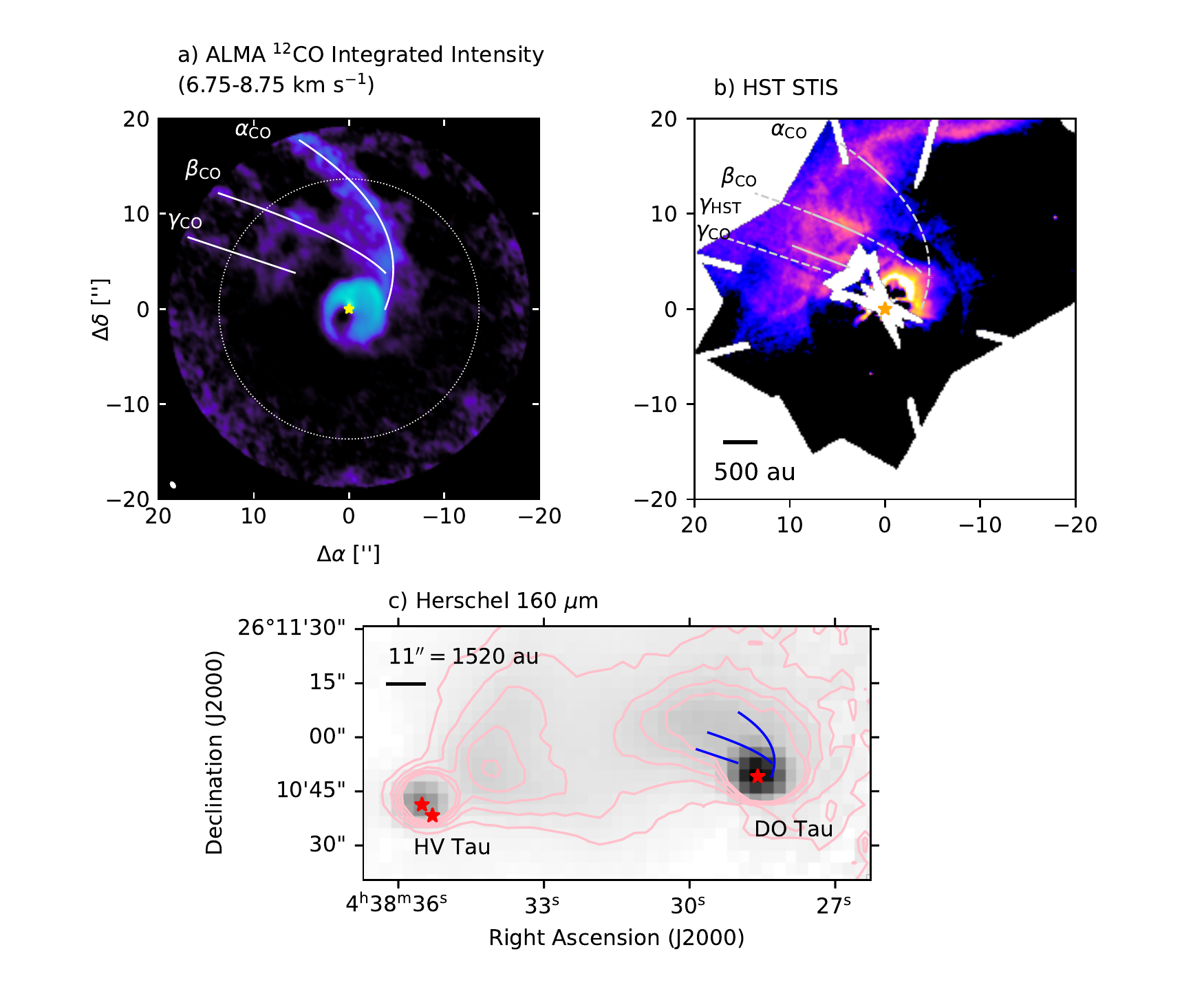}
\end{center}
\caption{a) ALMA $^{12}$CO integrated intensity map of DO Tau, annotated in white with the locations of the  $\alpha_\text{CO}$, $\beta_\text{CO}$, and $\gamma_\text{CO}$ streams. Only velocities from 6.75 to 8.75 km s$^{-1}$ are included. The gray star marks the position of DO Tau. Axes are marked with the angular offsets with respect to the star. b) HST STIS image of DO Tau, with the dashed gray curves denoting the locations of the $\alpha_\text{CO}$, $\beta_\text{CO}$, and $\gamma_\text{CO}$ streams. The solid gray curves mark the $\alpha$, $\beta$, and $\gamma$ features identified within the HST image from Section \ref{sec:scatteredlight}. The yellow star marks the location of DO Tau. The bright end of the color scale is saturated in order to show the contrast between the large-scale structures more clearly. c) Herschel 160 $\mu$m image showing the nebulosity surrounding HV Tau and DO Tau. The blue curves correspond to the white annotations drawn on the ALMA image denoting the $\alpha_\text{CO}$, $\beta_\text{CO}$, and $\gamma_\text{CO}$ streams. The pink contours correspond to intensity levels of [5, 10, 15, 20, 25] mJy pixel$^{-1}$. The red stars mark the locations of HV Tau and DO Tau. (Only one star is used to represent HV Tau A and B due to their close proximity. HV Tau C is to the northeast of HV Tau A and B, as shown in Figure \ref{fig:SDSS}). \label{fig:filamentcomparison}}
\end{figure*}

In projection, $\alpha_\text{CO}$, $\beta_\text{CO}$, $\delta_\text{CO}$, and $\epsilon_\text{CO}$ all appear to converge on the western side of DO Tau's northern arm. The $\alpha_\text{CO}$ and $\beta_\text{CO}$ streams can be traced up to the edge of the image, with a projected separation of $\sim19''$ ($\sim2600$ au) from DO Tau. The appearance of the streams is adversely affected by foreground contamination, loss of sensitivity away from the phase center, and spatial filtering (the maximum recoverable scale of the ALMA configuration used is $\sim4''$). However, the correspondence between the $\alpha_\text{CO}$ and $\beta_\text{CO}$ and the $\alpha$ and $\beta$ streams detected in the HST STIS image lends greater confidence to the characterization of their geometry (Figure \ref{fig:filamentcomparison}). The $\gamma_\text{CO}$ stream appears to be slightly offset to the southeast from the $\gamma$ stream identified with HST, but observations with better $uv$ coverage and sensitivity should be obtained to confirm whether this offset is genuine. 

A comparison of the stream locations with the Herschel PACS 160 $\mu$m image (Figure \ref{fig:filamentcomparison}) suggests that the $\alpha_\text{CO}$ and $\beta_\text{CO}$ (and perhaps $\gamma_\text{CO}$) streams form the base of an elongated structure that extends northeast of DO Tau. This elongated structure has previously been described as a tidal tail that forms part of a bridge structure connected to HV Tau \citep{2018MNRAS.479.5522W}. Since the Herschel PACS image has relatively low spatial resolution ($\sim11''$ at 160 $\mu$m), we are unable to identify individual infrared counterparts to the ALMA and HST streams.

\begin{figure*}
\begin{center}
\includegraphics{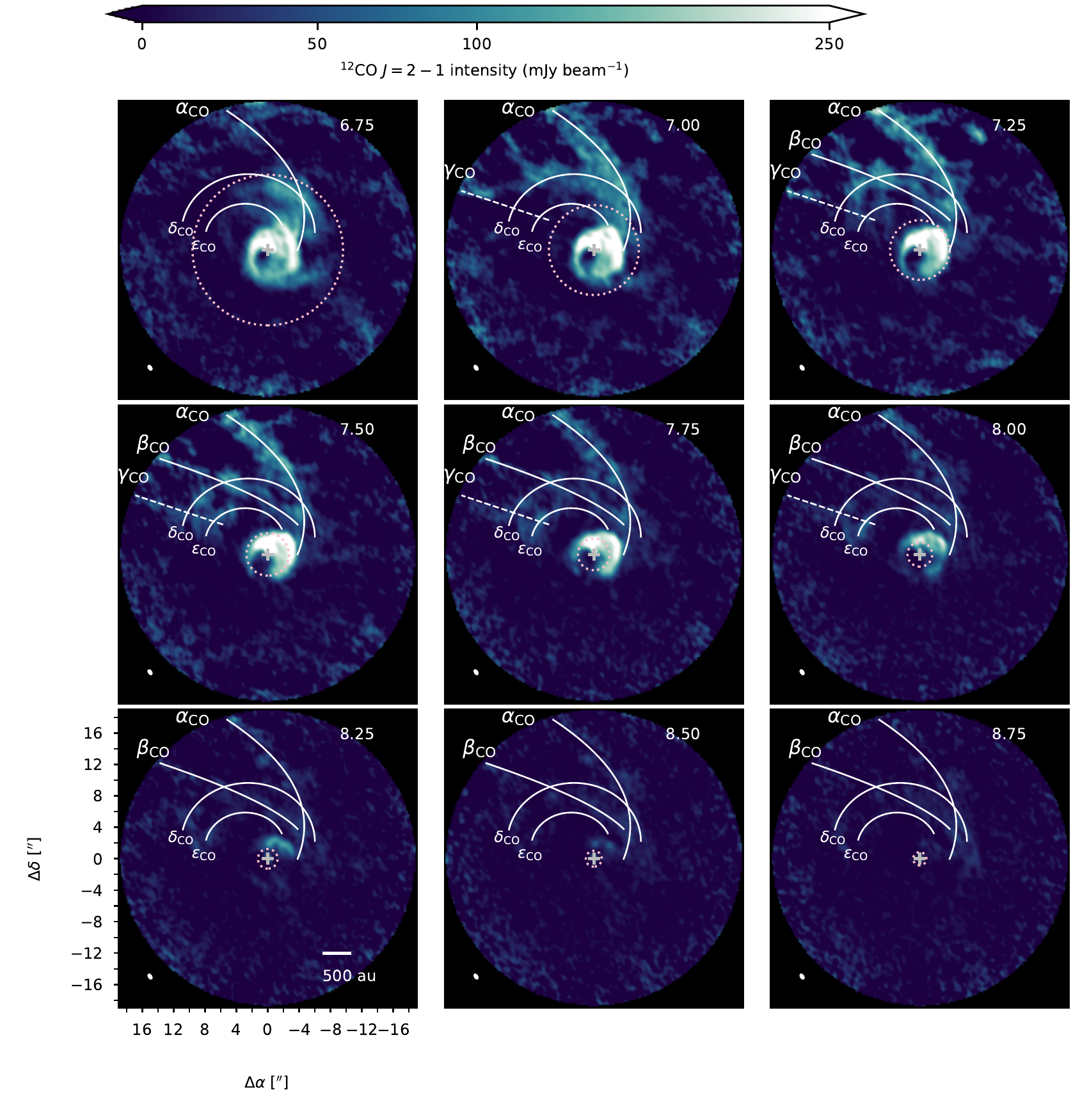}
\end{center}
\caption{$^{12}$CO $J=2-1$ channel maps annotated in white with the locations of the $\alpha_\text{CO}$, $\beta_\text{CO}$, $\gamma_\text{CO}$, $\delta_\text{CO}$, and $\epsilon_\text{CO}$ streams. The $\gamma_\text{CO}$ feature is marked with a dashed rather than a solid line because its identification is less certain. The dotted pink circle denotes the radius at which the line of sight velocity relative to DO Tau is equal to the escape velocity. The LSRK velocity (km s$^{-1}$) appears in the top right of each panel. The synthesized beam is shown as a solid white ellipse in the lower left corner. A gray cross marks the position of the millimeter continuum peak, which is located at the phase center.  Offsets from the phase center (in arcseconds) are marked in the lower left panel. The top end of the color bar is saturated in order to show the faint extended features in more detail. North is up and east is to the left.  \label{fig:filamentchanmaps}}
\end{figure*}

To determine whether the stream-like structures are gravitationally bound, we compute the radius $r$ at which the line of sight velocity relative to DO Tau is equal to the escape velocity, $\sqrt{\frac{2GM_\star}{r}}$. Structures outside this radius must be moving faster than the escape velocity and are thus gravitationally unbound. (Structures inside this radius may or may not be unbound depending on what the other velocity components are.) In making this calculation, we assume that the mass of the gas and dust surrounding DO Tau is small compared to the stellar mass of $0.54\pm0.07$ $M_\odot$ \citep{2021ApJ...908...46B}. The resulting escape velocity radii are plotted on Figure \ref{fig:filamentchanmaps}. Since the streams generally lie outside these radii, we conclude that they are not gravitationally bound.

\subsection{Mass constraints on structures surrounding DO Tau}
\begin{figure}
\begin{center}
\includegraphics{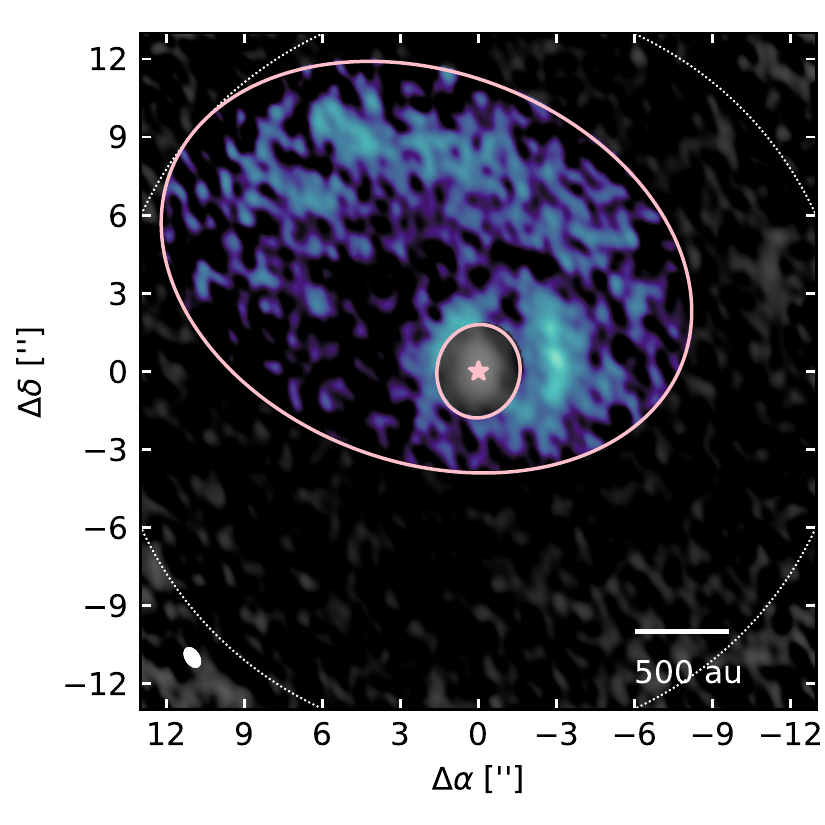}
\end{center}
\caption{$^{13}$CO integrated intensity map with the colored region between the two solid pink ellipses showing where the flux from the extended structures is measured. The star marks the location of DO Tau. The white dotted circle marks the FWHM of the primary beam. \label{fig:massarea}}
\end{figure}

Under the assumption that $^{13}$CO is optically thin outside DO Tau's Keplerian disk, we can use it to estimate a lower bound for the gas mass of the structures around the disk. However, it should be kept in mind that the flux may be significantly underestimated due to spatial filtering. Furthermore, the ALMA field of view does not capture the full extent of these structures. 

In the LTE limit, column density can be estimated from optically thin emission with: 

\begin{equation}\label{eq:coldensity}
    N = \frac{Q(T)}{g_u}\exp{\left(\frac{E_u}{T}\right)}\frac{4\pi}{A_{ul}hc}\int I_\nu dv,
\end{equation}
where $Q$ is the partition function, $g_u$ is the upper state degeneracy, $E_u$ is the upper state energy, $T$ is the excitation temperature,  $A_{ul}$ is the Einstein $A$ coefficient, and $\int I_\nu dv$ is the integral of the intensity over the velocity axis \citep[e.g.,][]{1999ApJ...517..209G}. For $^{13}$CO $J=2-1$, $g_u=10$ (note that the Cologne Database for Molecular Spectroscopy includes hyperfine splitting in its partition function and degeneracy), $E_u=15.9$ K, and $A_{ul}=6.08\times10^{-7}$ s$^{-1}$ \citep{2001AA...370L..49M, 2005JMoSt.742..215M}.  

To estimate the flux, we made a $^{13}$CO integrated intensity map without clipping in order to avoid biasing measurements upward. We then extracted the flux within the highlighted region shown in Figure \ref{fig:massarea}, finding a value of $\sim4.5$ Jy km s$^{-1}$. This value excludes emission from the disk, which we estimate has a radial extent of $\sim250$ au based on visual inspection. Although the boundary between the disk and the extended structures is not well-defined, the exact choice of value does not affect the order of magnitude of the mass estimate for the extended structures since they dominate the flux due to their large emitting area. The mean integrated intensity within the region of extraction is $\sim11$ mJy beam$^{-1}$ km s$^{-1}$. To derive a mean column density, we adopt a gas temperature of 30 K, based on temperatures that  \citet{2018MNRAS.479.5522W} estimated from Herschel dust maps of the region around DO Tau. At this temperature, $Q=23.4$ for $^{13}$CO \citep{2001AA...370L..49M, 2005JMoSt.742..215M}. We find a mean $^{13}$CO column density of $\sim3\times10^{14}$ cm$^{-2}$. We assume that the composition of the gas is similar to that of local molecular clouds, i.e., the gas is predominantly molecular hydrogen, $^{12}$CO:H$_2=10^{-4}$, and  $^{12}\text{CO}$:$^{13}\text{CO}=69$ \citep[e.g.,][]{1982ApJ...262..590F, 1999RPPh...62..143W}. We estimate a lower bound of $\sim4\times10^{-4}$ $M_\odot$ for the gas mass of the extended structures. Since $^{13}$CO is likely optically thick within the Keplerian disk, we do not use it to estimate the disk mass. However, \citet{2015ApJ...808..102K} estimated a total disk mass of 0.014 $M_\odot$ based on radiative transfer modeling of the millimeter continuum. Thus, the mass of the extended structures appears to be at least a few percent of the disk mass.    

Based on Herschel far-infrared maps of dust emission, \citet{2018MNRAS.479.5522W} derived a lower bound of $10^{-2}$ $M_\odot$ for the total mass of extended structures surrounding DO Tau and HV Tau. This is much higher than our lower bound of $\sim4\times10^{-4}$ $M_\odot$, which is due at least in part to the Herschel measurements being taken over a much larger area. DO Tau and HV Tau have a projected separation of $\sim91''$, whereas the FWHM of the primary beam of the ALMA $^{13}$CO image is only $\sim29''$. Spatial filtering of $^{13}$CO is almost certainly responsible for part of the discrepancy as well. Other possibilities are that the gas-to-dust ratio or $^{12}$CO:H$_2$ ratio differs from ISM values. This could be the case if the material originated from within the protoplanetary disks, since the gas-to-dust and $^{12}$CO:H$_2$ ratios in disks are both expected to decrease over time \citep[e.g.,][]{2015ApJ...804...29G, 2019AA...631A..69M}. \citet{2022arXiv220104089S} find evidence from [C I] observations that carbon is depleted by a factor of 17 from ISM levels in the outer regions of the DO Tau disk, which would in turn reduce the $^{12}$CO:H$_2$ ratio.  Widefield observations with better $uv$ coverage will be necessary to robustly examine the consistency between the $^{13}$CO-derived mass and the Herschel-derived mass.

\subsection{Summary of structures associated with DO Tau}
\begin{figure*}
\begin{center}
\includegraphics{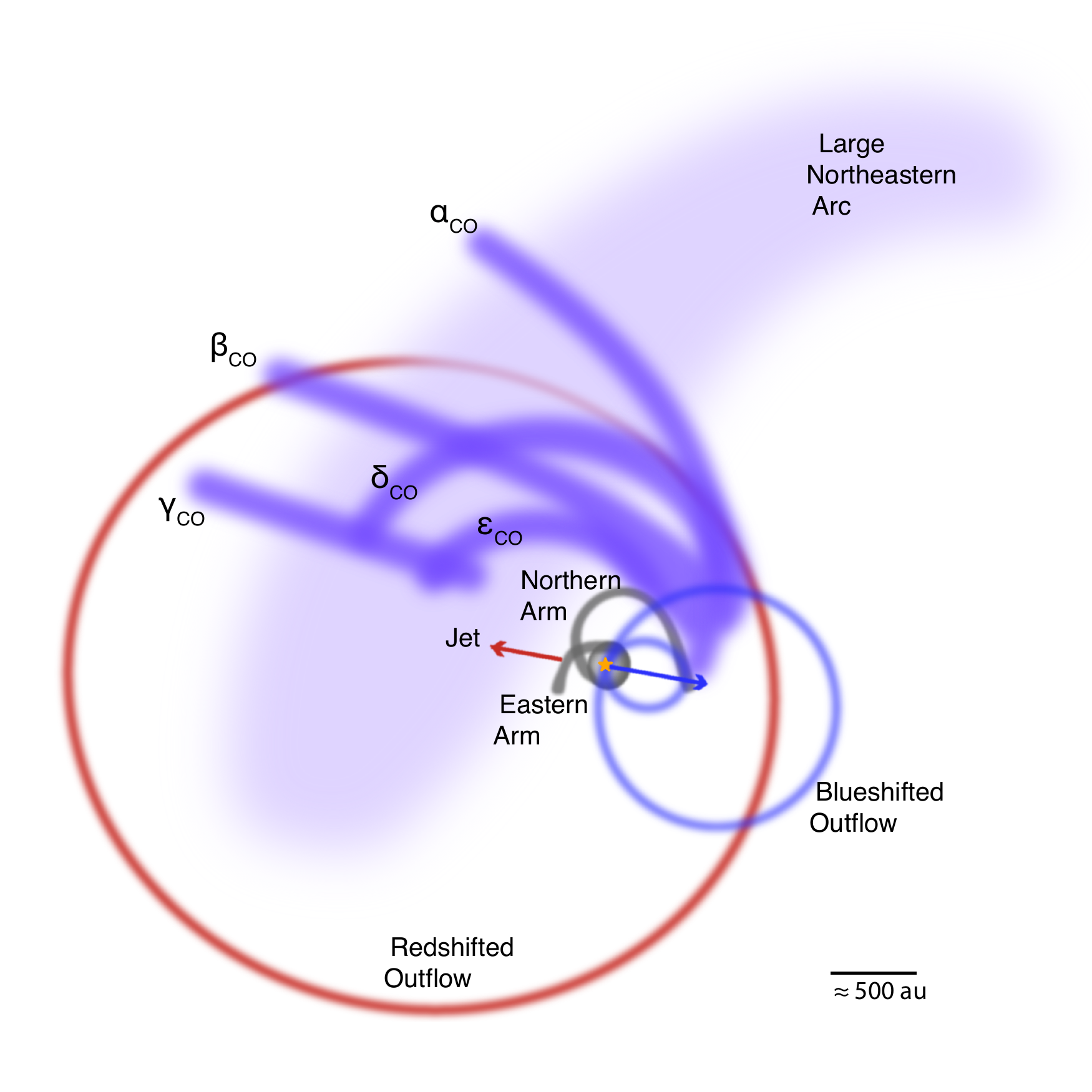}
\end{center}
\caption{A schematic of the structures surrounding DO Tau, based on SPHERE and HST scattered light observations and ALMA molecular line observations presented in this work, as well as jet observations from \citet{2021AA...650A..46E}. The yellow star marks the position of DO Tau. All other structures are labelled within the schematic. Note that the CO streams likely extend further eastward than depicted in this schematic, but we are limited by ALMA and HST's field of view.  \label{fig:schematic}}
\end{figure*}

Figure \ref{fig:schematic} presents a schematic showing the spatial relationship between the various structures around DO Tau identified in scattered light and molecular line emission. The smallest structure characterized is the circumstellar disk itself, with a radial extent of $\sim80$ au in scattered light. The northern and eastern arms, both connected to the disk, trace structures up to several hundred au from DO Tau. The streams and the large northeastern arc are much larger in scale, with extents of several thousand au. The scales of the molecular outflow shells range from hundreds to thousands of au.


\section{Discussion \label{sec:discussion}}
Here we discuss potential origins for the arms, stream-like structures, and large northeastern arc near DO Tau. We limit discussion of the jet and molecular outflow, since they have already recently been analyzed at length by \citet{2020AJ....159..171F} and \citet{2021AA...650A..46E}.

\subsection{Evidence for an encounter between DO Tau and HV Tau}
\citet{2018MNRAS.479.5522W} presented stellar encounter simulations reproducing the general morphology of the dust ``bridge'' connecting DO Tau and HV Tau, as imaged by Herschel. They inferred that DO Tau and HV Tau began as a quadruple system, but then subsequently became either unbound or only loosely bound. However, the simulation resolution was not high enough for comparisons with the complex network of structures detected by HST, SPHERE, and ALMA. Arm-like structures reminiscent of DO Tau's northern and eastern arms, though, do appear in high-resolution stellar flyby simulations of other systems \citep[e.g.,][]{1993MNRAS.261..190C,2008AA...487L..45P, 2015MNRAS.449.1996D, 2019MNRAS.483.4114C}. 

The DO Tau system also bears notable similarities to the RW Aur system, which has been hypothesized to have undergone a recent stellar encounter \citep[e.g.,][]{2015MNRAS.449.1996D, 2018ApJ...859..150R}. Like DO Tau, RW Aur features multiple arm-like structures in CO emission \citep{2006AA...452..897C, 2018ApJ...859..150R}. Both systems have also been noted for their unusually strong photometric variability on timescales of several weeks or longer \citep{2016AJ....151...29R, 2017MNRAS.465.3889R}. Among other explanations, a misaligned disk has been hypothesized to be responsible for RW Aur's dimming events \citep{2016AA...596A..38F, 2019AA...625A..49K}. A similar link may exist for DO Tau, which has been hypothesized to have a misaligned inner disk based on evidence of a precessing jet \citep{2021AA...650A..46E}. Hydrodynamical simulations have demonstrated that stellar encounters can lead to disk warps and misalignments, albeit with rapid realignment timescales \citep[e.g.,][]{1993MNRAS.261..190C, 2019MNRAS.483.4114C, 2020MNRAS.491.4108N}. In general, disk misalignments have been more commonly attributed to the presence of a misaligned embedded companion, which simulations suggest can yield more extreme and longer-lasting disk warps compared to stellar encounters \citep[e.g.,][]{2018MNRAS.473.4459F, 2019MNRAS.483.4221Z, 2020MNRAS.491.4108N}. Nevertheless, the evidence for misalignments in both the RW Aur A and DO Tau disks motivates further examination of the long-term impact of external perturbations on the evolution of the inner disk. Characterizing the mechanisms that can lead to disk misalignment is key for determining the extent to which processes that occur during the gas-rich protoplanetary disk phase may be responsible for the misaligned orbits observed in mature planetary systems \citep[e.g.,][]{2012Natur.491..418B, 2014MNRAS.440.3532L, 2016ApJ...830....5S}. 

One characteristic that sets DO Tau apart from most other systems that have been hypothesized to have undergone a stellar encounter (e.g., RW Aur, AS 205, UX Tau, FU Ori) is that DO Tau's current projected separation from HV Tau is $\sim12,600$ au, whereas the current projected separations between the other hypothesized stellar encounter pairs are typically only on the order of a few hundred au \citep[e.g.,][]{2010MNRAS.402.1349F, 2018ApJ...869L..44K, 2020ApJ...896..132Z, 2020AA...639L...1M, 2022MNRAS.510L..37B}.  \citet{2022NatAs...6..331D} recently identified a more extreme case in which Z CMa appears to have undergone an encounter with a source now separated in projection by $\sim4700$ au, but this is still considerably smaller than the projected separation between DO Tau and HV Tau. A difficulty with the stellar encounter scenario for DO Tau and HV Tau is that in general, simulations suggest that arm-like structures produced by stellar encounters should dissipate within several thousand years \citep[e.g.,][]{2019MNRAS.483.4114C}. However, a kinematic analysis in \citet{2018MNRAS.479.5522W} indicates that an encounter between DO Tau and HV Tau would have had to take place $\sim0.1$ Myr ago to match their current separation. (The simulated bridge structure between DO Tau and HV Tau in \citet{2018MNRAS.479.5522W} starts to dissipate before the stars reach their present-day separation, but this behavior was due to numerical limitations of the SPH simulation setup). 

\citet{2018MNRAS.479.5522W} raised the caveat that the proper motions of DO Tau and HV Tau AB suggest that rather than moving away from each other (as one would expect if DO Tau had been ejected after an encounter), the systems may be moving toward one another. The \textit{Gaia} EDR3 \citep{2021AA...649A...1G} proper motion of DO Tau is $\mu_{\alpha^*} = 6.27\pm0.04$ mas yr$^{-1}$, $\mu_\delta = -21.05\pm0.03$ mas yr$^{-1}$ and the proper motion of HV Tau AB is $\mu_{\alpha^*} = 4.78\pm0.06$ mas yr$^{-1}$, $\mu_\delta = -21.4\pm0.05$ mas yr$^{-1}$. Proper motion information is not available for HV Tau C. However, \citet{2018MNRAS.479.5522W} also point out that since HV Tau AB is not resolved by \textit{Gaia}, its derived proper motion does not necessarily correspond to that of its center of mass and could therefore give an inaccurate impression of its velocity relative to DO Tau (see also \citet{2020MNRAS.491L..72C}). 

Stellar ejections have been hypothesized as a way to power molecular outflows \citep[e.g.,][]{2011ApJ...727..113B}. Based on the the dynamical ages of the outflow shells estimated in this work and \citet{2020AJ....159..171F}, though, the launch of these outflow shells appears to be a more recent event than the hypothesized encounter between DO Tau and HV Tau, which \citet{2018MNRAS.479.5522W} estimated took place 0.1 Myr ago. Nevertheless, it is worth considering how the after-effects of the stellar encounter might have affected the behavior of the outflow, which is noticeably asymmetric. One possible cause for jet and outflow asymmetries is environmental heterogeneity \citep[e.g.,][]{1994ApJ...427L..99H, 2002ApJ...575..911A}. DO Tau's local environment is quite complex, which may be due in part to the disruptive influence of a stellar encounter. However, molecular clouds are also intrinsically heterogeneous, so it is not necessarily straightforward to isolate the influence of an encounter, as discussed below.  

\subsection{Evidence for remnant envelope material}
\begin{figure}
\begin{center}
\includegraphics[scale=0.5]{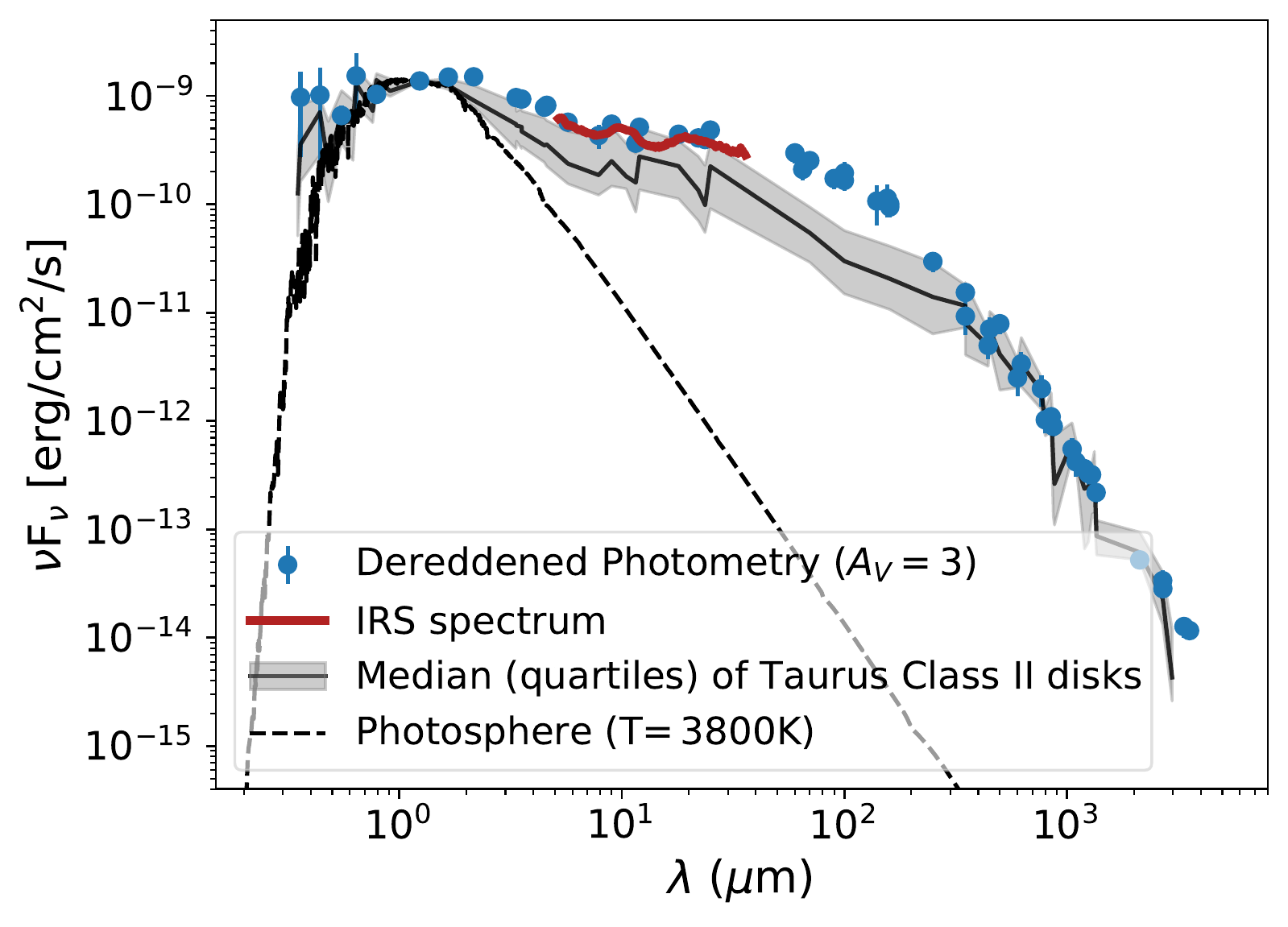}
\end{center}
\caption{A comparison between the spectral energy distribution of DO Tau and other Class II disks in Taurus. The blue points (with $1\sigma$ error bars) denote dereddened photometric measurements presented in \citet{2013ApJ...771..129A} and \citet{2017ApJ...849...63R}. An $A_v$ value of 3 is adopted based on \citet{2013ApJ...771..129A}. The red curve corresponds to the Spitzer/IRS spectrum retrieved from the Cornell Atlas of Spitzer/IRS Sources \citep{2011ApJS..196....8L, 2015ApJS..218...21L}. The dashed black curve shows the estimated stellar photosphere. The solid black curve corresponds to the median dereddened Taurus Class II SED from  \citet{2017ApJ...849...63R}, while the gray shaded area corresponds to the interquartile range. \label{fig:SED}}
\end{figure}

DO Tau has traditionally been identified as a Class II system based on its SED \citep[e.g.,][]{1995ApJS..101..117K, 2010ApJS..186..111L}. Because of the disk's bright gas and dust emission, it has frequently been included in studies aimed at characterizing trends in Class II disk properties \citep[e.g.,][]{2013ApJ...766..134N, 2013ApJ...776...21H, 2014ApJ...788...59W, 2019ApJ...882...49L, 2019ApJ...876...25B}. Stellar age estimates for DO Tau have spanned a wide range, from $\sim0.4-6$ Myr \citep[e.g.,][]{2018ApJ...865..157A,  2019ApJ...882...49L}, but the system exhibits several attributes more characteristic of the younger side of the estimated age range. DO Tau's large northeastern arc resembles the arc-like reflection nebulae that have been observed around the partially embedded HL Tau and DG Tau systems \citep[e.g.][]{1995AJ....109.1181N, 1995ApJ...449..888S}. In addition, CO emission maps of HL Tau and DG Tau have revealed arcs and streams on scales of hundreds to thousands of au, reminiscent of those detected in the CO maps of DO Tau \citep[e.g.,][]{2017AA...608A.134Y, 2018AA...620L...1G}. DO Tau's strong CO outflow is also suggestive of youth, although molecular outflows have on rare occasions been detected in other Class II systems \citep[e.g.,][]{2014prpl.conf..451F, 2018AA...618A.120L}. Another potential indication of relative youth is DO Tau's spectral energy distribution (Figure \ref{fig:SED}), which exhibits higher mid-IR and far-IR emission than most other Class II systems in Taurus. (While a value of $A_v=3$ from \citet{2013ApJ...771..129A} was assumed in order to de-redden DO Tau's SED, the discrepancy with other Taurus Class II sources is still present if we instead adopt $A_v=0.75$ from \citet{2014ApJ...786...97H}.) This excess may be due to the presence of residual envelope material and/or to DO Tau's disk being comparatively less settled \citep[e.g.,][]{1989ApJ...340..823W, 2004AA...421.1075D}. Both characteristics are often (though not exclusively) associated with younger systems. 

While the simulations in \citet{2018MNRAS.479.5522W} could qualitatively reproduce the large bridge structure connecting DO Tau and HV Tau solely through ejection of disk material following a close encounter, the morphological similarities between DO Tau and sources known to be partially embedded raise the question of whether some or perhaps even most of the material surrounding DO Tau actually originated from outside the disk. The formation of bridge-like structures between stars is not uniquely an outcome of tidal stripping by stellar encounters. Bridge-like structures appear in simulations of binary formation through turbulent fragmentation, although separations exceeding $10^4$ au are uncommon past 100 kyr \citep[e.g.,][]{2016ApJ...827L..11O, 2019ApJ...887..232L}. Simulations by \citet{2019AA...628A.112K} demonstrate that bridge-like structures can form between protostars in multiple systems due to compression of their natal filament. However, given that bridge structures that form in this manner only survive for a few tens of thousands of years in the simulations, this mechanism does not provide a fully satisfactory explanation for the bridge structure connecting DO Tau and HV Tau. 

The possibilities that DO Tau underwent a stellar encounter and that the system is partially embedded are not mutually exclusive. Stellar interactions are expected to be more probable during the embedded phase because protostars often form in close groups and because the presence of an envelope increases the cross-section of a system \citep[e.g.,][]{2005ApJ...632..397G, 2018MNRAS.475.5618B}. The large-scale structures around DO Tau could therefore represent some combination of disk material stripped by a stellar encounter and disturbed remnant envelope material. 

At any rate, the observations of DO Tau provide further motivation to re-examine the environments of other disks traditionally considered to be Class II disks to determine if they likewise are (partially) embedded or interacting with cloud material. Star formation simulations from \citet{2014ApJ...797...32P} suggest that Class II sources experience infall from nearby cloud material even up to ages of at least several million years. In this case, DO Tau's complex large-scale structures would not necessarily be indicative of youth. A growing number of Class II disks have been identified as potentially experiencing ``late infall,'' at ages of roughly a Myr or more  \citep[e.g.,][]{2012AA...547A..84T, 2019AA...628A..20D, 2020AA...642A.119G, 2021ApJS..257...19H, 2022AA...658A..63M}. Ongoing accretion of external material during the Class II phase has been suggested as a way to increase the budget of material available to form planets within a given disk \citep{2018AA...618L...3M, 2019AA...628A..20D}.

\subsection{Avenues for clarifying the origins of DO Tau's extended structures}

Determining the extent to which the material surrounding DO Tau is primordial is key for understanding how disks evolve. Under the assumption that all the material surrounding DO Tau and HV Tau was stripped from their disks, the simulations in \citet{2018MNRAS.479.5522W}'s indicated that the stellar encounter caused the disks to lose half their mass. However, if some of the external material is primordial, then the impact of stellar encounters on disk mass may not be so extreme. Furthermore, if DO Tau is still interacting with its envelope, it may not be appropriate to group it with envelope-free Class II disks in population studies. 

One limitation in our analysis is that some of the structures detected by ALMA extend beyond its field of view, so we do not know their complete morphology. While the low-resolution Herschel dust maps suggest that the streams extend toward HV Tau, mosaicked ALMA observations would be needed to check if they are connected. Meanwhile, the kinematics of the large northeastern arc are unknown because it does not have a clear counterpart in the molecular emission maps, which could be due to spatial filtering. Mosaicked maps with better $uv$ coverage could be used to investigate whether the structure is gravitationally bound to DO Tau. 

High resolution hydrodynamical simulations of stellar encounters with disks both with and without envelopes could help to clarify where DO Tau's extended structures might have originated. Attempting to reproduce the morphology and kinematics of DO Tau's structures in detail would be a very computationally formidable task, though, because of the wide range of size scales involved and the large space of initial conditions that would need to be explored. Simulations spanning disk to sub-parsec scales have generally focused on the first couple hundred thousand years of a protostellar system's life \citep[e.g.,][]{2017ApJ...846....7K, 2018MNRAS.475.5618B, 2018MNRAS.475.2642K,2021ApJ...917L..10L}, but observations of complex large-scale structures associated with more evolved systems are highlighting the need to extend these types of simulations to later stages.
 
Another potential avenue to distinguish between material ejected from the disk and remnant material from the star-formation process is to compare the composition of the gas around DO Tau to that of (partially) embedded systems, isolated Class II disks, and Class II stellar flyby candidate systems. While spatially resolved molecular observations of Class I and II disks have been steadily growing \citep[e.g.,][]{2020ApJ...890..142P, 2020ApJ...898...97B, 2020ApJ...901..166V}, published spatially resolved molecular line observations of stellar flyby candidates have largely been limited to CO \citep[e.g.,][]{2018ApJ...859..150R, 2018ApJ...869L..44K, 2020ApJ...896..132Z}. Further molecular line observations of the latter type of object, in conjunction with astrochemical modeling, will be needed to determine whether material stripped from the disk has a chemical signature distinct from envelope material. For example, CS emission in the embedded DG Tau system exhibits arm-like structures reminiscent of those detected around DO Tau \citep{2022AA...658A.104G}. However, it remains to be seen whether other stellar flyby candidate systems have similar CS emission arms. Other species worth targeting in DO Tau and other stellar flyby candidates include SO and SO$_2$, which often appear to be associated with accretion shocks in embedded systems but are rarely detected in envelope-free Class II systems \citep[e.g.,][]{2014Natur.507...78S,2019AA...626A..71A, 2022AA...658A.104G}. It will also be useful to improve constraints on DO Tau's C/O ratio, which is expected to increase between the Class I and II stages \citep[e.g.,][]{2018ApJ...865..155C,2019AA...631A..69M}. Results have thus far been mixed; \citet{2022arXiv220104089S} derived a relatively high C/O ratio of $\sim1$ based on spatially unresolved [C I] obervations but found that the C$_2$H upper limit from \citet{2019ApJ...876...25B} was better explained by an ISM-like ratio of 0.47. Obtaining deeper and higher resolutions of [C I] and C$_2$H, or estimating the C/O ratio from SO/CS, can help to resolve this ambiguity.

While DO Tau is not itself an FU Ori object, \citet{2004AA...420..975M} suggested that the presence of nearby Herbig-Haro objects may be evidence of a past FU Ori-like outburst. The complex system of extended structures around DO Tau are also reminiscent of scattered light and molecular emission images of FU Ori systems \citep{2016SciA....2E0875L, 2017MNRAS.465..834Z}. The mechanism responsible for triggering the strong outbursts in FU Ori-type objects (and potentially sculpting complex large-scale structures around them) is still debated; hypotheses include gravitational instability, accretion from surrounding cloud material, binary interactions, and stellar encounters \citep[e.g.,][]{1992ApJ...401L..31B, 2005ApJ...633L.137V, 2010MNRAS.402.1349F, 2014prpl.conf..387A, 2019AA...628A..20D, 2022NatAs...6..331D, 2022MNRAS.510L..37B}. Further observational and theoretical investigation of FU Ori-type systems may provide insight into DO Tau's history as well.

\section{Summary\label{sec:summary}}
We analyzed new and archival scattered light and molecular line observations to map the extent, relative orientation, and kinematics of the series of complex structures surrounding DO Tau. Our findings are as follows: 
\begin{enumerate}
\item We detect DO Tau's circumstellar disk in polarized scattered light for the first time. Like the millimeter continuum emission presented in \citet{2019ApJ...882...49L}, the disk does not exhibit any clear substructure in scattered light. 
\item The SPHERE scattered light and ALMA molecular line observations show that the disk is connected to two arm-like structures, one on the eastern side and one on the northern side. They extend up to several hundred au away from the disk. 
\item The HST scattered light and ALMA CO emission trace multiple stream-like structures extending northeast of DO Tau over scales of at least a couple thousand au. In projection, these streams mostly appear to converge at DO Tau's northern arm. The streams are redshifted with respect to DO Tau's systemic velocity and are not gravitationally bound to DO Tau. Based on their orientation, the streams appear to be part of the bridge-like structure connecting DO Tau and HV Tau as seen in Herschel far-infrared images. 
\item We also detect a faint redshifted counterpart to the blueshifted outflow previously detected in CO emission. The redshifted outflow appears to be misaligned with respect to the jet (as traced by [Fe II]), but wider-field CO mapping will be necessary to confirm the geometry of the redshifted outflow.  
\item While some of DO Tau's complex structures are compatible with a previously hypothesized stellar encounter, the system also shows signposts of still being partially embedded. Future high resolution simulations of protostellar evolution up to the Class II stage and observational characterization of the composition of the gas in and around DO Tau will be key for clarifying the origins of these structures.
\end{enumerate}

This case study of DO Tau highlights the utility of spatially resolved panchromatic observations for probing the environments of protoplanetary disks. For DO Tau, no single tracer reveals all of the  associated large-scale structures. Similar joint analyses of scattered light and molecular line observations of other systems will be key for understanding how the disk population is affected by interactions with their environments. 

\section*{Acknowledgments}
This work is based on observations collected at the European Southern Observatory under ESO programme(s) 0104.C-0850(A) and 1104.C-0415(E), and based on observations made with the NASA/ESA Hubble Space Telescope, obtained from the data archive at the Space Telescope Science Institute. STScI is operated by the Association of Universities for Research in Astronomy, Inc. under NASA contract NAS 5-26555. This paper makes use of ALMA data\\
\dataset[ADS/JAO.ALMA\#2016.1.00627.S]{https://almascience.nrao.edu/aq/?project\_code=2016.1.00627.S}. ALMA is a partnership of ESO (representing its member states), NSF (USA) and NINS (Japan), together with NRC (Canada) and NSC and ASIAA (Taiwan), in cooperation with the Republic of Chile. The Joint ALMA Observatory is operated by ESO, AUI/NRAO and NAOJ. The National Radio Astronomy Observatory is a facility of the National Science Foundation operated under cooperative agreement by Associated Universities, Inc. Funding for SDSS-III has been provided by the Alfred P. Sloan Foundation, the Participating Institutions, the National Science Foundation, and the U.S. Department of Energy Office of Science. The SDSS-III web site is http://www.sdss3.org/. SDSS-III is managed by the Astrophysical Research Consortium for the Participating Institutions of the SDSS-III Collaboration including the University of Arizona, the Brazilian Participation Group, Brookhaven National Laboratory, Carnegie Mellon University, University of Florida, the French Participation Group, the German Participation Group, Harvard University, the Instituto de Astrofisica de Canarias, the Michigan State/Notre Dame/JINA Participation Group, Johns Hopkins University, Lawrence Berkeley National Laboratory, Max Planck Institute for Astrophysics, Max Planck Institute for Extraterrestrial Physics, New Mexico State University, New York University, Ohio State University, Pennsylvania State University, University of Portsmouth, Princeton University, the Spanish Participation Group, University of Tokyo, University of Utah, Vanderbilt University, University of Virginia, University of Washington, and Yale University. This research has made use of data reprocessed as part of the ALICE program, which was supported by NASA through grants HST-AR-12652 (PI: R. Soummer), HST-GO-11136 (PI: D. Golimowski), HST-GO-13855 (PI: E. Choquet), HST-GO-13331 (PI: L. Pueyo), and STScI Director's Discretionary Research funds, and was conducted at STScI which is operated by AURA under NASA contract NAS5-26555. This work has made use of the SPHERE Data Centre, jointly operated by OSUG/IPAG (Grenoble), PYTHEAS/LAM/CeSAM (Marseille), OCA/Lagrange (Nice), Observatoire de Paris/LESIA (Paris), and Observatoire de Lyon/CRAL, and is supported by a grant from Labex OSUG$@$2020 (Investissements d'avenir – ANR10 LABX56). We acknowledge the use of NASA's SkyView facility (http://skyview.gsfc.nasa.gov) located at NASA Goddard Space Flight Center. Herschel is an ESA space observatory with science instruments provided by European-led Principal Investigator consortia and with important participation from NASA. This research is based in part on data collected at the Subaru Telescope, which is operated by the National Astronomical Observatory of Japan. We are honored and grateful for the opportunity to use observations from Maunakea, which has cultural, historical, and natural significance in Hawaii. This research has made use of NASA's Astrophysics Data System. We thank Sean Andrews, Sarah Sadavoy, and Andrew Winter for useful discussions. We also thank the referee for comments improving the clarity of the manuscript. 

Support for J. H. was provided by NASA through the NASA Hubble Fellowship grant \#HST-HF2-51460.001-A awarded by the Space Telescope Science Institute, which is operated by the Association of Universities for Research in Astronomy, Inc., for NASA, under contract
NAS5-26555. M. B. acknowledges funding from the European Research Council (ERC) under the European Union's Horizon 2020 research and innovation programme (grant PROTOPLANETS No. 101002188). T.B. acknowledges funding from the European Research Council (ERC) under the European Union's Horizon 2020 research and innovation programme under grant agreement No 714769 and funding by the Deutsche Forschungsgemeinschaft (DFG, German Research Foundation) under grants 361140270, 325594231, and Germany's Excellence Strategy - EXC-2094 - 390783311. D. H. is supported by CICA through a grant and grant number 110J0353I9
from the Ministry of Education of Taiwan. A.Z. acknowledges support from the FONDECYT Iniciaci\'on en investigaci\'on project number 11190837. CHR is grateful for support from
the Max Planck Society. This project has received funding from the European Union's Horizon 2020 research and innovation programme under the Marie Sklodowska-Curie grant agreement No 823823 (DUSTBUSTERS).

\facilities{ALMA, VLT:Melipal, HST (NICMOS, STIS)}

\software{ \texttt{analysisUtils} (\url{https://casaguides.nrao.edu/index.php/Analysis_Utilities}), \texttt{AstroPy} \citep{2013AA...558A..33A}, \texttt{bezier} \citep{Hermes2017}, \texttt{CASA} \citep{2007ASPC..376..127M}, \texttt{centerRadon} \citep{2019ascl.soft06021R}, \texttt{cmasher} \citep{cmasher}, \texttt{IRDAP} \citep{2020ascl.soft04015V}, \texttt{matplotlib} \citep{Hunter:2007}, \texttt{SciPy} \citep{2020SciPy-NMeth}, \textit{SkyView} \citep{1998IAUS..179..465M}, Tiny Tim \citep{2011SPIE.8127E..0JK}}

\appendix
\section{Comparison of standard RDI and IDF-RDI images of DO Tau.\label{sec:RDIcomparison}}

\begin{figure*}
\begin{center}
\includegraphics[scale=1]{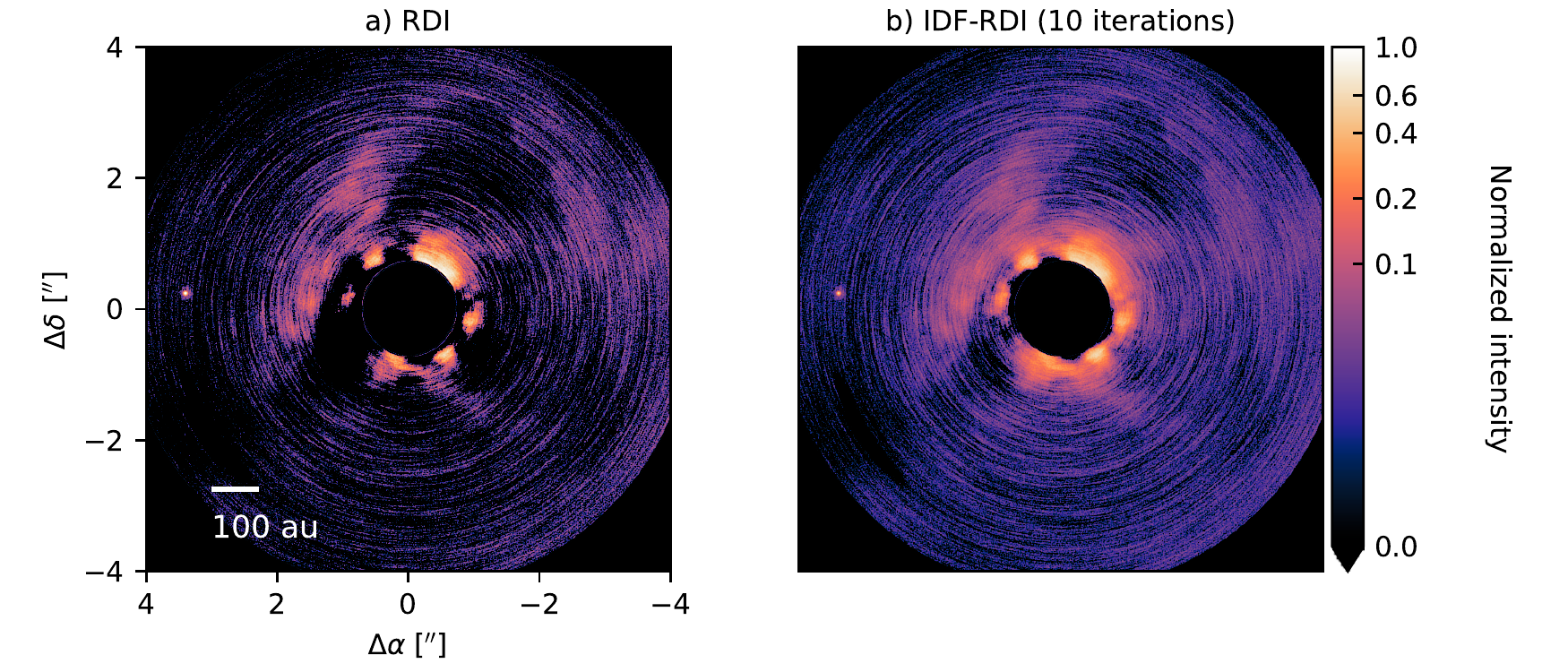}
\end{center}
\caption{Left: Total intensity image of DO Tau produced with standard RDI (equivalent to the zeroth iteration of IDF-RDI). Right: Total intensity image of DO Tau after 10 iterations of IDF-RDI. \label{fig:RDIcomparison}}
\end{figure*}

Figure \ref{fig:RDIcomparison} compares total intensity images of DO Tau produced through standard RDI with KLIP and through IDF-RDI. While the extended structures around DO Tau appear similar in the two images, IDF-RDI mitigates the oversubtraction artifacts visible in the RDI image.

\section{Details of ALMA Observational Setup\label{sec:ALMAdetails}}

Details of the ALMA observations of DO Tau are provided in Tables \ref{tab:observations}, \ref{tab:1.3 mm}, and \ref{tab:1.1 mm}. 

\begin{deluxetable*}{ccccccc}[!h]
\tablecaption{Summary of ALMA Observational Setup \label{tab:observations}}
\tablehead{\colhead{Date} &\colhead{Antennas} & \colhead{Baselines}&\colhead{Time on source}&\colhead{Flux}&\colhead{Bandpass}&\colhead{Phase}\\
&&\colhead{(m)}&\colhead{(min)}&\colhead{calibrator}&\colhead{calibrator}&\colhead{calibrator}}
\startdata
\multicolumn{7}{c}{1.3 mm setting} \\
\hline
2016 Dec 01 & 44 & $15-704$&10&J0423-0120&J0237+2848 &J0426+2327\\
2016 Dec 01 & 44 & $15-704$&10&J0423-0120&J0510+1800 & J0426+2327\\
\hline
\multicolumn{7}{c}{1.1 mm setting} \\
\hline
2016 Dec 01 & 44 & $15-704$&9 & J0510+1800 & J0510+1800& J0426+2327\\
2016 Dec 02 & 45 &$15-704$&9 &J0510+1800&J0510+1800& J0426+2327 \\
2016 Dec 03 & 44 &$15-704$&9&J0423-0120&J0237+2848& J0426+2327
\enddata
\end{deluxetable*}

\begin{deluxetable}{ccc}[!h]
\tablecaption{1.3 mm spectral setting \label{tab:1.3 mm}}
\tablehead{
\colhead{Center Frequency\tablenotemark{a}} &\colhead{Bandwidth} &\colhead{Native channel spacing}\\
\colhead{(GHz)}&\colhead{(MHz)}&\colhead{(MHz)}}
\startdata
217.227 & 58.6 & 0.061\\
218.210 & 58.6 & 0.061\\
218.313 & 58.6 & 0.061\\
218.721& 58.6 & 0.061\\
219.548 & 117.2 & 0.061\\
220.387 & 117.2 & 0.061\\
230.525 & 117.2 & 0.061\\
231.309 & 117.2 & 0.061\\
232.406 & 937.5 & 0.244 \\
\enddata
\tablenotetext{a}{Topocentric frequency}
\end{deluxetable}

\begin{deluxetable}{ccc}[!h]
\tablecaption{1.1 mm spectral setting \label{tab:1.1 mm}}
\tablehead{
\colhead{Center Frequency\tablenotemark{a}} &\colhead{Bandwidth} &\colhead{Native channel spacing}\\
\colhead{(GHz)}&\colhead{(MHz)}&\colhead{(MHz)}}
\startdata
242.903& 937.5 & 0.244 \\
244.212 & 117.2 & 0.061\\
244.925 & 117.2 & 0.061\\
258.146  & 117.2 & 0.061\\
259.001 & 117.2 & 0.061\\
260.507 & 117.2 & 0.061\\
262.031  & 117.2 & 0.061
\enddata
\tablenotetext{a}{Topocentric frequency}
\end{deluxetable}

\section{Non-annotated versions of scattered light images\label{sec:noannotations}}

\begin{figure*}
\begin{center}
\includegraphics[scale=1]{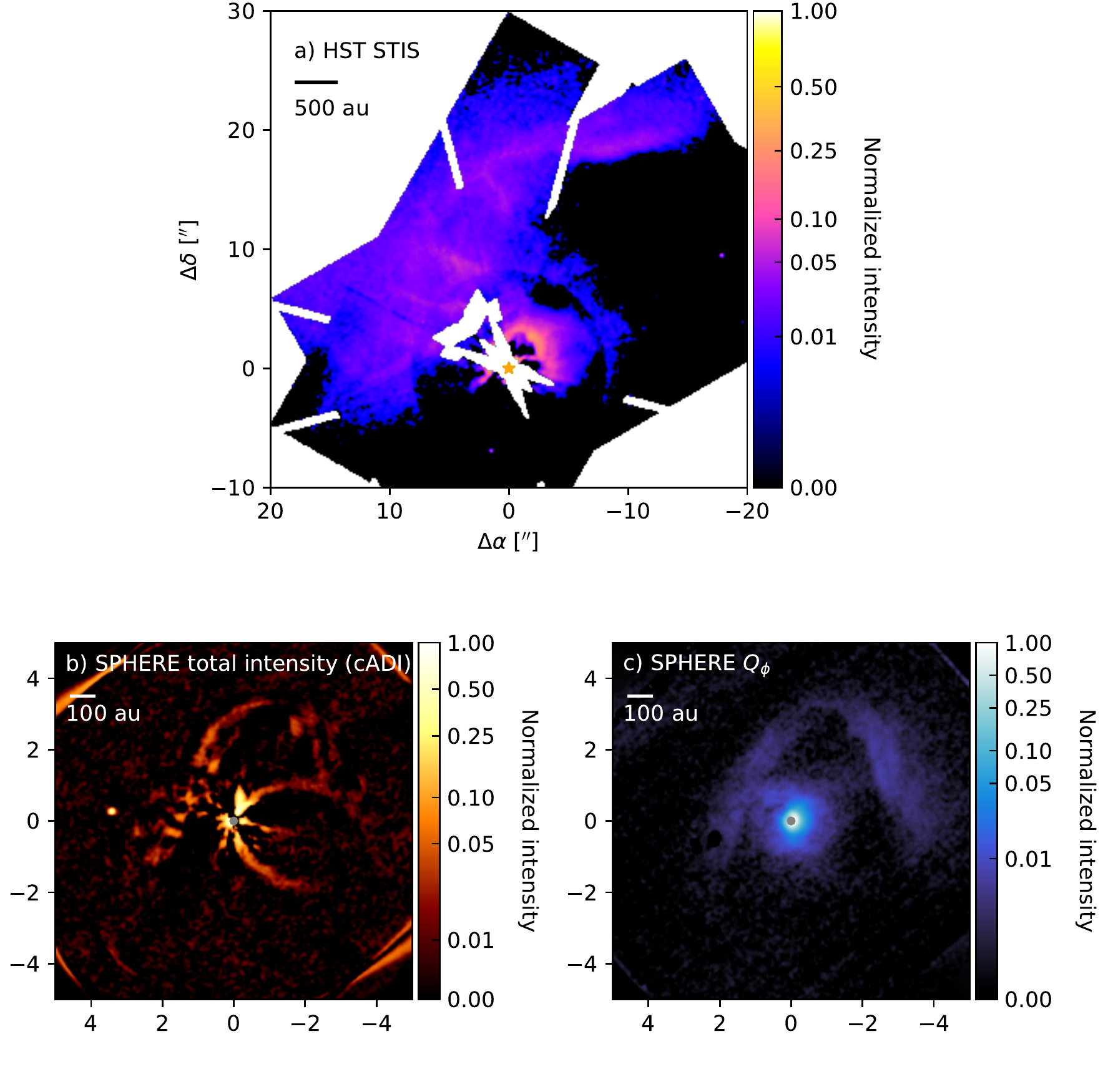}
\end{center}
\caption{Non-annotated versions of the HST STIS and SPHERE images presented in Figures \ref{fig:SDSS} and \ref{fig:diskinset}. \label{fig:noannotations}}
\end{figure*}

Non-annotated versions of the HST and SPHERE images presented in the main text are shown in Figure \ref{fig:noannotations}. 

\section{Re-detection of a point source in the field of DO Tau\label{sec:pointsource}}

\begin{deluxetable*}{ccccccc}
\tablecaption{Point source properties\label{tab:pointsource}}
\tablehead{
\colhead{Date} &\colhead{Instrument} &\colhead{Filter}&\colhead{Separation ($''$)}&\colhead{P.A. (deg.)}&\colhead{S/N}}
\startdata
1998 Dec 02 & NICMOS & F110W &$3.51\pm0.04$&$90.7\pm0.6$&2.5\\
1998 Dec 02 & NICMOS & F160W & $3.51\pm0.04$&$90.8\pm0.6$&5.6\\
2005 Nov 12 & CIAO & H & $3.449\pm0.020$&$87.85^{+0.33}_{-1.11}$&6.9 \\
2019 Dec 20 & SPHERE & BB\_H & $3.418\pm0.008$&$86.0\pm0.2$&61.9 \\
\enddata
\end{deluxetable*}

\citet{2008PASJ...60..223I} reported the detection of a point source east of DO Tau in $H$-band observations taken with the Subaru CIAO instrument on 2005 November 12. We re-detect this point source in both epochs of the SPHERE $H-$band observations as well as in archival NICMOS observations in the F110W and F160W filters.  Together, these observations provide a 21-year baseline to examine whether the point source is bound to DO Tau.

\subsection{Archival Data Reduction}
Since \citet{2008PASJ...60..223I} did not report a position angle for the point source in their Subaru $H$-band observations from program o05146 (PI: M. Tamura), we retrieved the raw data from Subaru's SMOKA archive for re-analysis. Details of the observations are provided in \cite{2008PASJ...60..223I}. We performed dark subtraction, filtered bad pixels, recentered the central star, and median combined the individual exposures. 

To obtain a longer baseline to analyze the motion of the point source, we checked the HST archive for earlier observations. DO Tau was imaged by HST NICMOS with the NIC2 coronagraph on 1998 December 02 in the F110W (central wavelength: 1.1 $\mu$m) and F160W (central wavelength: 1.6 $\mu$m) filters as part of program HST-GO-7418 (P. I.: D. Padgett). The exposure time in each filter was 512 seconds. The observations were processed with the ALICE pipeline as described in \citet{2014SPIE.9143E..57C, 2018AJ....155..179H}. The sizes of the DO Tau images included in the ALICE public data release\footnote{\url{https://archive.stsci.edu/doi/resolve/resolve.html?doi=10.17909/T9W89V}} are truncated to $6''\times6''$; however, since the point source reported by \citet{2008PASJ...60..223I} is outside the field of view of the public version of the images, we produced new $10''\times10''$ versions. In brief, the input science images and PSF reference library images come from the database of NICMOS observations re-calibrated by the LAPLACE project \citep{2010hstc.workE..15S}. PSF subtraction for DO Tau was performed with KLIP. We used a PSF library of 18 frames and subtracted two KL modes for the F110W image, while we used a PSF library of 100 frames and subtracted one KL mode for the F160W image. 

\begin{figure*}
\begin{center}
\includegraphics[scale=1]{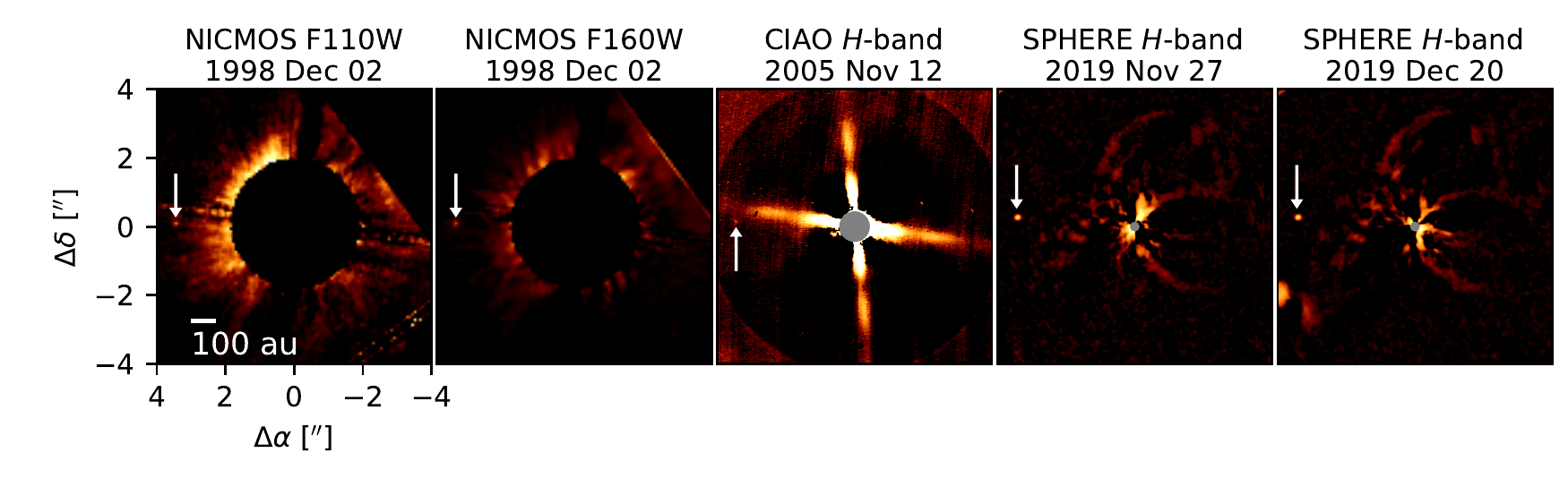}
\end{center}
\caption{A gallery of images showing the point source detected in the vicinity of DO Tau. Each panel is labeled with the instrument, filter, and date of observation. The white arrow points to the location of the point source. The axes of the leftmost image are labeled with the offset in arcseconds with respect to DO Tau. The black circles in the NICMOS images and gray circles in the CIAO and SPHERE images denote the extent of the coronagraphic mask. An azimuthally averaged radial profile has been subtracted from the CIAO image to make the point source stand out more clearly. The two SPHERE cADI images have been convolved with a Gaussian with a standard deviation of 3 pixels ($0.0368''$).\label{fig:pointsource}}
\end{figure*}

The NICMOS, CIAO, and SPHERE images of the point source are shown in Figure \ref{fig:pointsource}. The point source is not detected in the HST STIS optical image, for which \citet{2008PASJ...60..223I} estimated a detection limit of 28 mag. 

\subsection{Astrometric analysis}
For the NICMOS observations, the astrometry was performed following the methodology described in \citet{2015SPIE.9605E..1PC}. For each filter, a synthetic NICMOS PSF was generated using the Tiny Tim software \citep{2011SPIE.8127E..0JK}. The synthetic PSF was used as a matched filter template to find the position and flux values that maximize the correlation with the data. The 1$\sigma$ astrometric uncertainty was taken to be half the size of the NIC2 pixel. The S/N was estimated by measuring the standard deviation in an annulus around the point source. However, the background estimates may be skewed by the extended structures surrounding DO Tau.  

In the Subaru observations, the point source noted by \cite{2008PASJ...60..223I} is visible near the bright diffraction spike to the east of the central star. To extract the position of the point source in detector coordinates, a 2D Gaussian was fitted at the source position, which yields an uncertainty in the P.A. of the source relative to the DO Tau of 0.33$^\circ$. However, this uncertainty does not account for systematic offsets from a dedicated true north astrometric calibration. Following the discussion in \cite{2014MNRAS.444.2280G}, we incorporate an additional systematic offset of $\sim$1$^\circ$ into the uncertainties that we report.

The astrometric extraction of the SPHERE H-band data was performed at the SPHERE Data Center \citep{2017sf2a.conf..347D, 2018AA...615A..92G} by applying the TLOCI reduction algorithm \citep{2010SPIE.7736E..1JM,2018AA...615A..92G} to the intensity data and injecting negative point sources to null the source signal. While the astrometric calibration of SPHERE is very stable \citep{2016SPIE.9908E..34M}, we still used a dedicated astrometric calibration epoch with the cluster 47\,Tuc, which was acquired within three weeks of the observations of DO Tau. 
The initial TLOCI extraction yielded an uncertainty in the P.A. of 0.1$^\circ$. We then accounted for the uncertainty in the true north alignment of the detector, which is also 0.1$^\circ$.

\begin{figure}
\begin{center}
\includegraphics[width=0.48\textwidth]{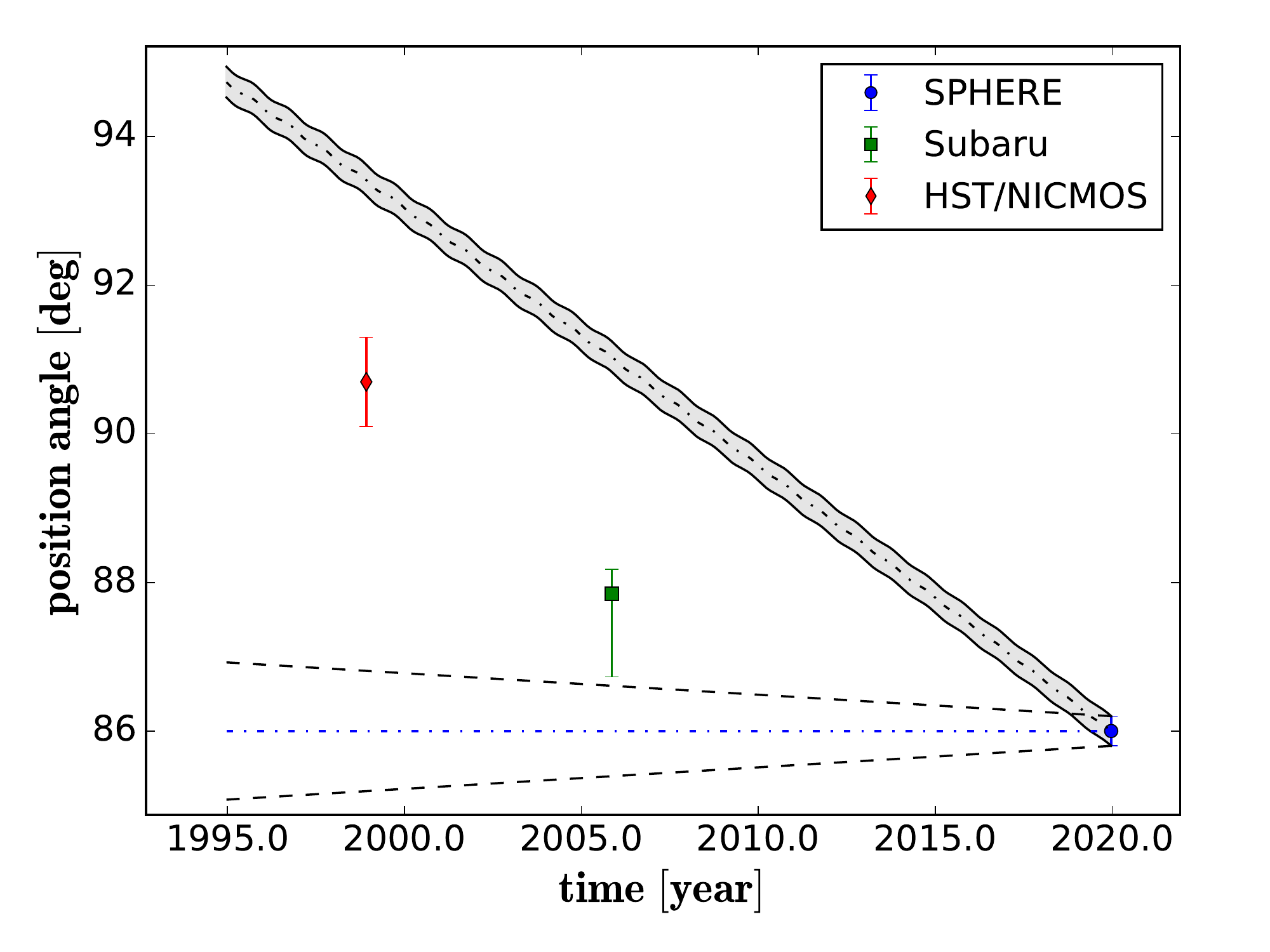}
\end{center}
\caption{Measurements of the P.A.  of the point source in the field of DO Tau, shown as a function of time. The gray ribbon shows the expected values for a stationary background source. The dashed black lines show the expected position angle as a function of time if the point source had a face-on circular orbit (i.e., the orientation at which the change in P.A. would be maximized). The blue dash-dot line marks the P.A. measured for the most recent observation. \label{fig:astrometry}}
\end{figure}

The astrometric measurements for the different epochs are given in Table \ref{tab:pointsource}. The position angles of the point source are plotted in Figure \ref{fig:astrometry} and compared to the expected P.A. of a bound source on a face-on circular orbit (i.e., the orientation at which the change in P.A. would be maximized) and of a stationary (distant) background source. The P.A. of the latter is calculated given a proper motion for DO Tau of $\mu_{\alpha^*} = 6.27\pm0.04$ mas yr$^{-1}$, $\mu_\delta = -21.05\pm0.03$ mas yr$^{-1}$ \citep{2021AA...649A...1G}. The change in P.A. of the observed point source is much larger than expected for a bound source, but smaller than for a stationary background source. Based on the SPHERE and NICMOS observations (and excluding the more uncertain Subaru observations), we estimate that the proper motion difference between the point source and DO Tau is $\Delta \mu_{\alpha^*}=-4.7\pm2.0$ mas yr$^{-1}$, $\Delta \mu_{\delta}=13.6\pm2.4$ mas yr$^{-1}$. To examine whether the object could be a background source that is still within Taurus, we retrieved Gaia DR2 proper motions of the Taurus members listed in \citet{2018AJ....156..271L} and computed their dispersions, which were 3.5 mas yr$^{-1}$ in the R.A. direction and 4.0 mas yr$^{-1}$ in declination. Since the relative proper motion of the point source deviates by more than $3\sigma$ in the declination direction from the dispersion in Taurus, we conclude that the point source is most likely not a member of Taurus, but not distant enough from the observer to appear stationary.

\section{Channel Maps of Molecular Line Observations\label{sec:chanmaps}}

Channel maps of $^{12}$CO $J=2-1$, $^{13}$CO $J=2-1$, C$^{18}$O $J=2-1$, and CS $J=5-4$ toward DO Tau are provided in Figure \ref{fig:chanmaps}. 

\figsetstart
\figsetnum{20}
\figsettitle{ALMA Channel Maps}
\figsetgrpstart
\figsetgrpnum{20.1}
\figsetgrptitle{$^{12}$CO $J=2-1$ channel maps, part 1}
\figsetplot{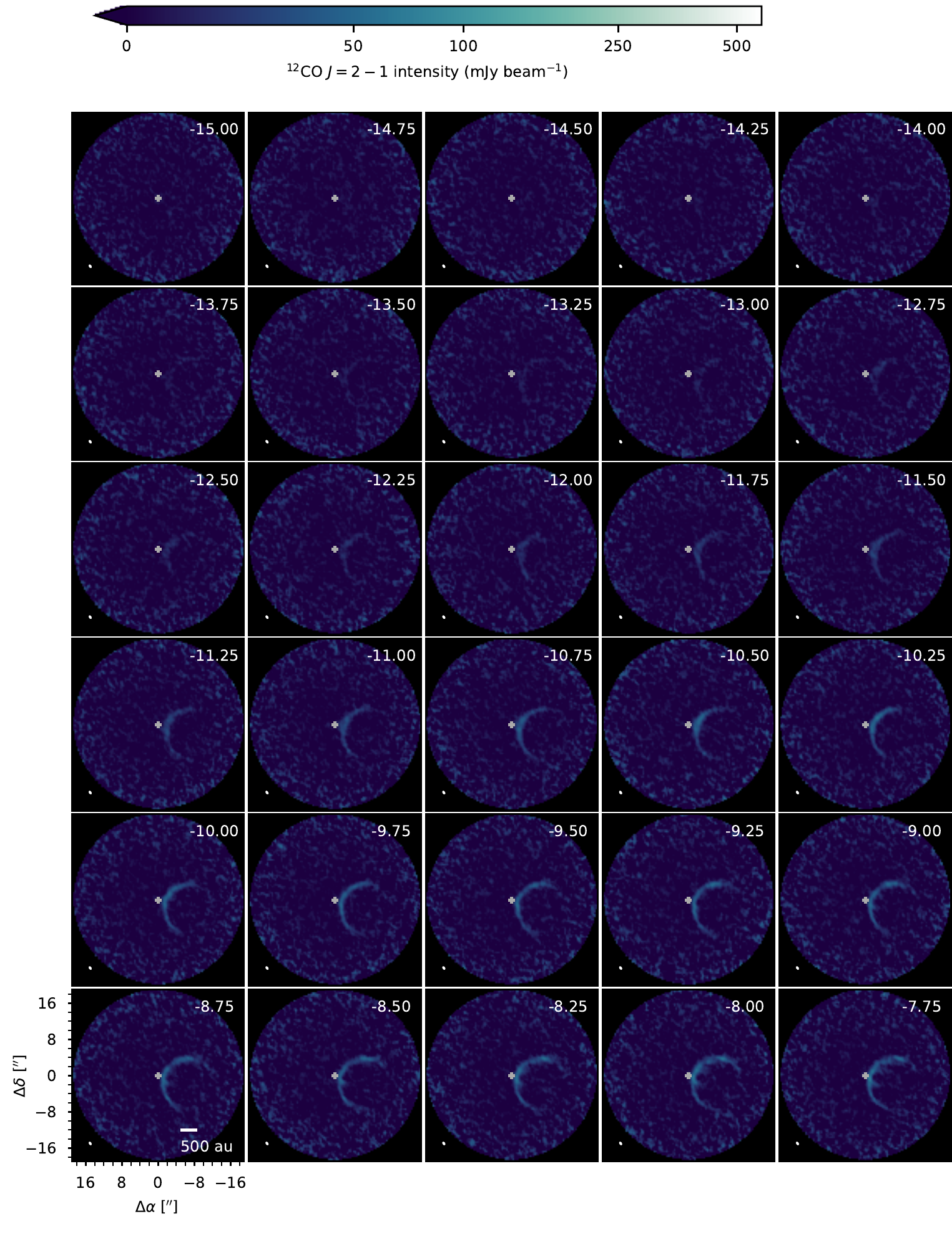}
\figsetgrpnote{Channel maps of $^{12}$CO $J=2-1$ emission toward DO Tau, part 1 of 5. The LSRK velocity (km s$^{-1}$) appears in the top right of each panel. The synthesized beam is shown in the lower left corner. A gray cross marks the position of the millimeter continuum peak, which is located at the phase center. Offsets from the phase center (in arcseconds) are marked in the lower left panel. North is up and east is to the left. An arcinsh stretch is applied to the color scale to highlight faint emission features.}
\figsetgrpend

\figsetgrpstart
\figsetgrpnum{20.2}
\figsetgrptitle{$^{12}$CO $J=2-1$ channel maps, part 2}
\figsetplot{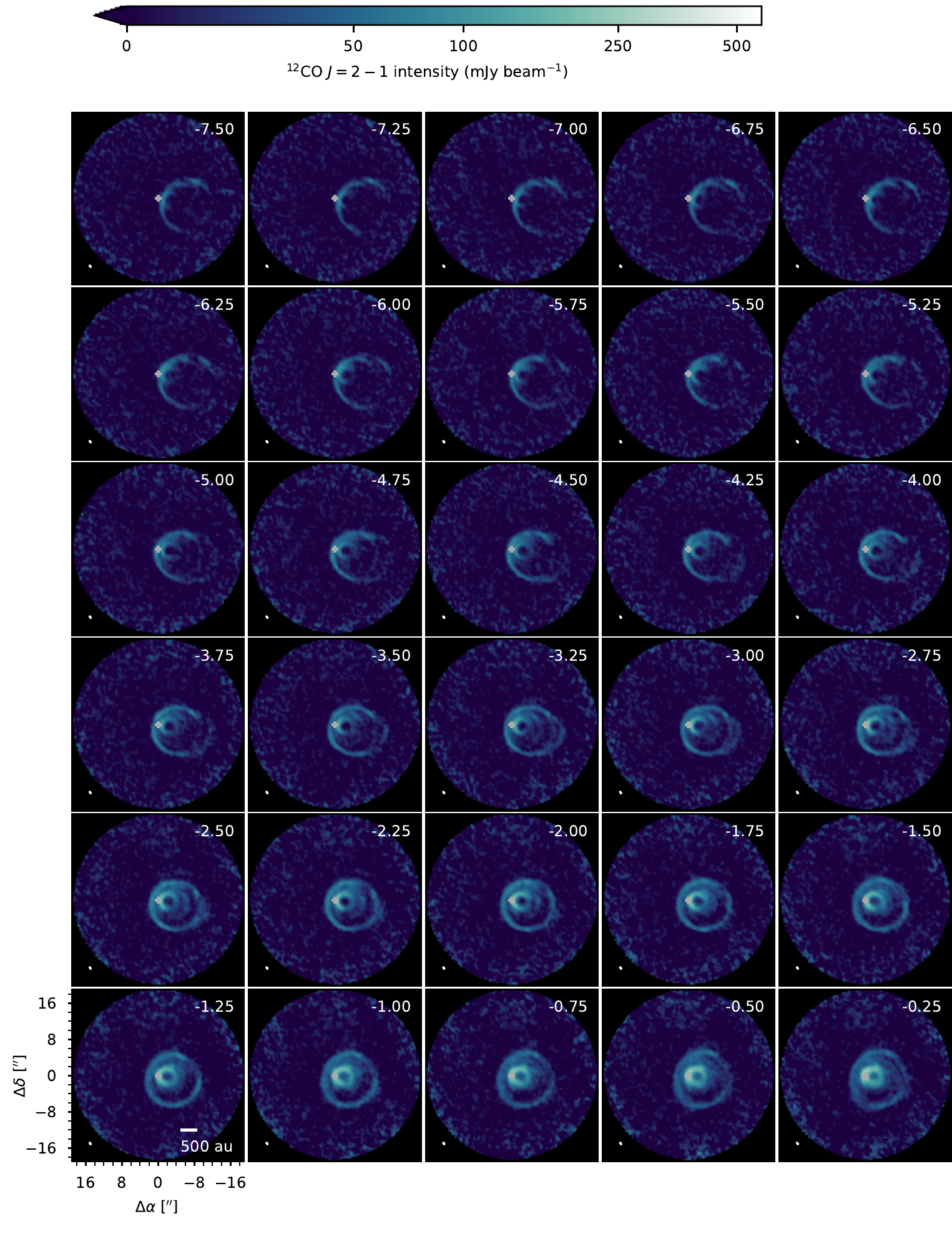}
\figsetgrpnote{Channel maps of $^{12}$CO $J=2-1$ emission toward DO Tau, part 2 of 5. The LSRK velocity (km s$^{-1}$) appears in the top right of each panel. The synthesized beam is shown in the lower left corner. A gray cross marks the position of the millimeter continuum peak, which is located at the phase center. Offsets from the phase center (in arcseconds) are marked in the lower left panel. North is up and east is to the left. An arcinsh stretch is applied to the color scale to highlight faint emission features.}
\figsetgrpend

\figsetgrpstart
\figsetgrpnum{20.3}
\figsetgrptitle{$^{12}$CO $J=2-1$ channel maps, part 3}
\figsetplot{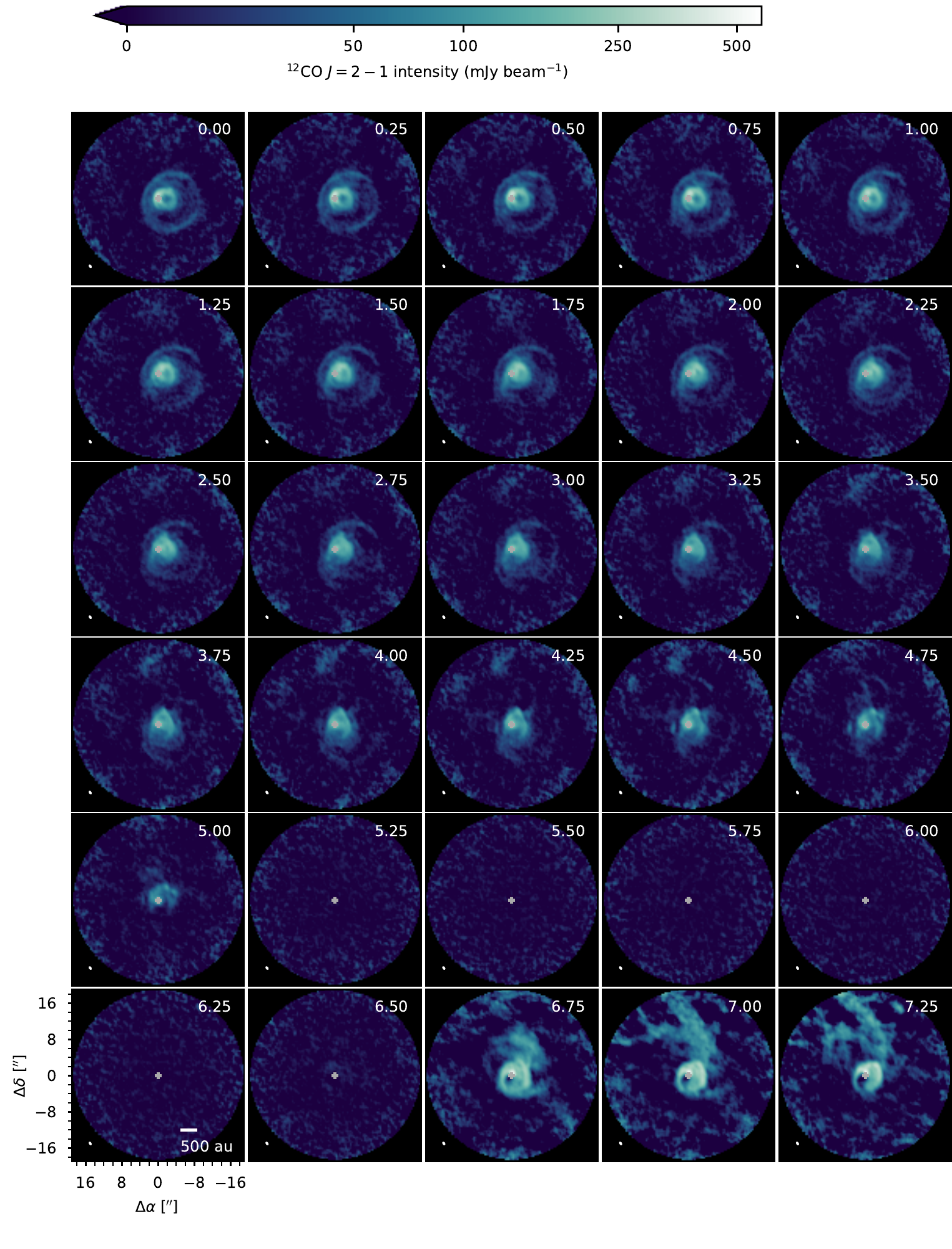}
\figsetgrpnote{Channel maps of $^{12}$CO $J=2-1$ emission toward DO Tau, part 3 of 5. The LSRK velocity (km s$^{-1}$) appears in the top right of each panel. The synthesized beam is shown in the lower left corner. A gray cross marks the position of the millimeter continuum peak, which is located at the phase center. Offsets from the phase center (in arcseconds) are marked in the lower left panel. North is up and east is to the left. An arcinsh stretch is applied to the color scale to highlight faint emission features.}
\figsetgrpend

\figsetgrpstart
\figsetgrpnum{20.4}
\figsetgrptitle{$^{12}$CO $J=2-1$ channel maps, part 4}
\figsetplot{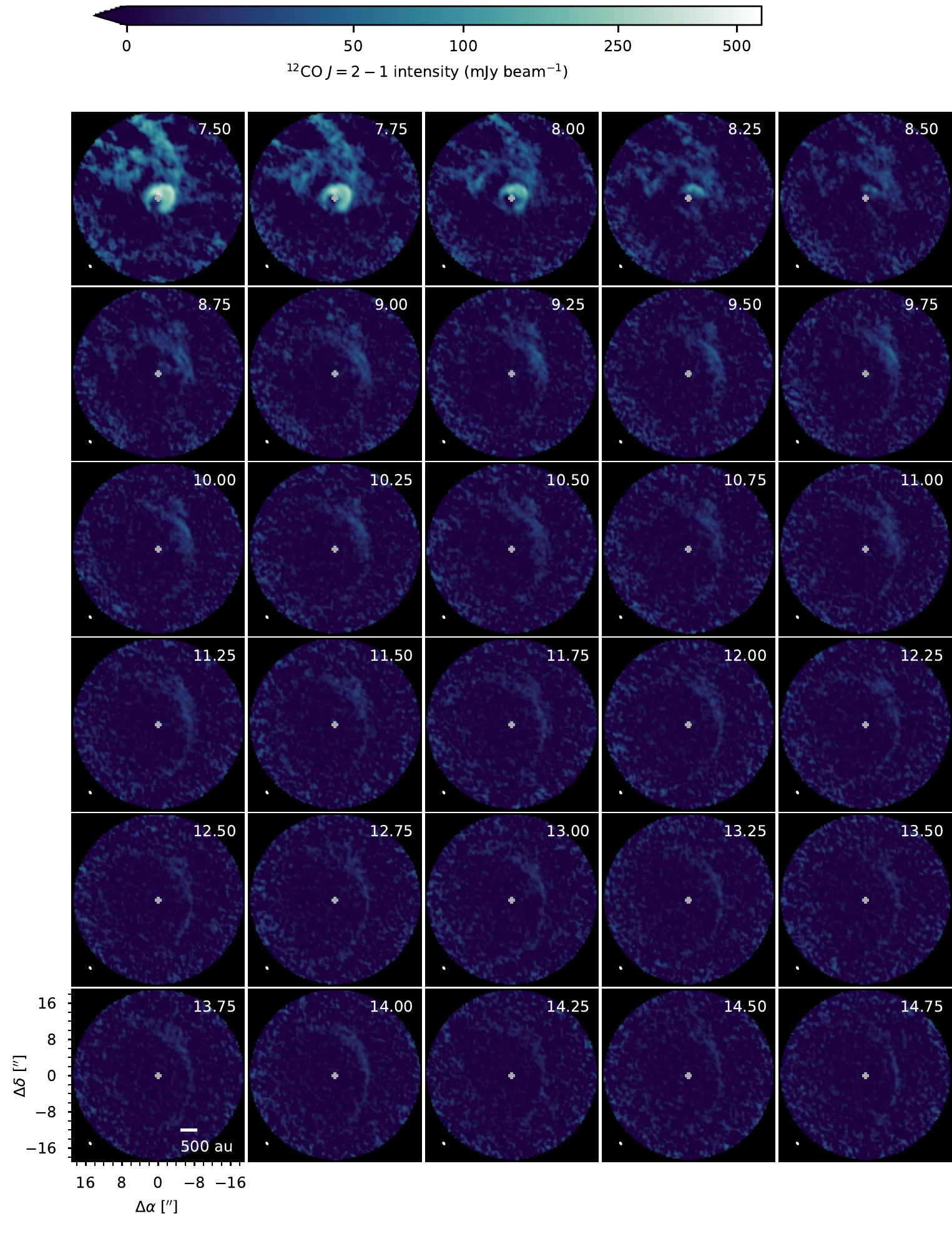}
\figsetgrpnote{Channel maps of $^{12}$CO $J=2-1$ emission toward DO Tau, part 4 of 5. The LSRK velocity (km s$^{-1}$) appears in the top right of each panel. The synthesized beam is shown in the lower left corner. A gray cross marks the position of the millimeter continuum peak, which is located at the phase center. Offsets from the phase center (in arcseconds) are marked in the lower left panel. North is up and east is to the left. An arcinsh stretch is applied to the color scale to highlight faint emission features.}

\figsetgrpstart
\figsetgrpnum{20.5}
\figsetgrptitle{$^{12}$CO $J=2-1$ channel maps, part 5}
\figsetplot{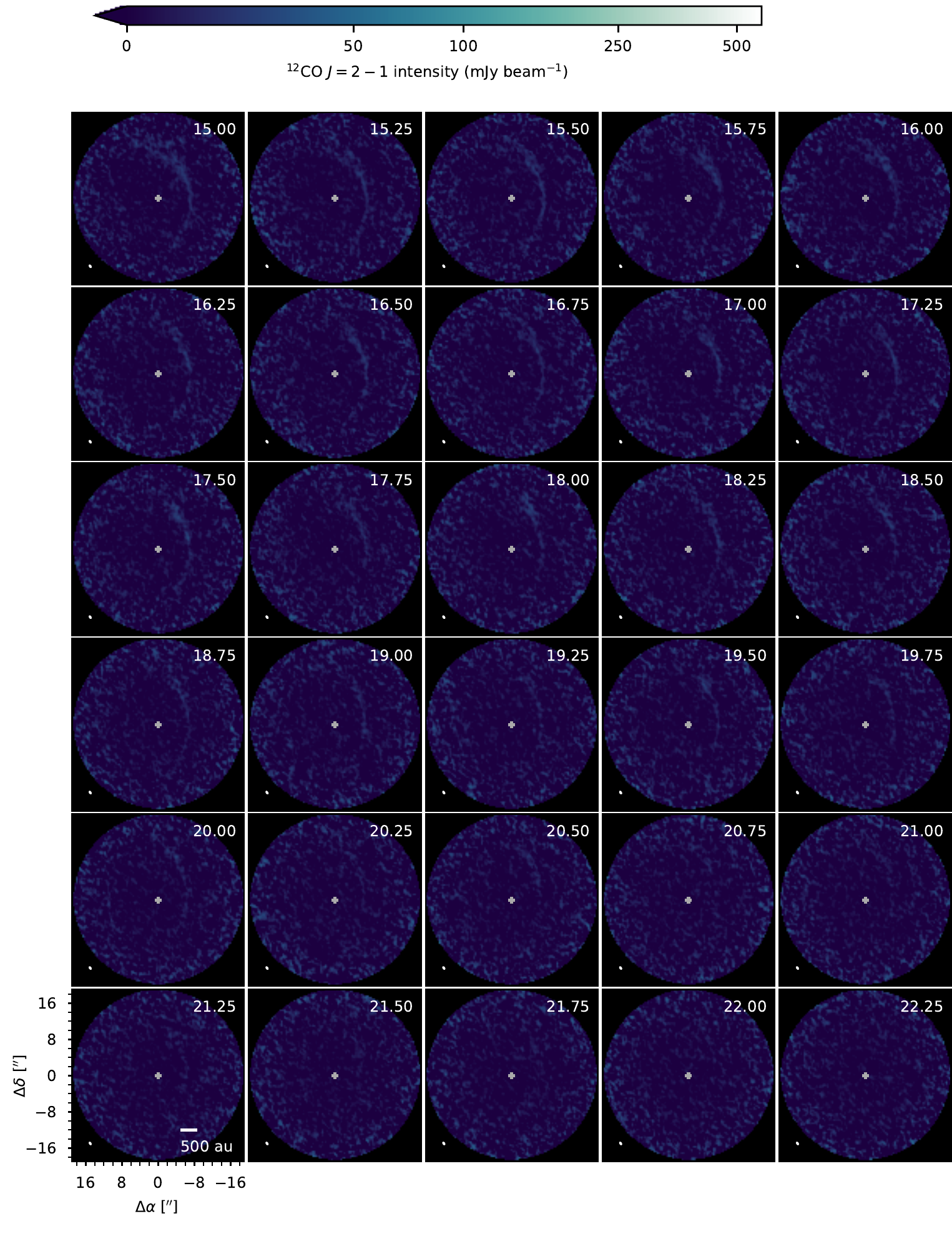}
\figsetgrpnote{Channel maps of $^{12}$CO $J=2-1$ emission toward DO Tau, part 5 of 5. The LSRK velocity (km s$^{-1}$) appears in the top right of each panel. The synthesized beam is shown in the lower left corner. A gray cross marks the position of the millimeter continuum peak, which is located at the phase center. Offsets from the phase center (in arcseconds) are marked in the lower left panel. North is up and east is to the left. An arcinsh stretch is applied to the color scale to highlight faint emission features.}

\figsetgrpstart
\figsetgrpnum{20.6}
\figsetgrptitle{$^{13}$CO $J=2-1$ channel maps}
\figsetplot{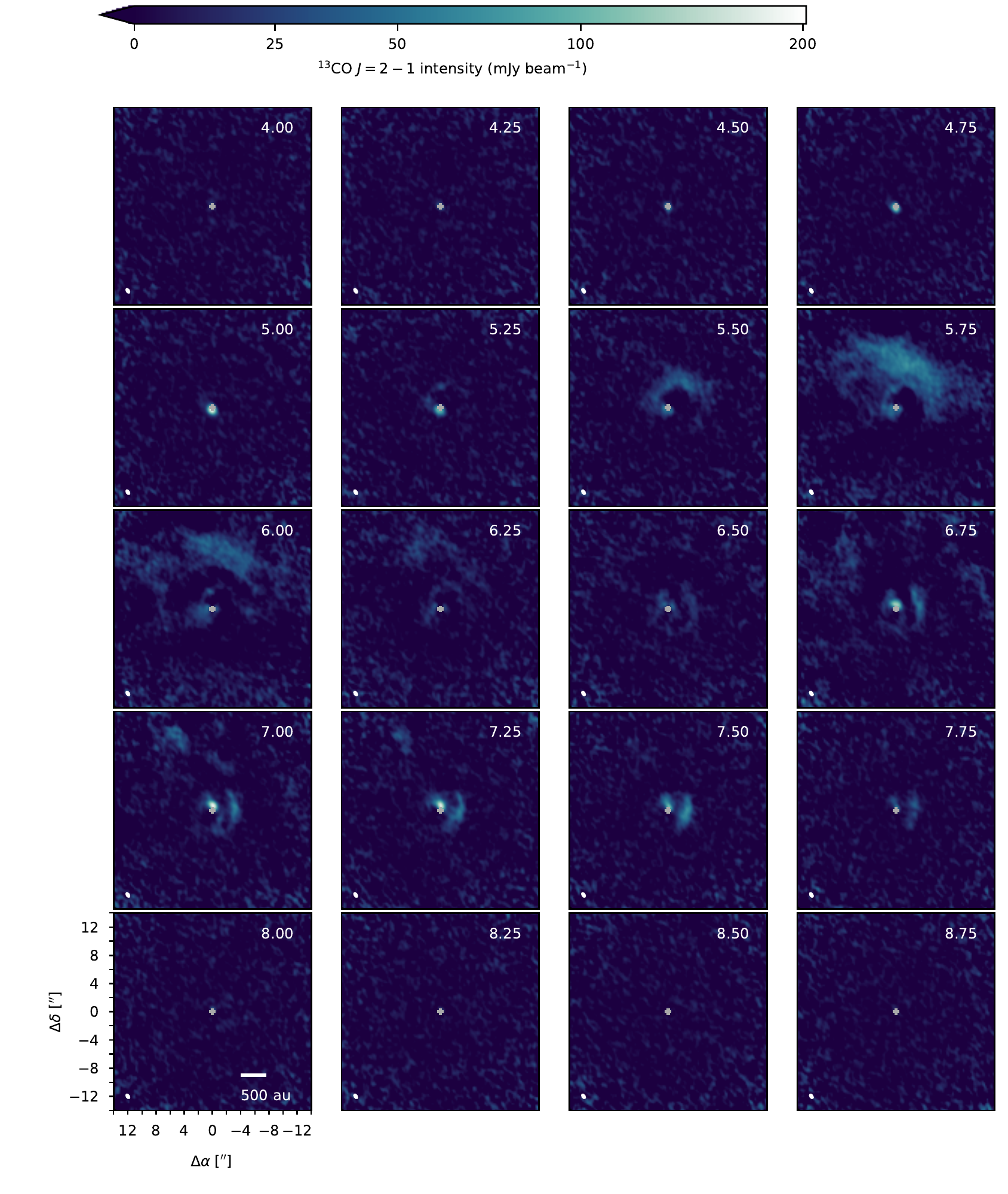}
\figsetgrpnote{Channel maps of $^{13}$CO $J=2-1$ emission toward DO Tau. The LSRK velocity (km s$^{-1}$) appears in the top right of each panel. The synthesized beam is shown in the lower left corner. A gray cross marks the position of the millimeter continuum peak, which is located at the phase center. Offsets from the phase center (in arcseconds) are marked in the lower left panel. North is up and east is to the left. An arcinsh stretch is applied to the color scale to highlight faint emission features.}
\figsetgrpend

\figsetgrpstart
\figsetgrpnum{20.7}
\figsetgrptitle{C$^{18}$O $J=2-1$ channel maps}
\figsetplot{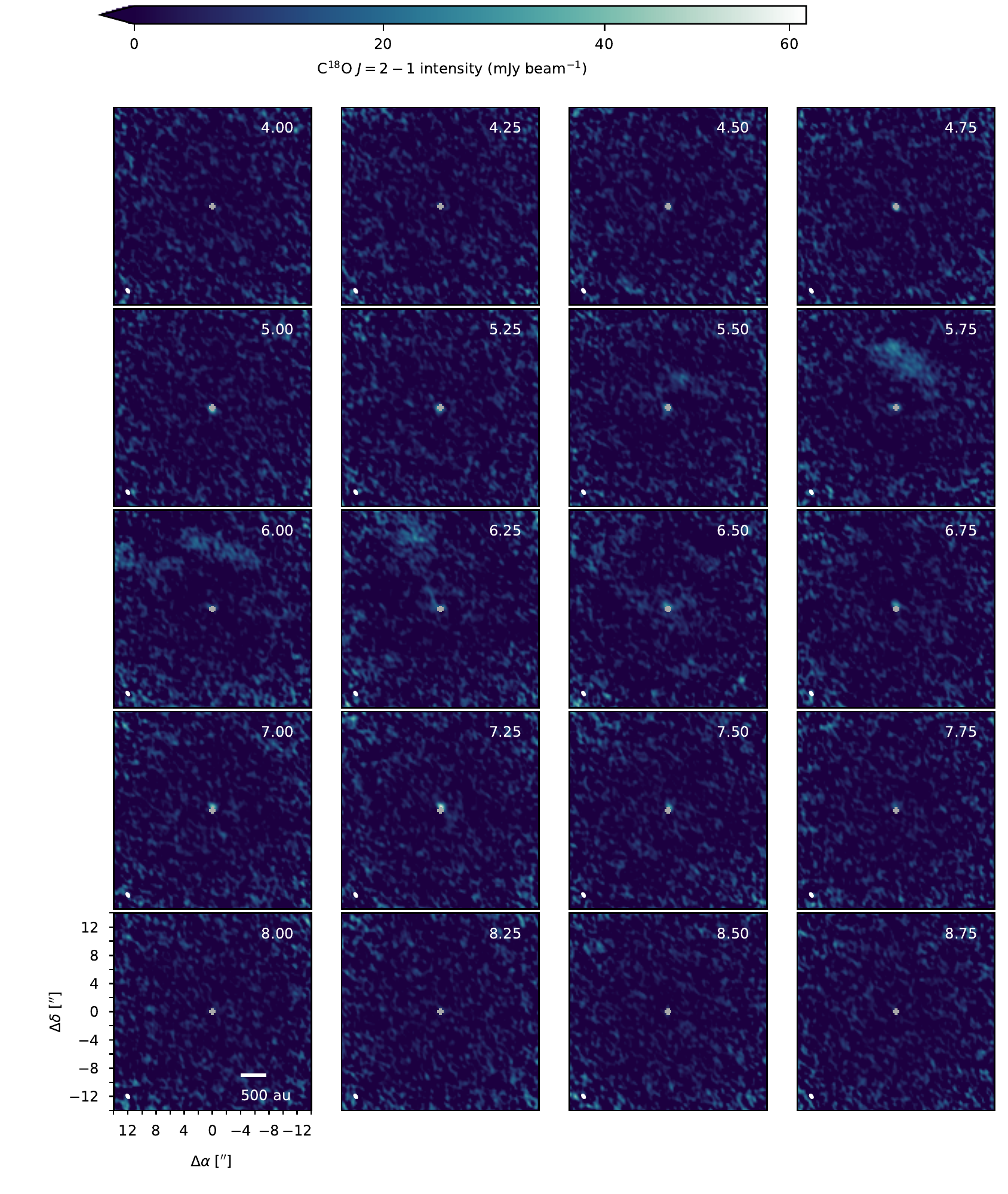}
\figsetgrpnote{Channel maps of C$^{18}$O $J=2-1$ emission toward DO Tau. The LSRK velocity (km s$^{-1}$) appears in the top right of each panel. The synthesized beam is shown in the lower left corner.  Offsets from the phase center (in arcseconds) are marked in the lower left panel. North is up and east is to the left. An arcinsh stretch is applied to the color scale to highlight faint emission features.}
\figsetgrpend

\figsetgrpstart
\figsetgrpnum{20.8}
\figsetgrptitle{CS $J=5-4$ channel maps}
\figsetplot{C18Ochanmap.pdf}
\figsetgrpnote{Channel maps of CS $J=5-4$ emission toward DO Tau. The LSRK velocity (km s$^{-1}$) appears in the top right of each panel. The synthesized beam is shown in the lower left corner. A gray cross marks the position of the millimeter continuum peak, which is located at the phase center. Offsets from the phase center (in arcseconds) are marked in the lower left panel. North is up and east is to the left. An arcinsh stretch is applied to the color scale to highlight faint emission features.}
\figsetgrpend
\figsetend

\begin{figure*}
\figurenum{20.1}
\begin{center}
\includegraphics{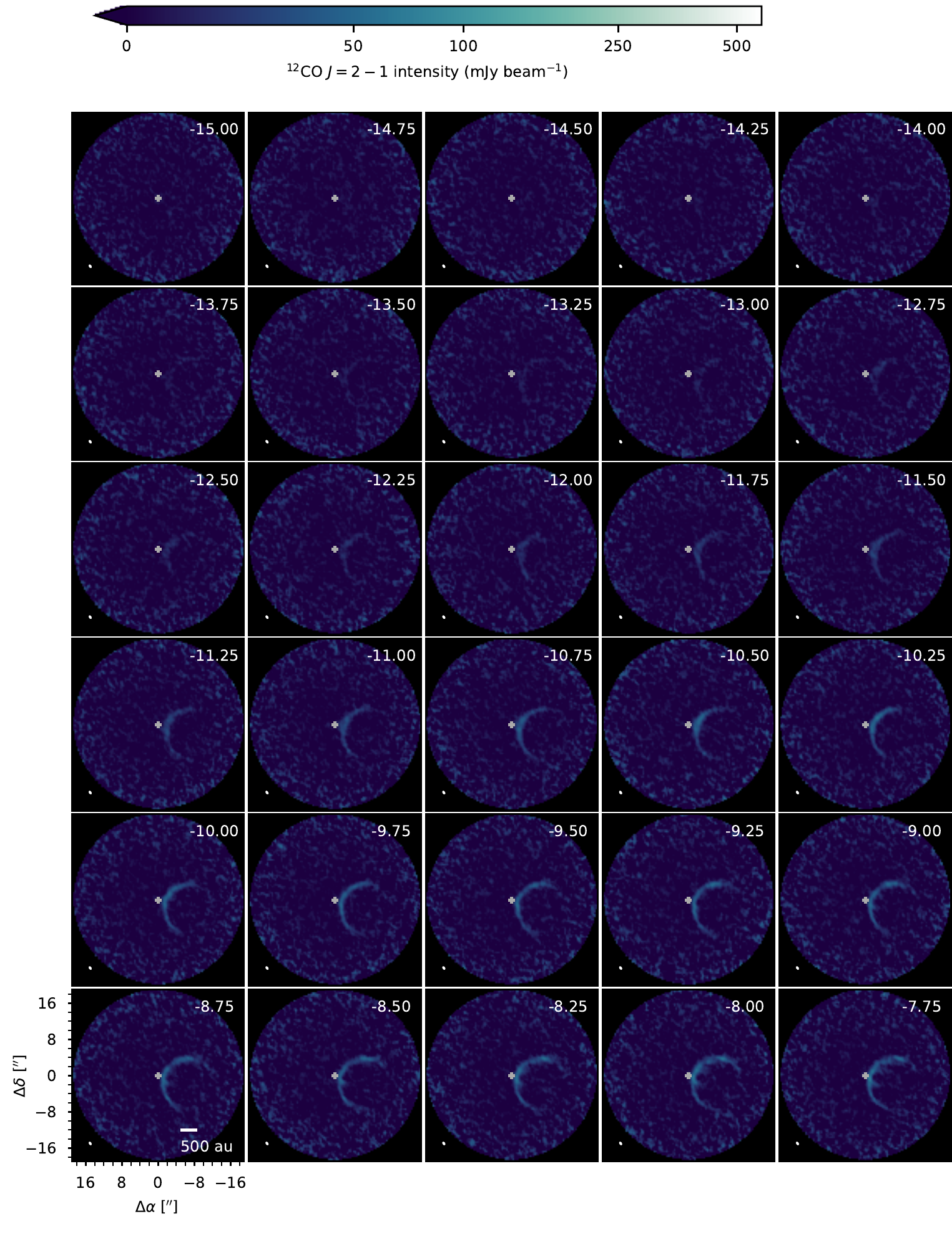}
\end{center}
\caption{Channel maps of $^{12}$CO $J=2-1$ emission toward DO Tau, part 1. The LSRK velocity (km s$^{-1}$) appears in the top right of each panel. The synthesized beam is shown as a solid white ellipse in the lower left corner. A gray cross marks the position of the millimeter continuum peak, which is located at the phase center.  Offsets from the phase center (in arcseconds) are marked in the lower left panel. North is up and east is to the left. An arcinsh stretch is applied to the color scale to highlight faint emission features. The complete figure set (8 images) is available in the online journal. \label{fig:chanmaps}}
\end{figure*}

 \begin{figure*}
 \figurenum{20.1, continued}
 \begin{center}
 \ContinuedFloat
 \includegraphics{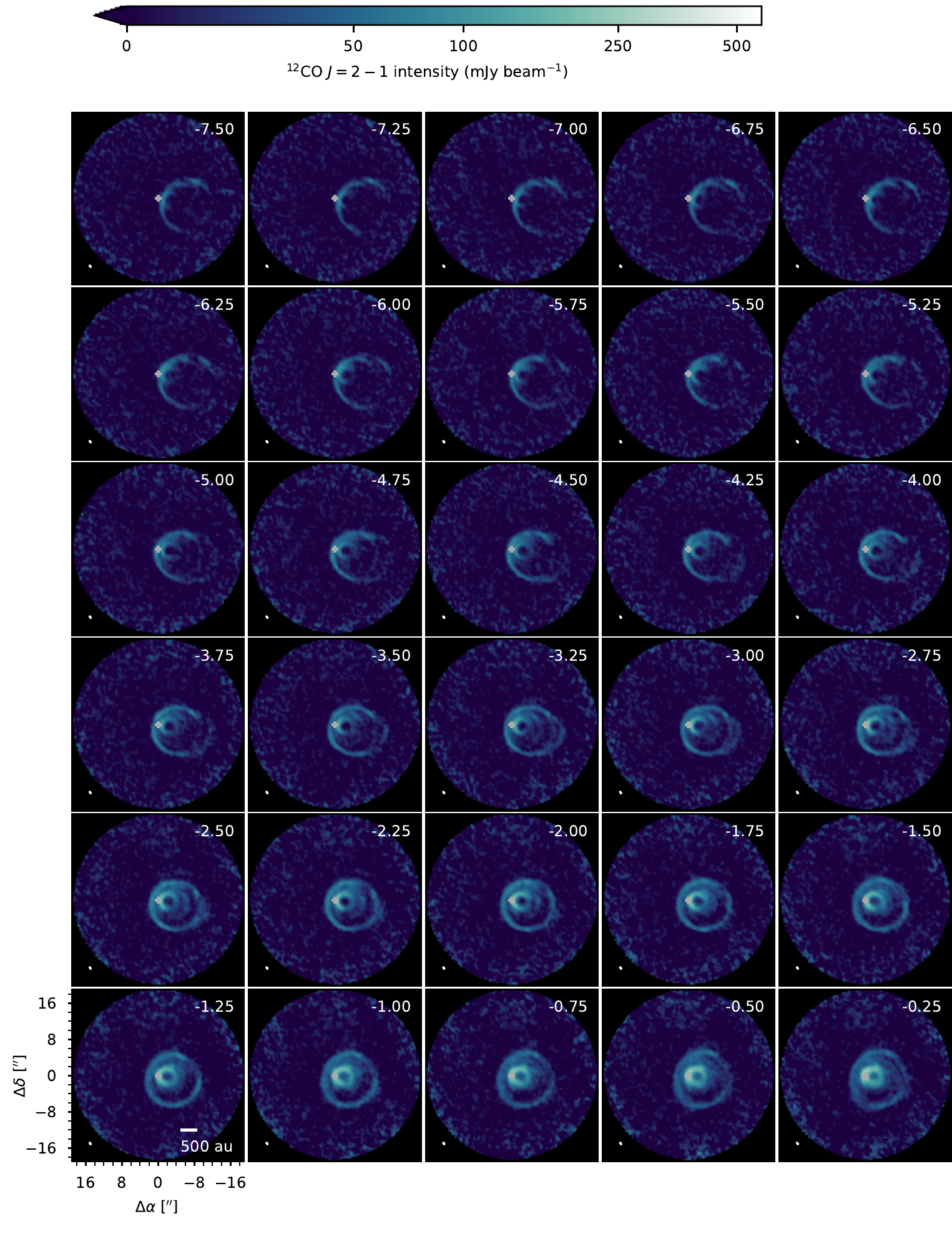}
 \end{center}
 \caption{Channel maps of $^{12}$CO $J=2-1$ emission toward DO Tau, part 2.}
 \end{figure*}

 \begin{figure*}
 \figurenum{20.1, continued}
 \begin{center}
 \ContinuedFloat
 \includegraphics{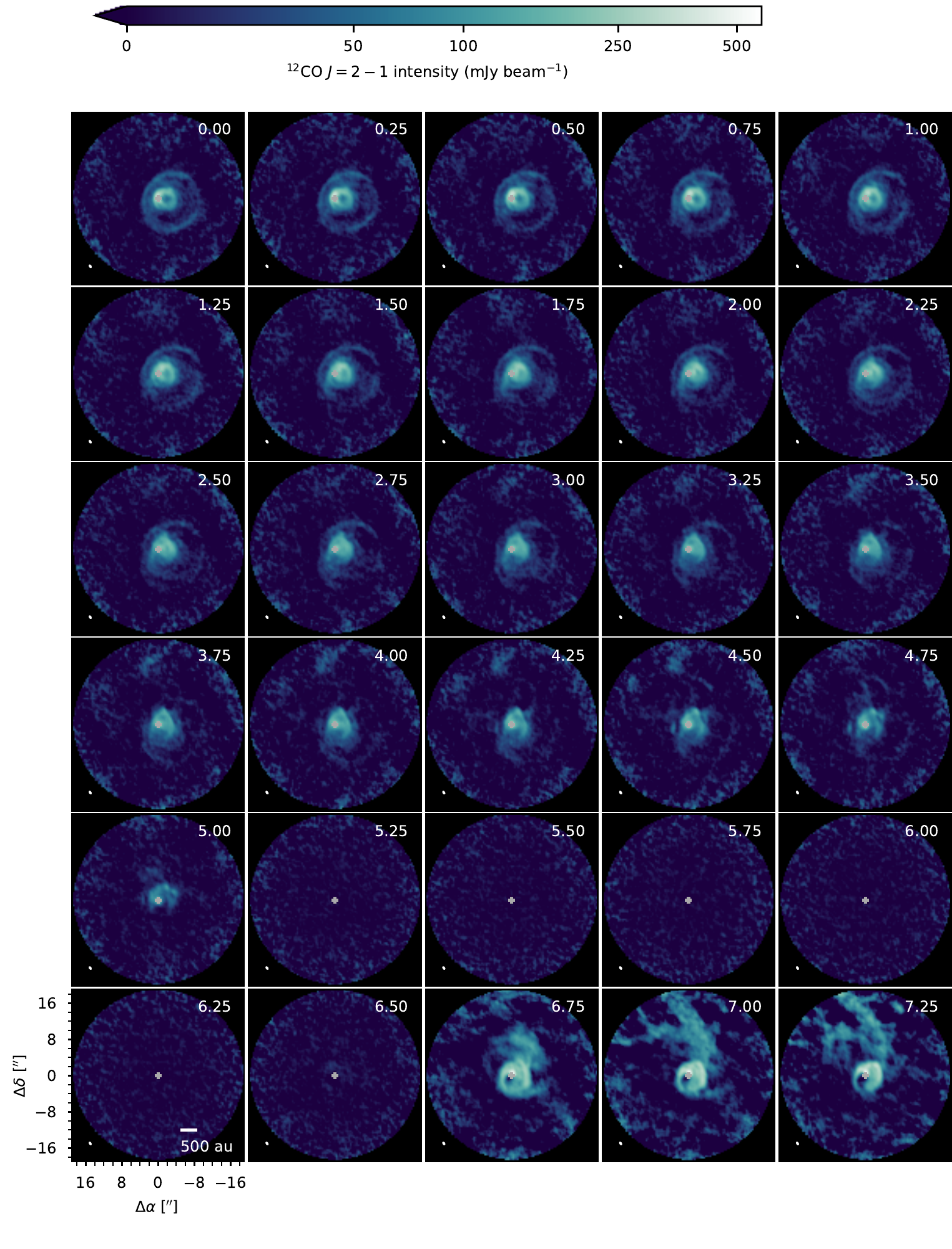}
 \end{center}
 \caption{Channel maps of $^{12}$CO $J=2-1$ emission toward DO Tau, part 3.}
 \end{figure*}

 \begin{figure*}
 \figurenum{20.1, continued}
 \begin{center}
 \ContinuedFloat
 \includegraphics{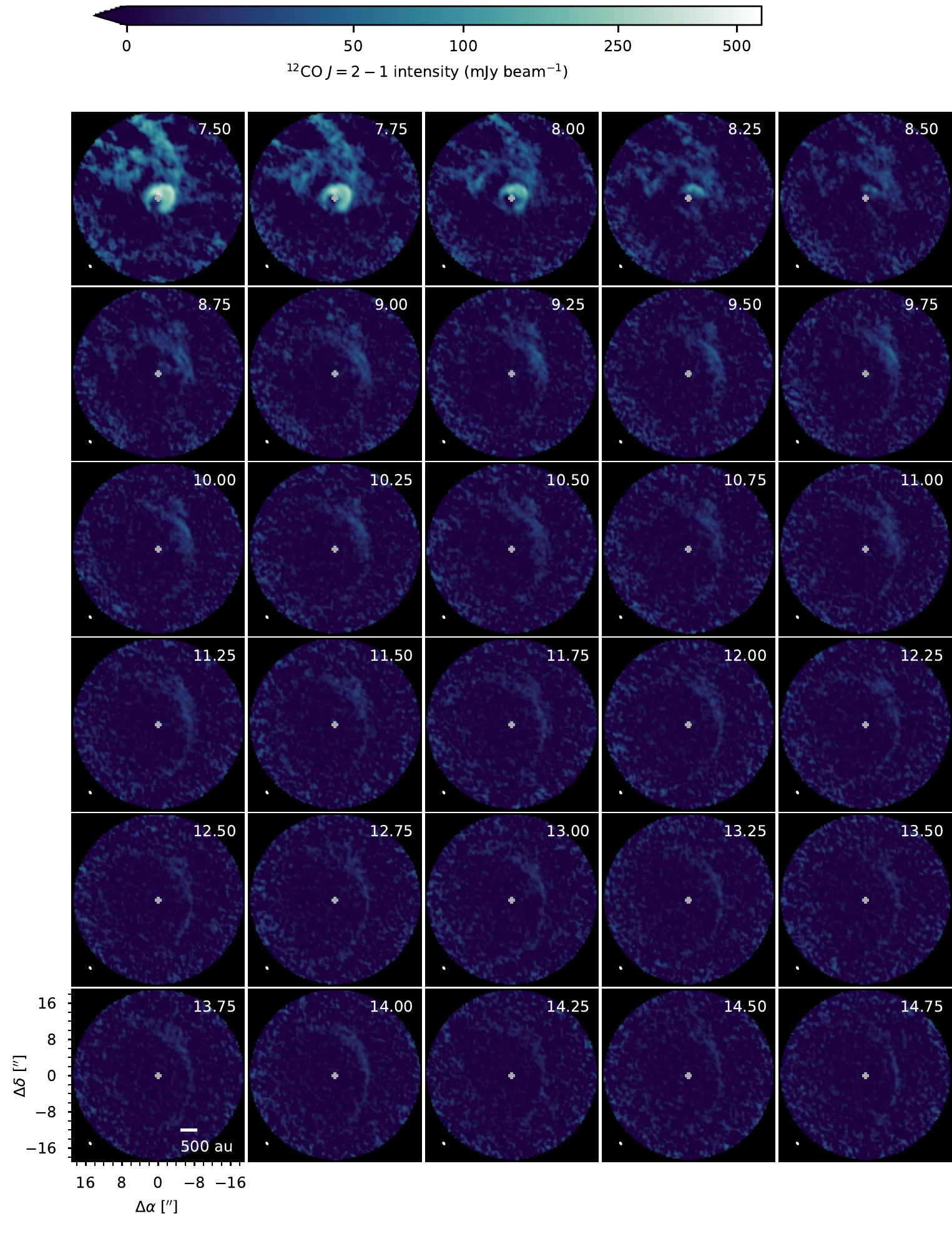}
 \end{center}
 \caption{Channel maps of $^{12}$CO $J=2-1$ emission toward DO Tau, part 4.}
 \end{figure*}

 \begin{figure*}
 \figurenum{20.1, continued}
 \begin{center}
 \ContinuedFloat
 \includegraphics{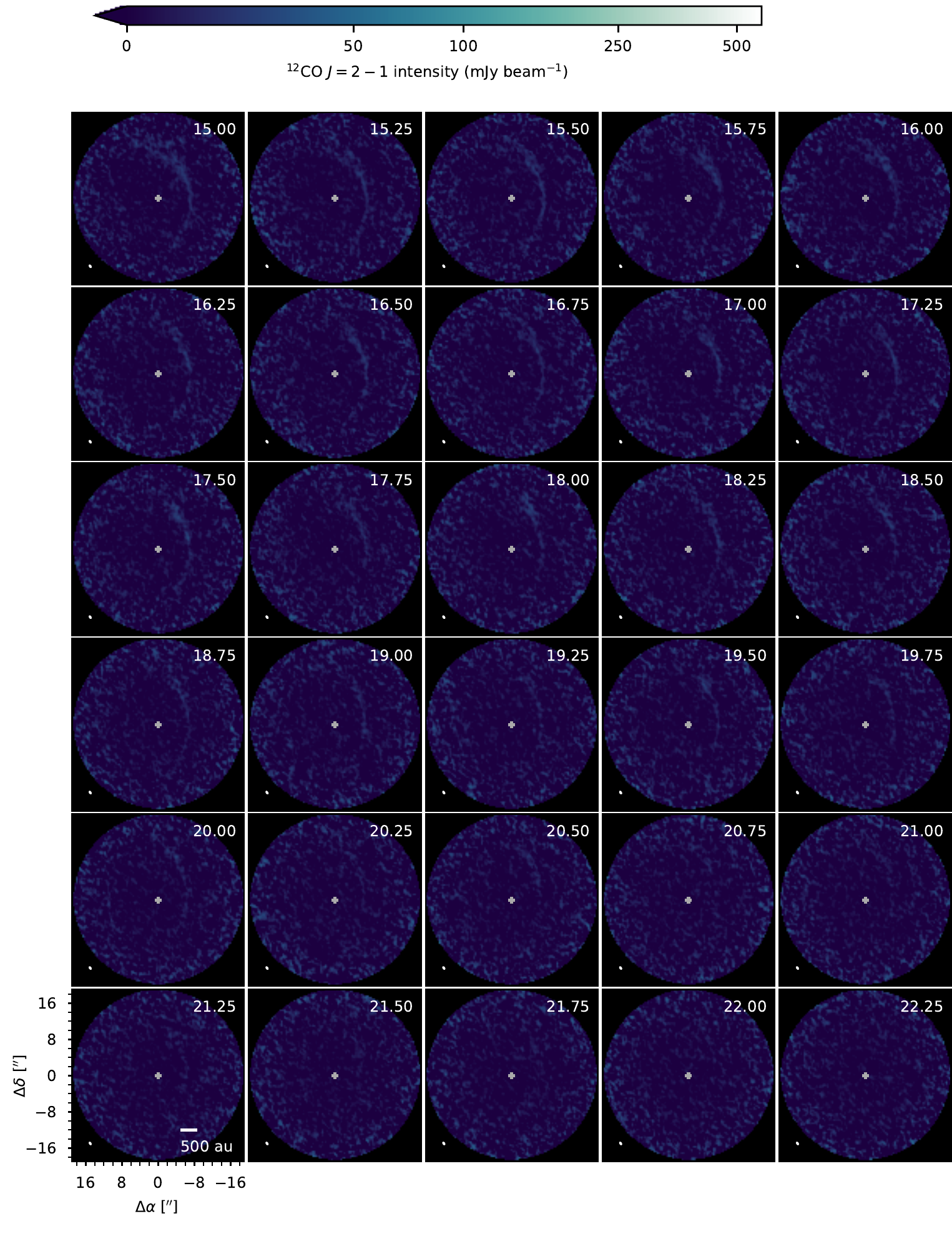}
 \end{center}
 \caption{Channel maps of $^{12}$CO $J=2-1$ emission toward DO Tau, part 5.}
 \end{figure*}

 \begin{figure*}
 \figurenum{20.2}
 \begin{center}
 \ContinuedFloat
 \includegraphics{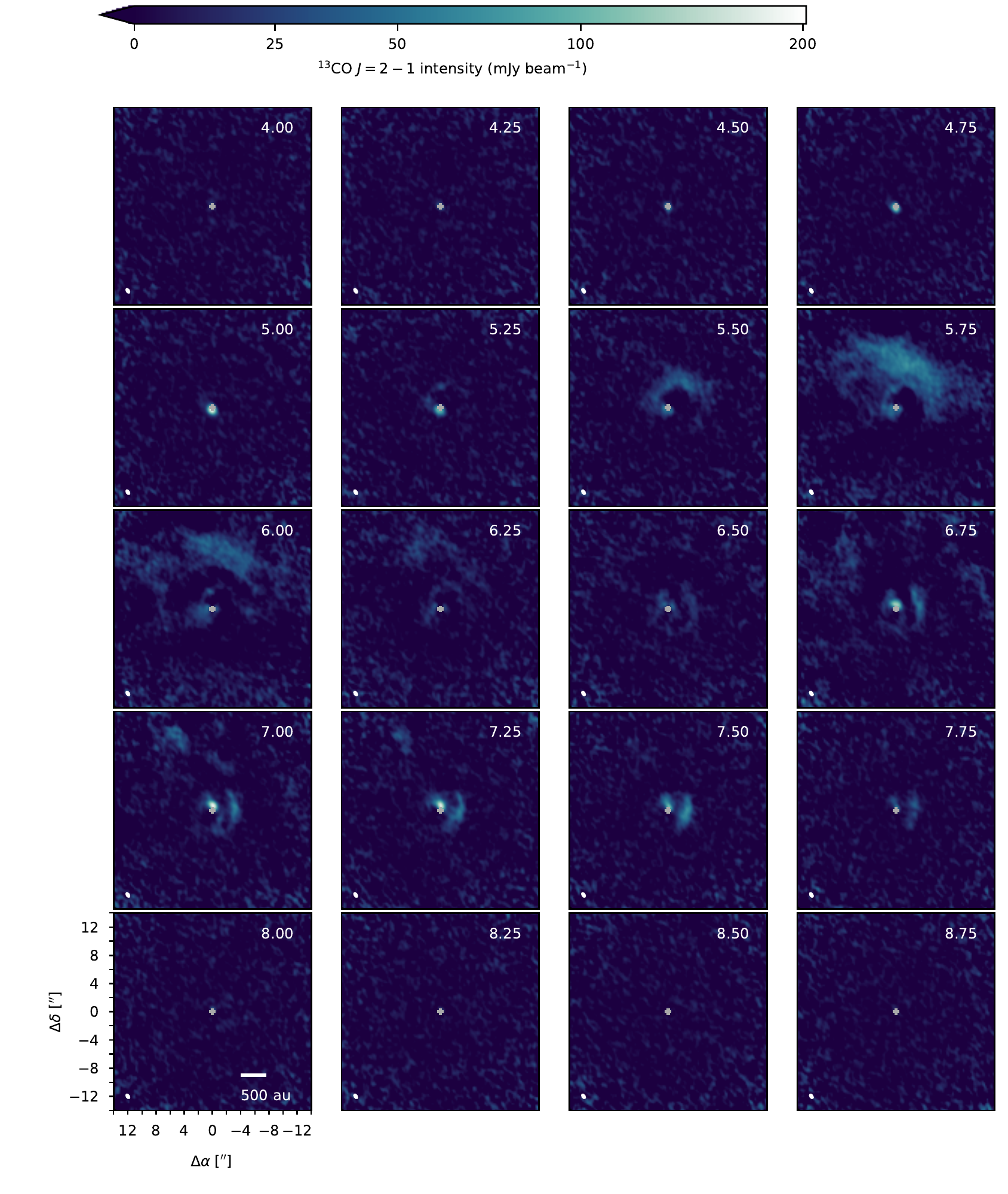}
 \end{center}
 \caption{Channel maps of $^{13}$CO $J=2-1$ emission toward DO Tau.}
 \end{figure*}

 \begin{figure*}
 \figurenum{20.3}
 \begin{center}
 \ContinuedFloat
 \includegraphics{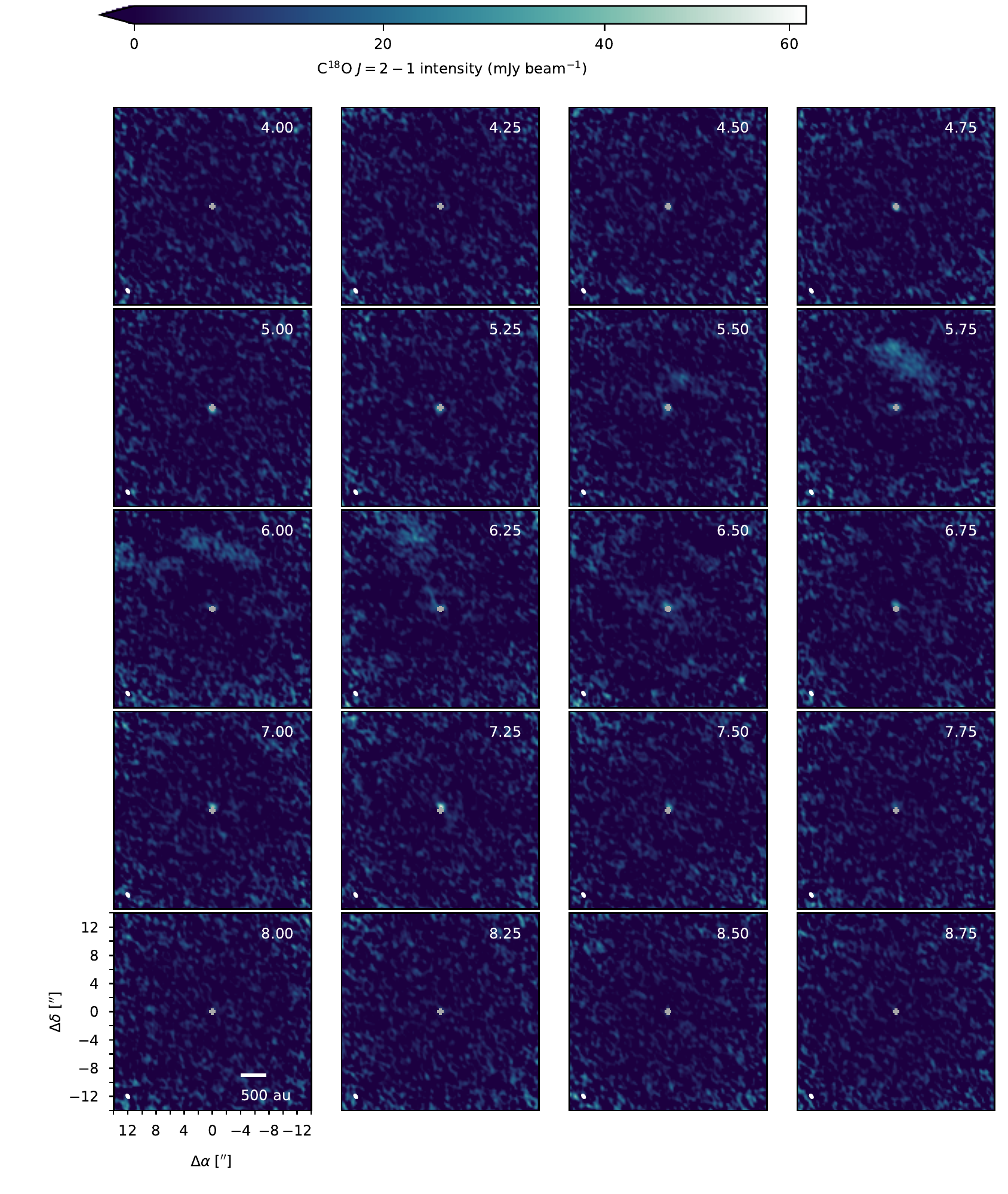}
 \end{center}
 \caption{Channel maps of C$^{18}$O $J=2-1$ emission toward DO Tau.}
 \end{figure*}

 \begin{figure*}
 \figurenum{20.4}
 \begin{center}
 \ContinuedFloat
 \includegraphics{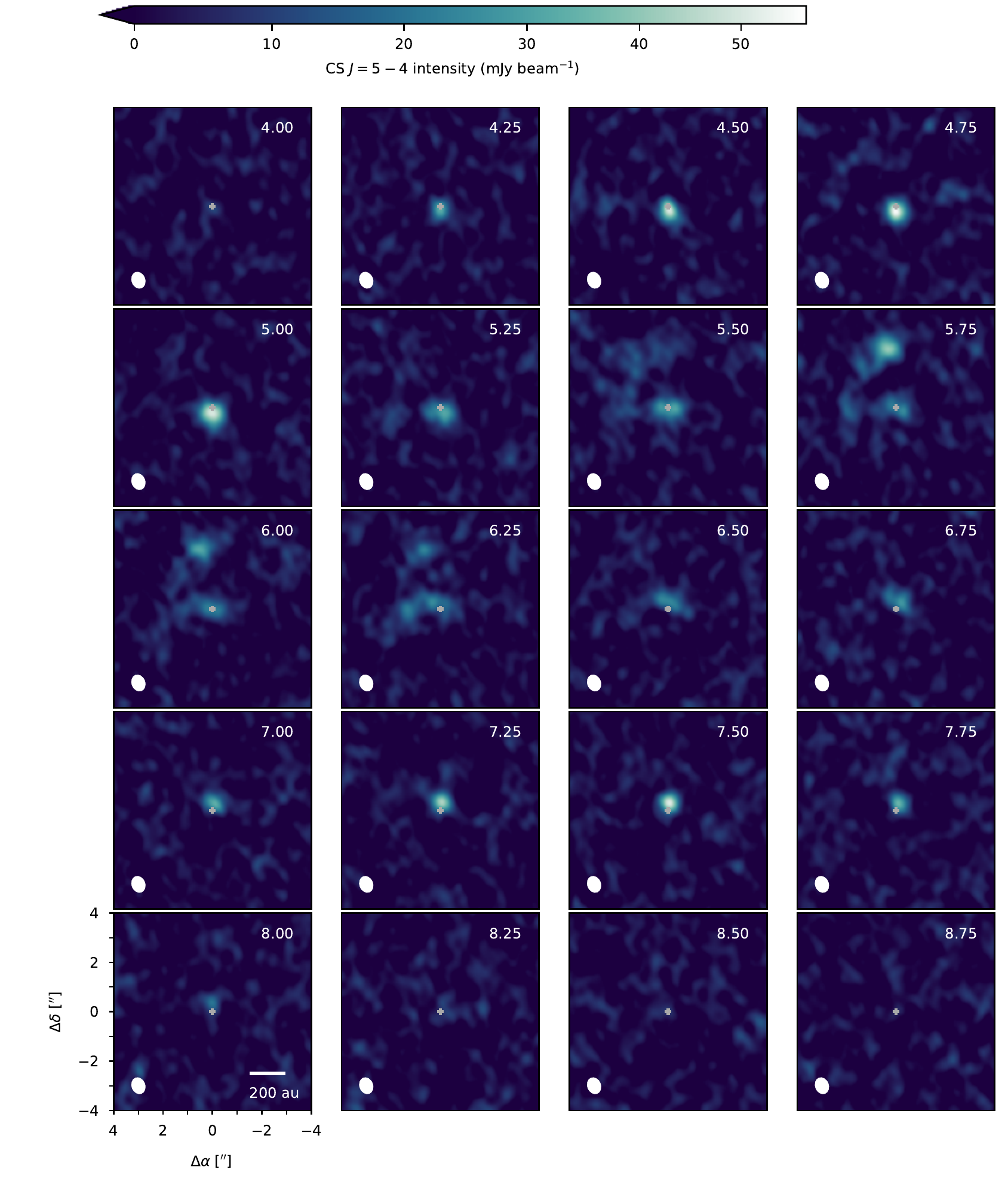}
 \end{center}
 \caption{Channel maps of CS $J=5-4$ emission toward DO Tau.}
 \end{figure*}

\end{CJK*}
\end{document}